\documentclass[12pt,a4paper]{article}
\usepackage{jcappub}
         
\usepackage{multicol,bbm,array}

\newcommand{\ud}{\mathrm{d}}
\newcommand{\ve}{\varepsilon}

\title{Cosmological dynamics with non-minimally coupled scalar field and a constant potential 
function}

\author[a]{Orest Hrycyna}
\author[b,c]{and Marek Szyd{\l}owski}
\affiliation[a]{Theoretical Physics Division, 
National Centre for Nuclear Research, \\ Ho{\.z}a 69, 00-681 Warszawa, Poland}
\affiliation[b]{Astronomical Observatory, Jagiellonian University, \\ Orla 171,
30-244 Krak{\'o}w, Poland}
\affiliation[c]{Mark Kac Complex Systems Research Centre, Jagiellonian
University, \\ {\L}ojasiewicza 11, 30-348 Krak{\'o}w, Poland}

\emailAdd{orest.hrycyna@ncbj.gov.pl}
\emailAdd{marek.szydlowski@uj.edu.pl}

\abstract{Dynamical systems methods are used to investigate global behaviour of the spatially flat Friedmann-Robertson-Walker 
cosmological model in gravitational theory with a non-minimally coupled scalar field and a constant 
potential function. We show that the system can be reduced to an autonomous three-dimensional 
dynamical system and additionally is equipped with an invariant manifold corresponding to an 
accelerated expansion of the universe. Using this invariant manifold we find an exact solution 
of the reduced dynamics.
We investigate all solutions for 
all admissible initial conditions using theory of dynamical systems to obtain a classification 
of all evolutional paths. The right-hand sides of the dynamical system depend crucially on the value of the non-minimal coupling 
constant therefore we study bifurcation values of this parameter under which the structure of 
the phase space changes qualitatively. We found a special bifurcation value of the non-minimal coupling constant which is distinguished by dynamics of the model and may suggest some additional symmetry in matter sector of the theory.
}

\arxivnumber{1506.03429}

\begin{document}

\maketitle

\section{Introduction}

In the modern cosmology there is an increasing interest in the description of a matter content of the universe in terms of a scalar field. The accelerated expansion of the universe \cite{Riess:1998cb,Perlmutter:1998np} stimulated investigations in the field of dynamical dark energy models \cite{Copeland:2006wr}. The quintessence idea was formulated \cite{Ratra:1987rm, Wetterich:1987fm} as a simplest model involving a scalar field with a potential function in order to describe the current accelerated expansion of the universe. The simplest candidate for the dark energy driving the current phase of the evolution of the universe seems to be a positive cosmological constant, however such a explanation suffers from the fine tuning problem \cite{Weinberg:1988cp, Sahni:1999gb} and the coincidence problem \cite{Sahni:1999gb, Zlatev:1998tr}. Nevertheless the straightforward application of this assumption within the general theory of relativity together with the Friedmann-Robertson-Walker symmetry leads to the $\Lambda$CDM model which is a standard model of the cosmological evolution \cite{Ade:2013zuv,Ade:2015xua} and is favoured by the observational data \cite{Kurek:2007tb,Kurek:2007gr,Szydlowski:2006ay}. Number of alternatives were proposed in order to alleviate and eliminate problems with a cosmological constant term like a phantom dark energy \cite{Caldwell:2003vq,Dabrowski:2003jm} or a extended quintessence \cite{Faraoni:2006ik, Faraoni:2000gx,Hrycyna:2007mq, Hrycyna:2007gd,Hrycyna:2008gk,Hrycyna:2009zj,Hrycyna:2010yv}. 

The inclusion into a matter sector of the gravitation theory of a non-minimal coupling term  between the gravity and the scalar field of the form $-\xi R\phi^{2}$ \cite{Chernikov:1968zm, Callan:1970ze, Birrell:1979ip} leads to natural and simplest generalisation of the scalar field Lagrangian. The motivations for this term come form different contexts. The general 
relativity has the methodological status of an effective theory with a given Lagrangian and such 
contribution naturally emerge in its expansion \cite{Donoghue:1994dn}. The 
non-minimal coupling between the scalar curvature and the scalar field appears as a result of quantum corrections to the scalar field in curved space \cite{Birrell:1984ix,Ford:1986sy,Parker:book} and is required by the renormalisation procedure \cite{Callan:1970ze}. The 
non-minimal coupling is also interesting in the context of 
superstring theory \cite{Maeda:1985bq} and induced gravity \cite{Accetta:1985du}. While the simplest inflationary model with a minimally coupled scalar field and a quadratic potential function is no longer favoured by the observational data \cite{Planck:2013jfk,Martin:2013nzq,Ade:2015lrj} there is an urgent need to extend this paradigm further beyond the minimally coupled case. 
From the theoretical point of view the coupling constant becomes a free parameter in the model which 
value should be estimated from the observational data \cite{Luo:2005ra, Szydlowski:2008zza, Atkins:2012yn, Hrycyna:2015a} or should  
be obtained from a more fundamental theory.
The model with a zero value of the non-minimal coupling was investigated in detail by Faraoni 
\cite{Faraoni:2000vg}. In this case Faraoni have found an exact solution which can be important in 
the context of an inflation description. Our investigation generalises Faraoni's results to the 
case of the constant non-minimal coupling.

The non-minimally coupled scalar field cosmology was investigated by many authors in the connection with an inflationary epoch as well as a description of the quintessence \cite{Spokoiny:1984bd, Belinsky:1985zd,
Salopek:1988qh,Amendola:1990nn, Demianski:1991zv,Fakir:1992cg,
Barvinsky:1994hx, Faraoni:1996rf, Barvinsky:1998rn, Barvinsky:2008ia, Setare:2008mb,
Setare:2008pc,Hrycyna:2008gk, Uzan:1999ch, Chiba:1999wt, Amendola:1999qq,
Perrotta:1999am, Holden:1999hm, Bartolo:1999sq, Boisseau:2000pr,
Gannouji:2006jm, Carloni:2007eu, Bezrukov:2007ep, Nozari:2007eq,
Kamenshchik:1995ib}. Uzan \cite{Uzan:1999ch} pointed out the 
significance of scaling solutions in generalised inflation scenarios. The question of integrability 
of non-minimally coupled scalar field cosmological models has been recently studied by Kamenshchik et al. \cite{Kamenshchik:2013dga} who looked 
for integrable cases with a different forms of scalar field potential function. The new dynamical system methods has been recently developed by Skugoreva et al. \cite{Skugoreva:2014gka} in the context of a more general form of coupling parametrised by the scalar field. In the standard model of particle physics a non-minimally coupled Higgs field plays also important role \cite{DeSimone:2008ei, Bezrukov:2008ej, Barvinsky:2009fy, Clark:2009dc}.

The dynamical systems methods are widely used in cosmological applications since seminal papers by Belinskii \cite{Belinskii:1985, Belinskii:1987} and the most widespread parameterisation of the phase space is by the so-called expansion normalised variables \cite{Copeland:1997et} (see collection of works on the dynamical system analysis of anisotropic models \cite{Wainwright:book}).

In the present paper we investigate a dynamical behaviour of the flat Friedmann-Robertson-Walker cosmological model with matter in the form of the non-minimally coupled scalar field with a constant potential function and a barotropic dust matter. This model is the simplest generalisation of the $\Lambda$CDM model and can be treated as a testing ground for evolutional paths of the universe.  
Adding the effects of the non-minimal coupling to the general relativity models exact 
solutions can be found only for special cases of the non-minimal coupling, namely for the minimal $\xi=0$ and the conformal $\xi=\frac{1}{6}$. The dynamical system is integrable and the exact solution can be obtained in 
the form of elliptic functions like for the standard Friedmann-Robertson-Walker models with dust, cosmological constant and radiation matter, in the case of the conformal coupling, or a stiff matter contribution in the case of minimal coupling. While there is no possibility to obtain exact solutions in general case one can apply the dynamical system 
methods for studying all evolutional paths, i.e. trajectories in the phase space. We demonstrate that for the flat Friedmann-Robertson-Walker model the dynamics can be reduced to the form of the three dimensional autonomous 
dynamical system in which the coupling constant becomes the free parameter. The main advantage of the dynamical system methods is that that without knowledge of an 
exact solution it is possible to investigated the properties of the solutions as well as their 
stability. If the right sides of the system vanish then we obtain solutions in the form of critical points representing, from the 
physical point of view, asymptotic states of the system. In the phase space, which is the space of states of the system, they are represented 
by critical points of different types. Global dynamics is 
represented by the global phase portrait which contains all information about behaviour of 
trajectories and stability of the asymptotic states. Important information about the behaviour of the system is encoded in  
the behaviour of trajectories on invariant manifolds of lower dimension, which we identify for the system and we find exact solutions for the reduced dynamics.

The complete analysis of the dynamics requires the investigation of trajectories at the 
infinity. For this aim the construction of the Poincar{\'e} sphere with a corresponding 
projection can be useful (see \cite{Perko:book,Wiggins:book}). An alternative construction which 
can be applied in the study of dynamics is to introduce the projective coordinates on the  
phase space. Then projective maps cover the dynamics at infinity. Such an analysis of the 
behaviour of trajectories at infinity is important if we are going to formulate of some 
conclusions about the global stability of the system.

 Our investigation explicitly shows a degree of complexity 
of the dynamics as the value of the non-minimal coupling constant changes. We construct the phase 
portraits for all representative cases of the non-minimal coupling constant. The global dynamics 
crucially changes if this constant assumes some critical value -- a bifurcation value. Therefore 
the global phase portraits offer us classification of a scalar field cosmology with the constant value of the non-minimal coupling and can be treated as generalisation of corresponding classification of the $\Lambda$CDM model. Of course due to conformal duality of the system under considerations to the minimally coupled case, any change of the non-minimal coupling parameter $\xi$ will lead to the change of the corresponding non-constant potential in the Einstein frame formulation of the theory. One can admit that more complicated potential functions lead to more complicated dynamics. In the Jordan frame, followed in this paper, we find that the degree of complexity changes with the value of the non-minimal coupling constant with respect to a simple minimally or conformally coupled case which from the other hand lead to simple integrable dynamics.

The analysed dynamical system poses a three dimensional phase space. In the case of a 
zero potential (the case without the cosmological constant) the right hand sides of the system are 
homogeneous functions of degree three which unable us to reduce the system to the 
two-dimensional one. In the three-dimensional phase space the dynamics is very complex and crucially depends on the constant value of the coupling. We point out that the value of $\xi=3/16$ 
is distinguished by the dynamical behaviour of the model and leads to physically interesting and appealing evolution of the universe. 

\section{The model}
\label{sec:2}

The working model consists of the spatially flat Friedmann-Robertson-Walker (FRW) universe filled with  the non-minimally coupled scalar field and a barotropic matter with the constant equation of state parameter $w_{m}$. The total action is in the following form
\begin{equation}
S =S_{g} + S_{\phi}+ S_{m}\,,
\end{equation}
where the gravitational part of the theory is described by the standard Einstein-Hilbert action integral
\begin{equation}
\label{eq:gravity}
S_{g} = \frac{1}{2\kappa^{2}}\int\ud^{4}x\sqrt{-g}\,R \,,
\end{equation}
where $\kappa^{2}=8\pi G$, and the matter part of the theory is described by the following 
action for a non-minimally coupled scalar field
\begin{equation}
\label{eq:scalar}
S_{\phi}=- \frac{1}{2}\int\ud^{4}x\sqrt{-g}\Big(\ve\nabla^{\alpha}\phi\,\nabla_{\alpha}\phi + \ve\xi R 
\phi^{2} + 2U(\phi)\Big)\,,
\end{equation}
where $\ve=+1,-1$ corresponds to the canonical and the phantom scalar field, respectively, and a barotropic matter
\begin{equation}
S_{m} = \int\ud^{4}x\sqrt{-g}\mathcal{L}_{m}\,.
\end{equation}
The field equations for the theory are
\begin{equation}
R_{\mu\nu}-\frac{1}{2}g_{\mu\nu}R=\kappa^{2}\left(T^{(\phi)}_{\mu\nu}+T^{(m)}_{\mu\nu}\right)\,,
\end{equation}
where the energy-momentum tensor for the non-minimally coupled scalar field is given by
\begin{equation}
\label{eq:energy_mom}
\begin{split}
T^{(\phi)}_{\mu\nu}= \,\,& \ve \nabla_{\mu}\phi\nabla_{\nu}\phi -
\ve\frac{1}{2}g_{\mu\nu}\nabla^{\alpha}\phi\nabla_{\alpha}\phi -
U(\phi)g_{\mu\nu} + \\ & + \ve\xi\Big(R_{\mu\nu}-\frac{1}{2}g_{\mu\nu}R\Big)\phi^{2} +
\ve\xi\Big(g_{\mu\nu}\Box\phi^{2}-\nabla_{\mu}\nabla_{\nu}\phi^{2}\Big)\,,
\end{split}
\end{equation}
and for the barotropic matter included in the model
\begin{equation}
T^{(m)}_{\mu\nu}= \big(\rho_{m}+p_{m}\big)u_{\mu}u_{\nu} + p_{m}g_{\mu\nu}\,.
\end{equation}

Let us note that when the non-minimal coupling between the scalar field and the curvature is present the field equations can be written in several nonequivalent ways. In the case adopted here the energy momentum tensor for the scalar field \eqref{eq:energy_mom} is covariantly conserved, which may not be true for other choices \cite{Faraoni:2000wk, Faraoni:book, Capozziello:book}. For example, the redefinition of
gravitational constant $\kappa_{\text{eff}}^{-2}=\kappa^{-2}-\ve\xi\phi^{2}$ makes it
time dependent. The effective gravitational constant can diverge for the critical
value of the scalar field $\phi_{c}=\pm(\ve\xi\kappa^{2})^{-1/2}$. Though the FRW
model remains regular at this point, the model is unstable with respect to
arbitrary small anisotropic and inhomogeneous perturbations which become
infinite there. This results in the formation of a strong curvature singularity
prohibiting a transition to the region $\kappa_{\text{eff}}^{2}<0$
\cite{Starobinsky:1981,Figueiro:2009mm}.

In cosmological applications one can obtain substantial mathematical simplification using conformal transformation techniques, especially in the absence of ordinary matter, in order to relate cosmological models with a non-minimally coupled scalar field with its conformal counterpart with a minimally coupled field. The Jordan frame action integral, where the scalar field is non-minimally coupled to the Ricci scalar curvature is mapped into an Einstein frame where now the transformed scalar field $\tilde{\phi}$ is minimally coupled. The two frames are physically nonequivalent unless variable units of time, length, and mass are adopted in the Einstein frame \cite{Faraoni:book, Capozziello:book}. 

In the theory under considerations using the conformal factor $\Omega^{2} = 1-\ve\xi\kappa^{2}\phi^{2}$ the Einstein frame counterpart of the action integrals \eqref{eq:gravity} and \eqref{eq:scalar} is
$$
S^{(E)}_{g+\phi} = \frac{1}{2}\int\ud^{4}x\sqrt{-\tilde{g}}\left(\frac{1}{\kappa^{2}}\tilde{R}-\ve\frac{1-\ve\xi(1-6\xi)\kappa^{2}\phi^{2}}{(1-\ve\xi\kappa^{2}\phi^{2})^{2}}\tilde{g}^{\mu\nu}\nabla_{\mu}\phi\,\nabla_{\nu}\phi-2\frac{U(\phi)}{(1-\ve\xi\kappa^{2}\phi^{2})^{2}}\right)\,,
$$
and the scalar field can be transformed as
$$
\ud\tilde{\phi}=\frac{\sqrt{\ve\big(1-\ve\xi(1-6\xi)\kappa^{2}\phi^{2}\big)}}{1-\ve\xi\kappa^{2}\phi^{2}}\ud\phi\,.
$$
The condition for the theory given by this action integral to be ghost free is 
$$
\ve\big(1-\ve\xi(1-6\xi)\kappa^{2}\phi^{2}\big)>0\,,
$$
provided that $\Omega^{2}=1-\ve\xi\kappa^{2}\phi^{2}>0$, in this sense the conformal transformation technique cannot provide solutions with the scalar $\phi$ crossing the barriers $\phi_{c}=\pm(\ve\xi\kappa^{2})^{-1/2}$ and has the same limitations of the form of the field equations using the effective coupling $\kappa_{\text{eff}}^{-2}=\kappa^{-2}-\ve\xi\phi^{2}$ \cite{Faraoni:book, Capozziello:book}.

In the present paper we work exclusively in the Jordan frame formulation of the theory leaving aside the contentious question concerning the physical equivalence between the Jordan frame and the Einstein frame \cite{Faraoni:1999hp, Kamenshchik:2014waa}.

From now on we assume that the barotropic matter in the form of a dust matter and taking the spatially flat FRW metric
$$
\ud s^{2}=-\ud t^{2}+a^{2}(t)\Big(\ud x^{2}+\ud y^{2}+\ud z^{2}\Big)\,,
$$
we obtain the energy conservation condition in the standard form
\begin{equation}
\label{eq:constr}
\frac{3}{\kappa^{2}}H^{2} = \rho_{\phi} + \rho_{m}
\end{equation}
where $\rho_{m}$ is the energy density of the barotropic dust matter and
$$
\rho_{\phi} = \ve\frac{1}{2}\dot{\phi}^{2}+ U_{0} + \ve3\xi H^{2}\phi^{2} +
\ve6\xi H\phi\dot{\phi}
$$
is the energy density of the non-minimally coupled scalar field and we have assumed a constant potential function $U(\phi)=U_{0}=\text{const}$. The dynamical equation for the scalar field is
$$
\ddot{\phi}+3H\dot{\phi}+\xi R \phi=0\,.
$$

Now, we divide the constraint equation \eqref{eq:constr} by the present value of the Hubble 
function $H_{0}=H(a_{0})$ and we obtain
\begin{equation}
\label{eq:Hsqr}
\left(\frac{H}{H_{0}}\right)^{2} = \Omega_{\Lambda,0}+
\Omega_{m,0}\left(\frac{a}{a_{0}}\right)^{-3} +
\ve(1-6\xi)x^{2} + \ve6\xi(x+z)^{2}
\end{equation}
where
$$
\Omega_{\Lambda,0}\equiv \frac{\kappa^{2} U_{0}}{3 H_{0}^{2}}\,,\qquad
\Omega_{m,0} \equiv \frac{\kappa^{2}\rho_{m,0}}{3 H_{0}^{2}}
$$
and we introduced new dimensionless variables, which are the energy phase space 
variables in a slightly different form,
\begin{equation}
\label{eq:ps_variables}
x \equiv \frac{\kappa}{\sqrt{6}}\frac{\dot{\phi}}{H_{0}}\,, \qquad
z \equiv \frac{\kappa}{\sqrt{6}}\frac{H}{H_{0}}\phi\,.
\end{equation}

The acceleration equation expressed in the new variables is
$$
\frac{\dot{H}}{H^{2}} = -2 +
\frac{-\ve(1-6\xi)x^{2}+2\Omega_{\Lambda,0} +
\frac{1}{2}\Omega_{m,0}\big(\frac{a}{a_{0}}\big)^{-3}}
{\big(\frac{H}{H_{0}}\big)^{2}-\ve6\xi(1-6\xi)z^{2}}
$$
and using the energy conservation condition one can eliminate the $\Omega_{m,0}$ term.

The dynamical system on variables $x$, $z$ and $h=\frac{H}{H_{0}}$ is in the
following form
\begin{equation}
\label{eq:dynsys1}
\begin{split}
\frac{\ud x}{\ud \ln{a}} &= -3x-6\xi z\bigg(\frac{\dot{H}}{H^{2}}+2\bigg)\,,\\
\frac{\ud z}{\ud\ln{a}} &= x+z\frac{\dot{H}}{H^{2}}\,, \\
\frac{\ud h}{\ud\ln{a}} &= h \frac{\dot{H}}{H^{2}}\,,
\end{split}
\end{equation}
where
\begin{equation}
\label{eq:accel1}
	\frac{\dot{H}}{H^{2}} = -2
	+\frac{3}{2}\frac{\Omega_{\Lambda,0}+\frac{1}{3}h^{2}-
	\ve(1-6\xi)x^{2}-\ve2\xi(x+z)^{2}}{h^{2}-\ve6\xi(1-6\xi)z^{2}} \,.
\end{equation}
From the last equation in the dynamical system \eqref{eq:dynsys1}, one can easily notice, that 
the system has two invariant manifolds
$h=H/H_{0}=0$ and $\dot{H}/H^{2} =0$. This second property will be used later in the
paper in order to obtain an explicit solution of a restricted dynamical system.

We clearly see that the right hand sides of the dynamical system
\eqref{eq:dynsys1} are rational functions of their arguments due to form of the acceleration 
equation \eqref{eq:accel1}. To remove this degeneracy, we perform the following time transformation
\begin{equation}
\label{eq:time1}
(h^{2}-\ve6\xi(1-6\xi)z^{2})\frac{\ud}{\ud\ln{a}} = \frac{\ud}{\ud\tau}
\end{equation}
and the dynamical system describing the model under investigations is now in the following form
\begin{equation}
\label{eq:dynsys2}
\begin{split}
\frac{\ud x}{\ud \tau} & =   - 3x\big(h^{2}-\ve6\xi(1-6\xi)z^{2}\big) -\\  & - 9\xi
z\Big(\Omega_{\Lambda,0}+\frac{1}{3}h^{2}-\ve(1-6\xi)x^{2}
-\ve2\xi(x+z)^{2} \Big)\,,\\
\frac{\ud z}{\ud \tau} & =   x\big(h^{2}-\ve6\xi(1-6\xi)z^{2}\big) + \\ &  +
\frac{3}{2}z\Big(\Omega_{\Lambda,0} - h^{2}-\ve(1-6\xi)x^{2}-\ve2\xi(x+z)^{2}+
\ve8\xi(1-6\xi)z^{2}\Big)\,,\\
\frac{\ud h}{\ud \tau} & =   \frac{3}{2}h \Big(\Omega_{\Lambda,0} - h^{2}
-\ve(1-6\xi)x^{2}-\ve2\xi(x+z)^{2} + \ve8\xi(1-6\xi)z^{2}\Big)\,
\end{split}
\end{equation}
where all functions on the right hand side are polynomials and the time
transformation \eqref{eq:time1} removes singularities of the system
\eqref{eq:dynsys1}, where $h^{2} - \ve6\xi(1-6\xi)z^{2}=0$. 

The dynamical system \eqref{eq:dynsys2} holds a very special property. For the vanishing 
potential of the scalar field $\Omega_{\Lambda,0}=0$ the right hand sides of this system are 
homogeneous polynomial of degree $3$ in the phase space variables. In such case the projective 
coordinates can be used in order to lower dimension of the dynamical system and decouple one of 
the equations \cite{Perko:book,Wiggins:book}. This property will be widely used in subsection 
\ref{subsec:atinfty} while analysing dynamical behaviour of the system at infinity of the phase 
space.
 
Before we proceed to dynamical analysis of the model, we need to consider two special values of 
the non-minimal coupling constant $\xi$, namely, minimal $\xi=0$ and conformal coupling $\xi=\frac{1}{6}$ values.

Let consider first the case of the minimal coupling. For $\xi=0$ from \eqref{eq:Hsqr} we have
\begin{equation}
\label{eq:H_min}
\left(\frac{H(a)}{H(a_{0})}\right)^{2}\bigg|_{\xi=0} = \Omega_{\Lambda,0} +
\Omega_{m,0}\left(\frac{a}{a_{0}}\right)^{-3} + \ve\, x(a)^{2}
\end{equation}
and the system \eqref{eq:dynsys1} reduces to single equation
$$
\frac{\ud x}{\ud \ln{a}} = -3 x\,,
$$
which can be easily integrated resulting in
$$
x(a) = x(a_{0})\left(\frac{a}{a_{0}}\right)^{-3}\,.
$$
where the initial condition $x(a_{0})$ is taken at the present epoch.
Now, inserting this solution in to \eqref{eq:H_min} we obtain the following Hubble function
\begin{equation}
	\label{eq:Hmin}
	\left(\frac{H(a)}{H(a_{0})}\right)^{2}\bigg|_{\xi=0} = \Omega_{\Lambda,0} +
	\Omega_{m,0}\left(\frac{a}{a_{0}}\right)^{-3} + \ve\,
	x_{0}^{2}\left(\frac{a}{a_{0}}\right)^{-6}
\end{equation}
which describes the background dynamics of the spatially flat FRW model with the minimally coupled scalar field. The last term indicated contribution from the scalar field and has a similar form as a stiff matter with the equation of state parameter $w_{m}=1$. Depending on the type of scalar field (canonical $\ve=+1$ or phantom $\ve=-1$) this contribution has positive or negative energy density. The observational constraints on the such modified Hubble function were previously investigated in \cite{Szydlowski:2008zz} leading to the conclusion that this model is indistinguishable from the $\Lambda$CDM model.

The second case considered is the conformal coupling. For $\xi=\frac{1}{6}$ from \eqref{eq:Hsqr} we have
\begin{equation}
\label{eq:H_conf}
\left(\frac{H(a)}{H(a_{0})}\right)^{2}\bigg|_{\xi=\frac{1}{6}} = \Omega_{\Lambda,0} +
\Omega_{m,0}\left(\frac{a}{a_{0}}\right)^{-3} + \ve\,(x(a)+z(a))^{2}
\end{equation}
and the system \eqref{eq:dynsys1} reduces to
$$
\frac{\ud(x+z)}{\ud \ln{a}} = -2(x+z)\,,
$$
which can be explicitly integrated resulting in
$$
x(a) + z(a) = (x(a_{0})+z(a_{0}))\left(\frac{a}{a_{0}}\right)^{-2}\,,
$$
where the initial conditions $x(a_{0})$ and $z(a_{0})$ are taken at the present epoch. Inserting this solution in to \eqref{eq:H_conf} we obtain
\begin{equation}
	\label{eq:Hconf}
	\left(\frac{H(a)}{H(a_{0})}\right)^{2}\bigg|_{\xi=\frac{1}{6}} = \Omega_{\Lambda,0} +
        \Omega_{m,0}\left(\frac{a}{a_{0}}\right)^{-3} + \ve\,
	(x_{0}+z_{0})^{2}\left(\frac{a}{a_{0}}\right)^{-4}\,.
\end{equation}
The resulting Hubble function describes the standard $\Lambda$CDM model with additional term coming from the conformally coupled scalar field which behaves like a radiation in the model. The energy density of this radiation-like term is positive for a canonical scalar field and is negative for a phantom scalar field. Recently, Boisseau et al. found a bouncing solution with conformal coupling tending to dS space and containing dark radiation \cite{Boisseau:2015hqa}.

Using specific assumptions regarding vanishing of the scalar field potential function or the barotropic matter in the model the both of Hubble functions \eqref{eq:Hmin} and \eqref{eq:Hconf} can be integrated in exact form in terms of elliptic functions \cite{Chen:2014fqa,Chen:2015kza}.

\section{The non-minimal coupling and the structure of the phase space}

\subsection{Critical points and stability}

In dynamical system theory phase space of a system is organised by asymptotic states, i.e. 
critical points and trajectories joining them. In section \ref{sec:2} we were able to reduce 
dynamics of the model under investigations to the $3-$dimensional dynamical system 
\eqref{eq:dynsys2} where right hand sides are polynomials in the phase space variables. 
Redefinition of a time variable made possible to move from rational functions to polynomial 
functions of the phase space variables and remove singularities of the dynamics. The price to 
pay is a new time function along the phase space trajectories. 

In this subsection we are interested in a local behaviour of the dynamics which we want to 
directly connect with the Hubble function describing the background evolution of the 
cosmological model. The most important structures in the dynamical system theory are asymptotic states of the dynamics in the form of critical points. Our analysis we begin from the system \eqref{eq:dynsys1} with the natural logarithm of the scale factor as time variable along the phase space trajectories
\begin{equation}
\label{eq:dynsys_new}
\begin{split}
\frac{\ud x}{\ud \ln{a}} &= -3x-6\xi z\bigg(\frac{\dot{H}}{H^{2}}+2\bigg)\,,\\
\frac{\ud z}{\ud\ln{a}} &= x+z\frac{\dot{H}}{H^{2}}\,, \\
\frac{\ud h}{\ud\ln{a}} &= h \frac{\dot{H}}{H^{2}}\,,
\end{split}
\end{equation}
where
\begin{equation}
\label{eq:accel_new}
	\frac{\dot{H}}{H^{2}} = -2
	+\frac{3}{2}\frac{\Omega_{\Lambda,0}+\frac{1}{3}h^{2}-
	\ve(1-6\xi)x^{2}-\ve2\xi(x+z)^{2}}{h^{2}-\ve6\xi(1-6\xi)z^{2}} \,.
\end{equation}

In what follows we assume that the non-minimal coupling constant $\xi$ is different from the 
minimal and conformal coupling values $\xi\ne0,\frac{1}{6}$ and that the phase space variables 
$z$ and $h$ are such that there is no singularity in the denominator of the acceleration 
equation \eqref{eq:accel_new}.

Dynamical system \eqref{eq:dynsys_new} admits four non-singular critical points at a finite domain 
of the phase space. Two of them located at 
\begin{equation}
(x^{*}=0\,,z^{*}=0\,,h^{*}=\pm\sqrt{\Omega_{\Lambda,0}})\,,
\end{equation}
correspond to the de Sitter and anti-de Sitter states respectively, with the vanishing acceleration 
equation \eqref{eq:accel_new}. The only condition for the existence of those points is $
\Omega_{\Lambda,0}>0$ and it does not depend on a value of the non-minimal coupling constant $\xi$.

The remaining two asymptotic states located at
\begin{equation}
\left(x^{*}=\mp4\xi\sqrt{-\frac{\Omega_{\Lambda,0}}{\ve2\xi(1-6\xi)(3-10\xi)}}\,,
z^{*}=\pm(1-2\xi)\sqrt{-\frac{\Omega_{\Lambda,0}}{\ve2\xi(1-6\xi)(3-10\xi)}}\,, h^{*}=0\right)\,
\end{equation}
correspond to the Einstein static solutions with the vanishing Hubble function and the asymptotic value of the effective equation of state parameter $w_{\text{eff}}=-1-\frac{2}{3}\frac{4\xi}{1-2\xi}$.
The conditions for existence are $-\frac{\Omega_{\Lambda,0}}{\ve2\xi(1-6\xi)(3-10\xi)}>0$ 
together with the non-minimal coupling constant different form $\xi\ne\{0,\frac{1}{6},\frac{3}{10},\frac{1}{2}\}$. Explicitly this leads to the following conditions for values of the cosmological constant and the non-minimal coupling 
\begin{equation}
\label{eq:exis_2}
{\renewcommand{\arraystretch}{1.5}
\begin{matrix}
\Omega_{\Lambda,0}>0\,:&\quad\ve=-1\,,&\quad 0<\xi<\frac{1}{6}& \quad \vee \quad &\quad\xi>
\frac{3}{10}\,,\\
	                   &\quad\ve=+1\,, &\quad\xi<0 & \quad \vee \quad & \quad\frac{1}{6}<\xi<
\frac{3}{10}\,,\\
\Omega_{\Lambda,0}<0\,:&\quad\ve=-1\,,&\quad\xi<0 & \quad  \vee \quad &\quad\frac{1}{6}<\xi<
\frac{3}{10}\,,\\
	                   &\quad\ve=+1\,, &\quad0<\xi<\frac{1}{6} & \quad \vee \quad&\quad\xi>
\frac{3}{10}\,.
\end{matrix}}
\end{equation}

Now we can proceed to finding linearised solutions in the vicinity of those points and to obtain a physical interpretation of the dynamics. 
The de Sitter and anti-de Sitter states are given by the following critical points
\begin{equation}
(x^{*}=0\,,z^{*}=0\,,(h^{*})^{2}=\Omega_{\Lambda,0})\,,
\end{equation}
in terms of the original scalar field variables it is
$$
(\dot{\phi}^{*}=0\,, \phi^{*}=0\,, H^{*}=\text{const.})\,.
$$ 
The linearisation matrix calculated at this point lead to the following eigenvalues
\begin{equation}
\lambda_{1,2}=-\frac{3}{2}\mp\frac{3}{2}\sqrt{1-\frac{16}{3}\xi}\,,\quad \lambda_{3}=-3\,.
\end{equation}
The stability condition for this critical point is that the real parts of $\lambda_{1}$ and $
\lambda_{2}$ eigenvalues must be negative $\mathrm{Re}(\lambda_{1})<0$ and $\mathrm{Re}
(\lambda_{2})<0$. This leads to the following condition $\xi>0$, the non-minimal coupling 
constant $\xi$ must be positive in order to obtain the asymptotically stable de Sitter state. In 
details we have
\begin{equation}
\label{eq:stab_1}
{\renewcommand{\arraystretch}{1.5}
\begin{matrix}
\textrm{a stable node :} & 0<\xi<\frac{3}{16}\,,\\
\textrm{a stable focus (a sink) :} & \xi>\frac{3}{16}\,,
\end{matrix}
}
\end{equation}
while for $\xi<0$ the de Sitter state is a transient state of a saddle type critical point with $
\lambda_{1}<0$\,, $\lambda_{2}>0$ and $\lambda_{3}<0$.
The linearised solutions to the dynamics in the vicinity of the de Sitter and anti-de Sitter states are
\begin{equation}
\begin{split}
x(a) & = \frac{1}{\lambda_{1}-\lambda_{2}}\left((\lambda_{1}\,\Delta x-12\xi\,\Delta z)
\left(\frac{a}{a^{(i)}}\right)^{\lambda_{1}} - (\lambda_{2}\,\Delta x-12\xi\,\Delta z)
\left(\frac{a}{a^{(i)}}\right)^{\lambda_{2}}\right)\,,\\
z(a) & = \frac{1}{\lambda_{1}-\lambda_{2}}\left((\Delta x-\lambda_{2}\,\Delta z)\left(\frac{a}
{a^{(i)}}\right)^{\lambda_{1}} - (\Delta x-\lambda_{1}\,\Delta z)\left(\frac{a}{a^{(i)}}
\right)^{\lambda_{2}}\right)\,,\\
h(a) & = h^{*}+\Delta h \left(\frac{a}{a^{(i)}}\right)^{-3}\,,
\end{split}
\end{equation}
where $\lambda_{1}$ and $\lambda_{2}$ are the two out of three eigenvalues of the linearisation matrix and $\Delta x= x^{(i)}$, $\Delta z = z^{(i)}$ and $\Delta h = h^{(i)}-h^{*}$ are the initial conditions. 

From the last equation of this system we directly obtain that up to linear terms in initial conditions the Hubble function is
\begin{equation}
\left(\frac{H(a)}{H(a_{0})}\right)^{2} \approx \Omega_{\Lambda,0}+\Omega_{m,0}\left(\frac{a}{a_{0}}\right)^{-3}\,,
\end{equation}
and is indistinguishable from the Hubble function for the standard $\Lambda$CDM model.

The critical points corresponding to the static Einstein solutions are give by
\begin{equation}
\left(x^{*}=-\frac{4\xi}{1-2\xi}z^{*}\,, (z^{*})^{2}=-\ve(1-2\xi)^{2}\frac{\Omega_{\Lambda,0}}
{2\xi(1-6\xi)(3-10\xi)}\,, h^{*}=0\right)\,,
\end{equation}
and in the original scalar field variables this is equivalent to
$$
(\dot{\phi}^{*}=\text{const.}\,,\phi^{*}=\pm\infty\,, H^{*}=0)\,.
$$
The linearisation matrix gives the eigenvalues
\begin{equation}
\lambda_{1}=-3\,, \quad \lambda_{2}=-5+\frac{2}{1-2\xi}\,,\quad \lambda_{3}=\frac{4\xi}{1-2\xi}
\,,
\end{equation}
which constitute stability conditions for the state. Matching the existence conditions \eqref{eq:exis_2} we obtain the following stability conditions for
a stable node :  $\lambda_{1}<0\,, \lambda_{2}<0\,, \lambda_{3}<0$
\begin{equation}
\label{eq:stab_2}
{\renewcommand{\arraystretch}{1.5}
\begin{matrix}
\Omega_{\Lambda,0}>0\,:&\quad\ve=-1\,,&\quad \xi>\frac{1}{2}\,,\\
	                   &\quad\ve=+1\,,&\quad\xi<0\,,\\
\Omega_{\Lambda,0}<0\,:&\quad\ve=-1\,,&\quad\xi<0\,,\\
	                   &\quad\ve=+1\,,&\quad\xi>\frac{1}{2}\,,                   
\end{matrix}}
\end{equation}
for a saddle : $\lambda_{1}<0\,, \lambda_{2}>0\,, \lambda_{3}>0$
$$
{\renewcommand{\arraystretch}{1.5}
\begin{matrix}
\Omega_{\Lambda,0}>0\,:&\quad\ve=-1\,,&\quad \frac{3}{10}<\xi<\frac{1}{2}\,,\\
\Omega_{\Lambda,0}<0\,:&\quad\ve=+1\,,&\quad\frac{3}{10}<\xi<\frac{1}{2}\,,                   
\end{matrix}}
$$
and for a different type of saddle : $\lambda_{1}<0\,, \lambda_{2}<0\,, \lambda_{3}>0$
$$
{\renewcommand{\arraystretch}{1.5}
\begin{matrix}
\Omega_{\Lambda,0}>0\,:&\quad\ve=-1\,,&\quad 0<\xi<\frac{1}{6}\,,\\
	                   &\quad\ve=+1\,,&\quad \frac{1}{6}<\xi<\frac{3}{10}\,,\\
\Omega_{\Lambda,0}<0\,:&\quad\ve=-1\,,&\quad \frac{1}{6}<\xi<\frac{3}{10}\,,\\
	                   &\quad\ve=+1\,,&\quad 0<\xi<\frac{1}{6}\,.                  
\end{matrix}}
$$
Finally, the linearised solutions in the vicinity of the critical point corresponding to the Einstein static solution are
\begin{equation}
\begin{split}
x(a) & = x^{*} - \frac{1}{2}(\Delta x + 3 \Delta z)\left(\frac{a}{a^{(i)}}\right)^{\lambda_{1}} 
+ \frac{3}{2}(\Delta x + \Delta z)\left(\frac{a}{a^{(i)}}\right)^{\lambda_{2}}\,,\\
z(a) & = z^{*} + \frac{1}{2}(\Delta x + 3 \Delta z)\left(\frac{a}{a^{(i)}}\right)^{\lambda_{1}} 
- \frac{1}{2}(\Delta x + \Delta z)\left(\frac{a}{a^{(i)}}\right)^{\lambda_{2}}\,,\\
h(a) & = \Delta h \left(\frac{a}{a^{(i)}}\right)^{\lambda_{3}}\,,
\end{split}
\end{equation}
where $x^{*}$ and $z^{*}$ are the coordinates of the critical point, $\lambda_{1}$, $\lambda_{2}$, $\lambda_{3}$ are the eigenvalues of the linearisation matrix, and $\Delta x = x^{(i)}-x^{*}$, $\Delta z = z^{(i)}-z^{*}$, $\Delta h = h^{(i)}$ are the initial conditions taken in the close vicinity of the critical point.

From the last equation we directly obtain the evolution of the Hubble function
$$
\frac{H(a)}{H(a_{0})} = \frac{H(a^{(i)})}{H(a_{0})}\left(\frac{a}{a^{(i)}}\right)^{\lambda_{3}}\,
$$
which vanishes up to quadratic term in initial condition 
\begin{equation}
\left(\frac{H(a)}{H(a_{0})}\right)^{2} \approx 0\,.
\end{equation}

The critical points investigated in this section constitute non-singular asymptotic states in the form of de Sitter state and static Einstein solution. Examination of the stability conditions \eqref{eq:stab_1} and \eqref{eq:stab_2} indicates that the both states can be stable during expansion only for the phantom scalar field $\ve=-1$ with $\Omega_{\Lambda,0}>0$ and $\xi>\frac{1}{2}$. Most interestingly for the canonical scalar field $\ve=+1$ with $\Omega_{\Lambda,0}>0$ and negative values of the non-minimal coupling constant $\xi<0$ the de Sitter state is transient, in the form of a saddle type critical point, while the static Einstein state is asymptotically stable. 

\subsection{The analysis of dynamics at infinity}
\label{subsec:atinfty}

In order to obtain complete information about the structure of the phase space
of the dynamical system \eqref{eq:dynsys1} it is necessary to investigate the
behaviour of the trajectories of this system at infinity.

To study the dynamical system \eqref{eq:dynsys1} we need to introduce 
three charts $U_{(n)}$ of projective coordinates
\begin{equation}
\label{eq:charts}
\begin{split}
U_{(1)} :& \quad u_{(1)} = \frac{1}{x}\,,\,\, v_{(1)} = \frac{z}{x}\,,\,\, w_{(1)} =
	\frac{h}{x}\,,\\
U_{(2)} :& \quad u_{(2)} = \frac{x}{z}\,,\,\, v_{(2)} = \frac{1}{z}\,,\,\, w_{(2)} =
	\frac{h}{z}\,,\\
U_{(3)} :& \quad u_{(3)} = \frac{x}{h}\,,\,\, v_{(3)} = \frac{z}{h}\,,\,\, w_{(3)} =
	\frac{1}{h}\,.
\end{split}
\end{equation}
These three charts enable us to find a complete set of the critical points of the system \eqref{eq:dynsys2} at infinity of the phase space. Some critical points will be present in more than one chart.

The use of $U_{(1)}$ chart transforms the system \eqref{eq:dynsys2} in to the following one
\begin{equation}
\label{eq:sys_inf_1}
\begin{split}
	\frac{\ud u_{(1)}}{\ud\eta_{(1)}} & = 
	3u_{(1)}\Big( \big(w_{(1)}^{2}-\ve6\xi(1-6\xi)v_{(1)}^{2}\big) + \\ &
\hspace{1.5cm}
	+ 3\xi
	v_{(1)}\Big(\Omega_{\Lambda,0}u_{(1)}^{2}+\frac{1}{3}w_{(1)}^{2}-\ve(1-6\xi)
	-\ve2\xi(1+v_{(1)})^{2}\Big)\Big)\,,\\
\frac{\ud v_{(1)}}{\ud\eta_{(1)}} & = 
	(1+v_{(1)})\big(w_{(1)}^{2}-\ve6\xi(1-6\xi)v_{(1)}^{2}\big) + \\ & \quad
	+ \frac{3}{2}v_{(1)}(1+6\xi v_{(1)})
	\Big(\Omega_{\Lambda,0}u_{(1)}^{2}+\frac{1}{3}w_{(1)}^{2}-\ve(1-6\xi)-
\ve2\xi(1+v_{(1)})^{2}\Big)\,,\\
\frac{\ud w_{(1)}}{\ud\eta_{(1)}} & =  w_{(1)}\Big(
	\big(w_{(1)}^{2}-\ve6\xi(1-6\xi)v_{(1)}^{2}\big) + \\ &
\hspace{1.35cm}
	+\frac{3}{2}(1+6\xi v_{(1)})
	\Big(\Omega_{\Lambda,0}u_{(1)}^{2}+\frac{1}{3}w_{(1)}^{2}-\ve(1-6\xi)-
\ve2\xi(1+v_{(1)})^{2}\Big)
	\Big) \,,
\end{split}
\end{equation}
where the time function along the phase space trajectories is
\begin{equation}
\label{eq:time_inf_1}
\frac{\ud}{\ud\eta_{(1)}} = u_{(1)}^{2}\frac{\ud}{\ud\tau} =
\big(w_{(1)}^{2}-\ve6\xi(1-6\xi)v_{(1)}^{2}\big)\frac{\ud}{\ud\ln{a}}\,.
\end{equation}
The critical points of this system at the invariant manifold at infinity defined by the condition $u_{(1)}\equiv0$ are gathered in table \ref{tab:1}.

\begin{table}
	\centering
	\begin{tabular}{|>{\footnotesize}c|>{\footnotesize}l|>{\footnotesize}l|>{\footnotesize}c|}
		\hline
		& $v_{(1)}^{*}$ & $w_{(1)}^{*}$ & existence \\
		\hline
		A & $-\frac{1-4\xi}{2\xi}$ & $0$ & $\forall
		\xi\in\mathbf{R}\setminus\{0\}$ \\
		B & $-1\pm\sqrt{-\frac{1-6\xi}{6\xi}}$ & $0$ & $\xi<0$ or
		$\xi>\frac{1}{6}$ \\
		C & $-\frac{1}{6\xi}$ & $\pm\sqrt{\ve\frac{1-6\xi}{6\xi}}$ & $\ve=-1$: $\xi<0\lor\xi>	
		\frac{1}{6}$ or $\ve=+1$: $0<\xi<\frac{1}{6}$ \\
		D & $-\frac{1\pm\sqrt{-2(1-6\xi)}}{6\xi}$ &
		$\pm\sqrt{\ve6\xi(1-6\xi)}v_{(1)}^{*}$ & $\ve=-1$:
		$\xi>\frac{1}{6}$\\
		E & $0$ & $0$ & $\forall \xi\in\mathbf{R}$ \\
		\hline
	\end{tabular}
	\caption{Critical points of the system \eqref{eq:sys_inf_1} in the map $U_{(1)}$ at infinity 
of the phase space defined by the condition $u_{(1)}\equiv0$.}
\label{tab:1}
\end{table}

The energy conservation condition \eqref{eq:constr} expressed in the projective coordinates of the chart $U_{(1)}$ is given by
\begin{equation}
\left(\Omega_{\Lambda,0}+\Omega_{m,0}\left(\frac{a}{a_{0}}\right)^{-3}\right) u_{(1)}^{2} =
w_{(1)}^{2}-\ve(1-6\xi)-\ve6\xi(1+v_{(1)})^{2}\,,
\end{equation}
and at infinity of the phase space $u_{(1)}\equiv0$ reduces to the following condition
\begin{equation}
w_{(1)}^{2}\ge \ve(1-6\xi)+\ve6\xi(1+v_{(1)})^{2}\,.
\end{equation} 
The regions of the phase space where this condition is not fulfilled are nonphysical.

\begin{figure}
\centering
		\includegraphics[scale=0.5]{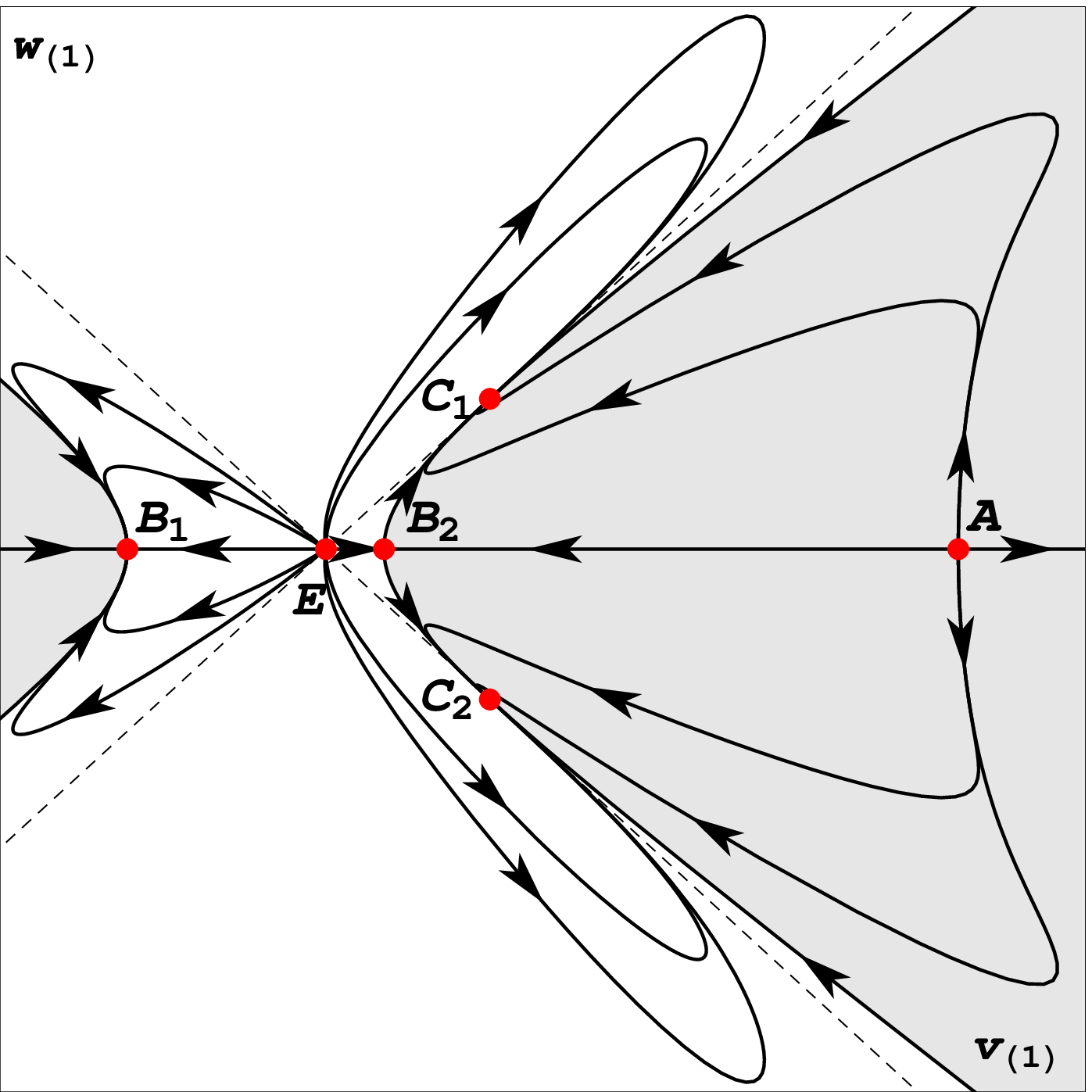}
		\includegraphics[scale=0.5]{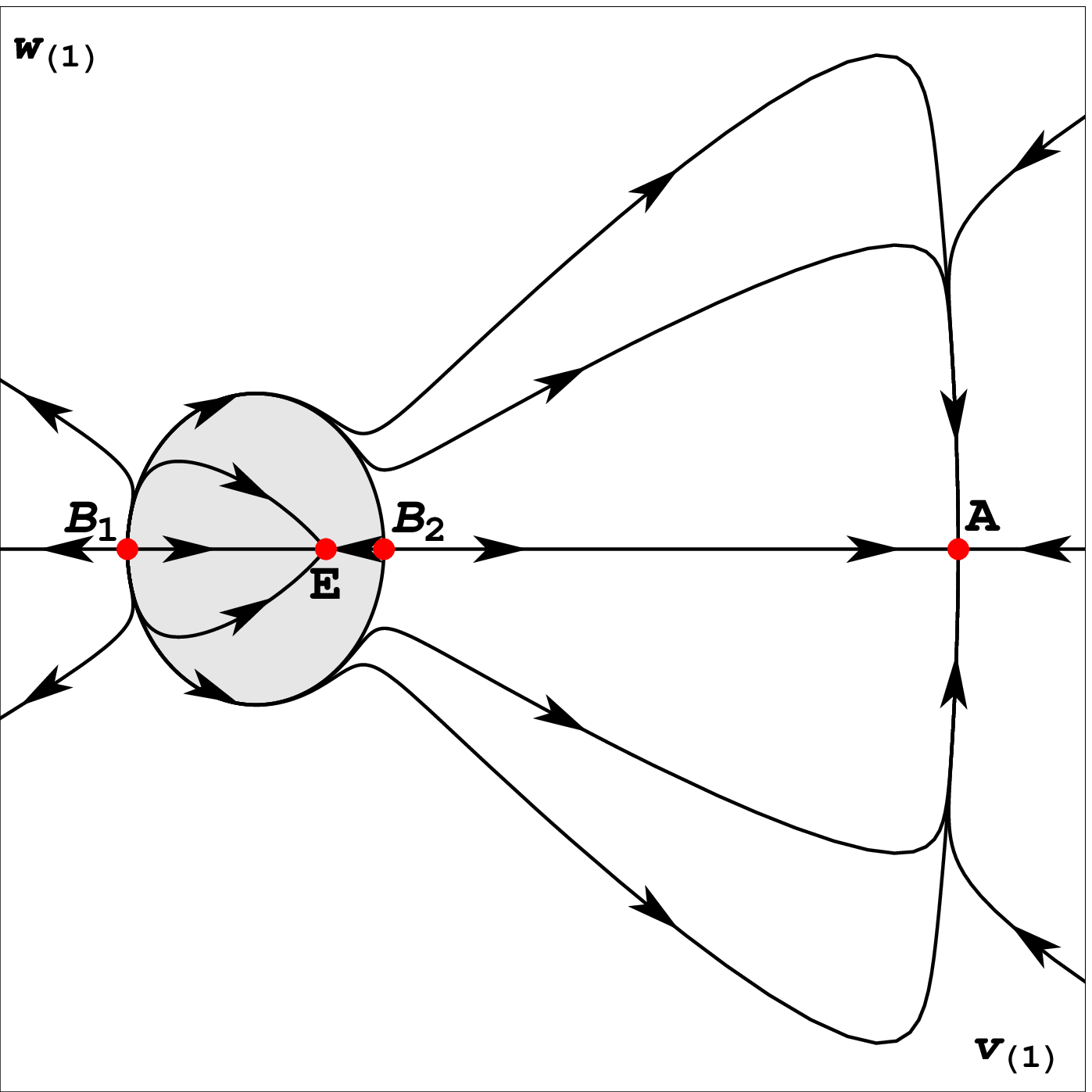}\\
		\includegraphics[scale=0.5]{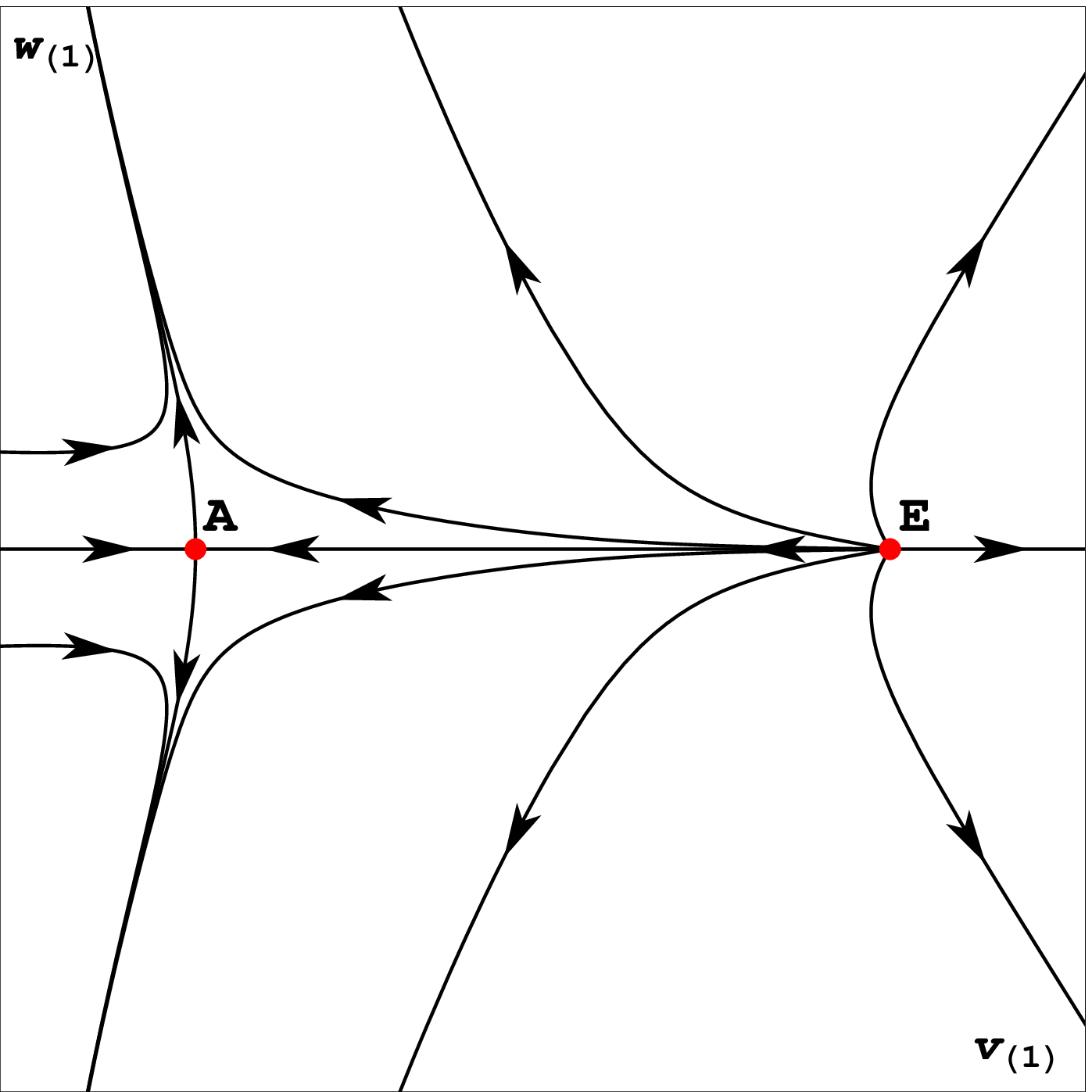}
		\includegraphics[scale=0.5]{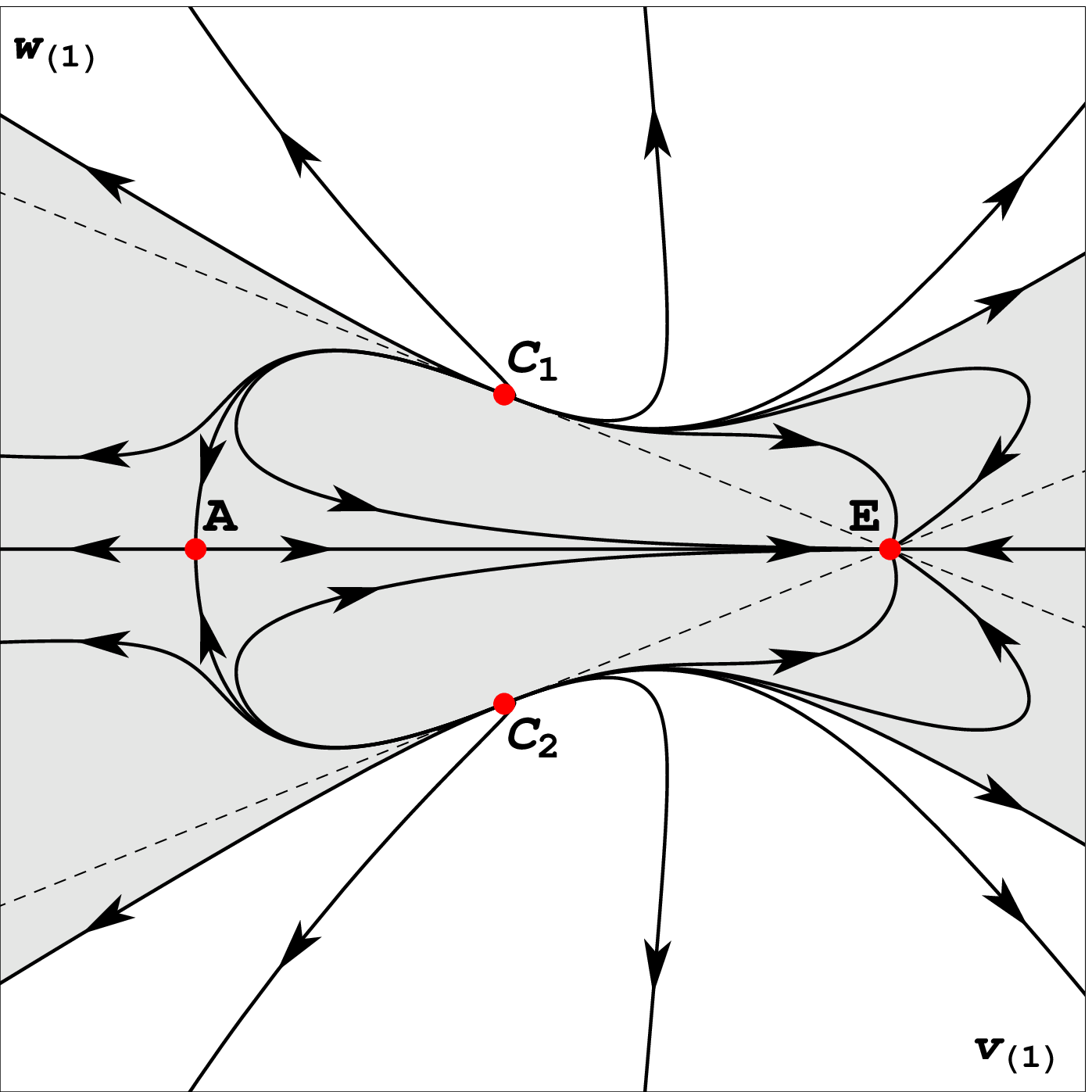}
		\caption{The phase space portraits for the dynamical system \eqref{eq:sys_inf_1} at the 
invariant manifold at infinity defined by $u_{(1)}\equiv0$. The following values of the model 
parameters were assumed: $\ve=-1$, $\xi<0$ (top left), 
						 $\ve=+1$, $\xi<0$ (top right), 
						 $\ve=-1$, $0<\xi<\frac{1}{6}$ (bottom left),
						 $\ve=+1$, $0<\xi<\frac{1}{6}$ (bottom right).} 
\label{fig:1}
\end{figure}

\begin{figure}
\centering
		\includegraphics[scale=0.5]{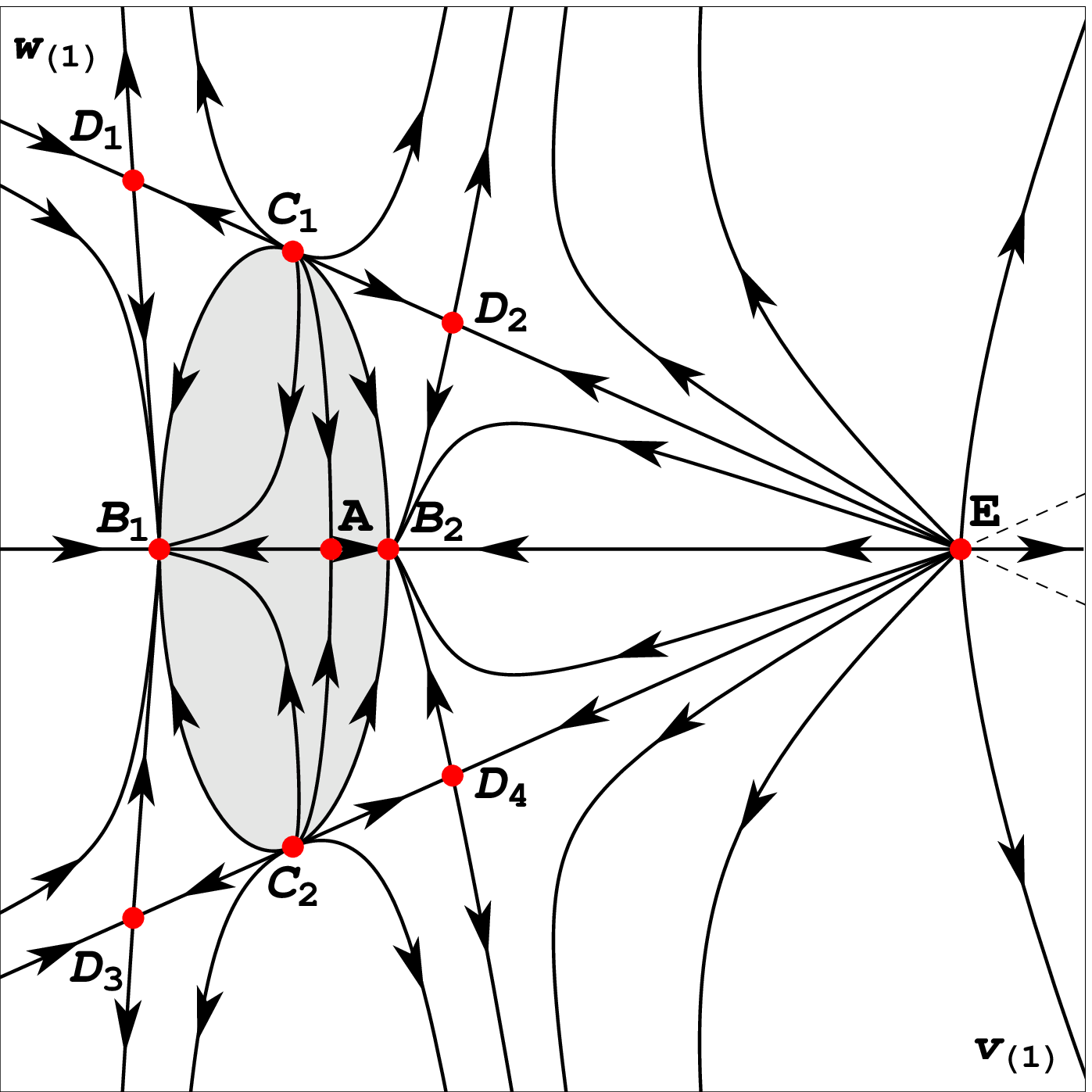}
		\includegraphics[scale=0.5]{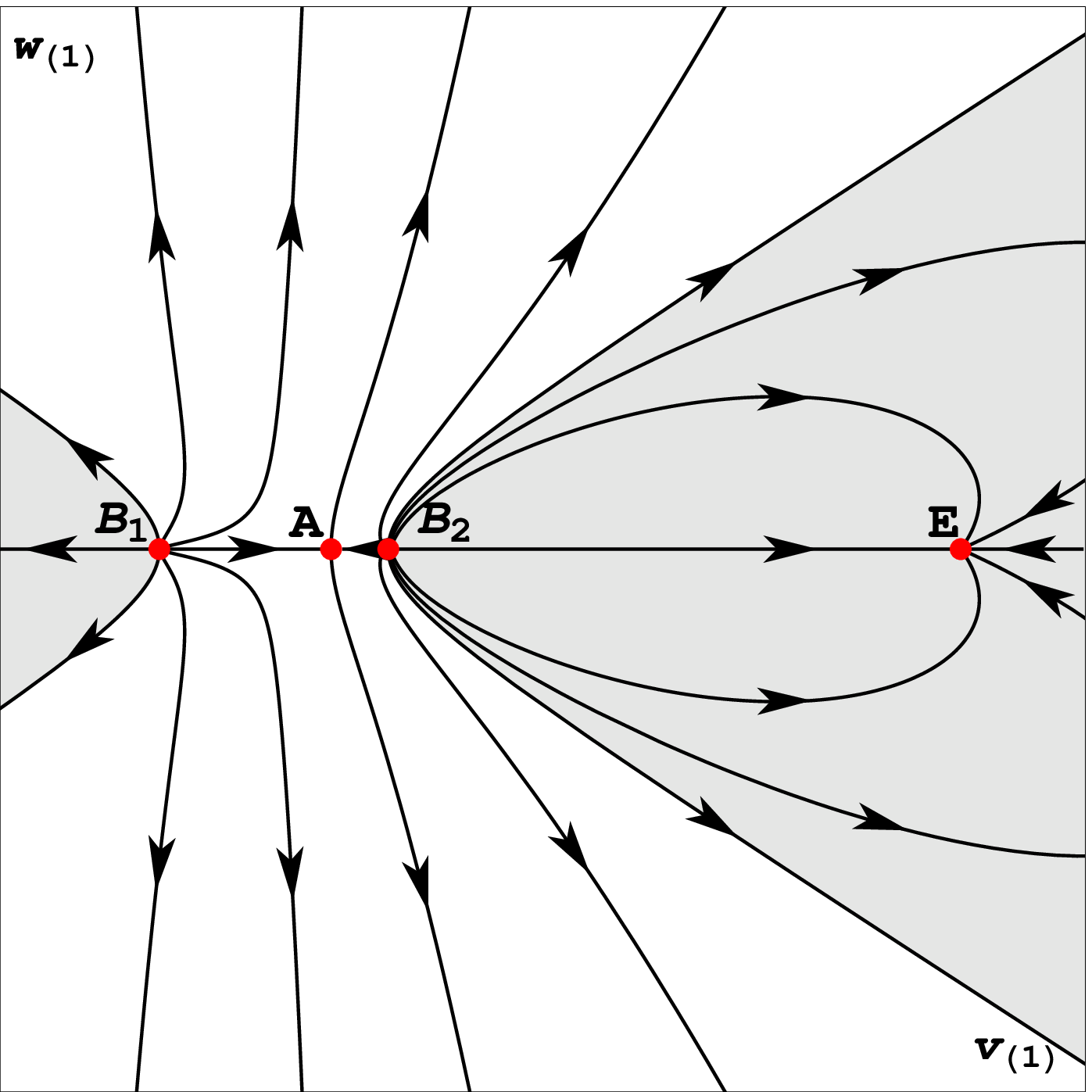}\\
		\includegraphics[scale=0.5]{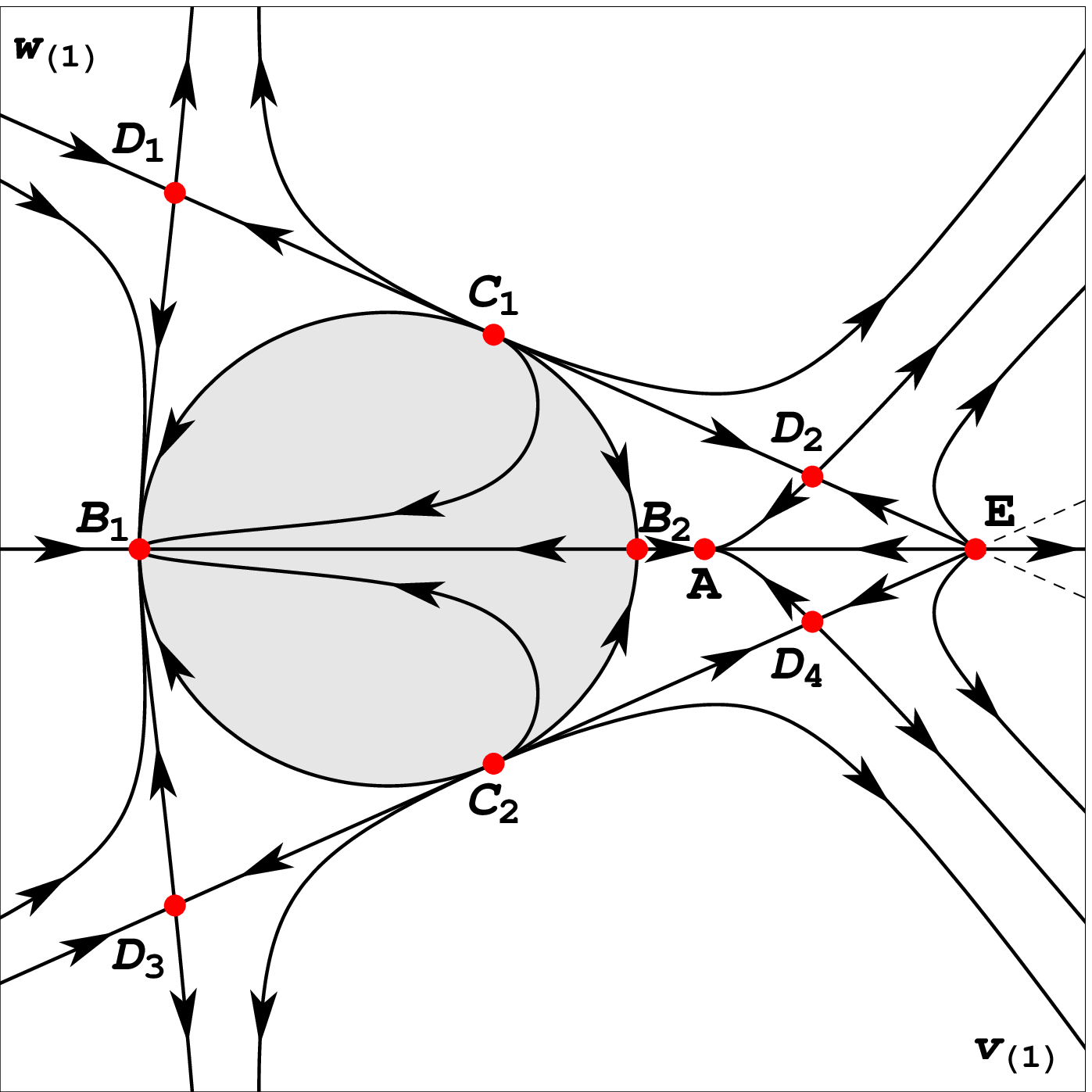}
		\includegraphics[scale=0.5]{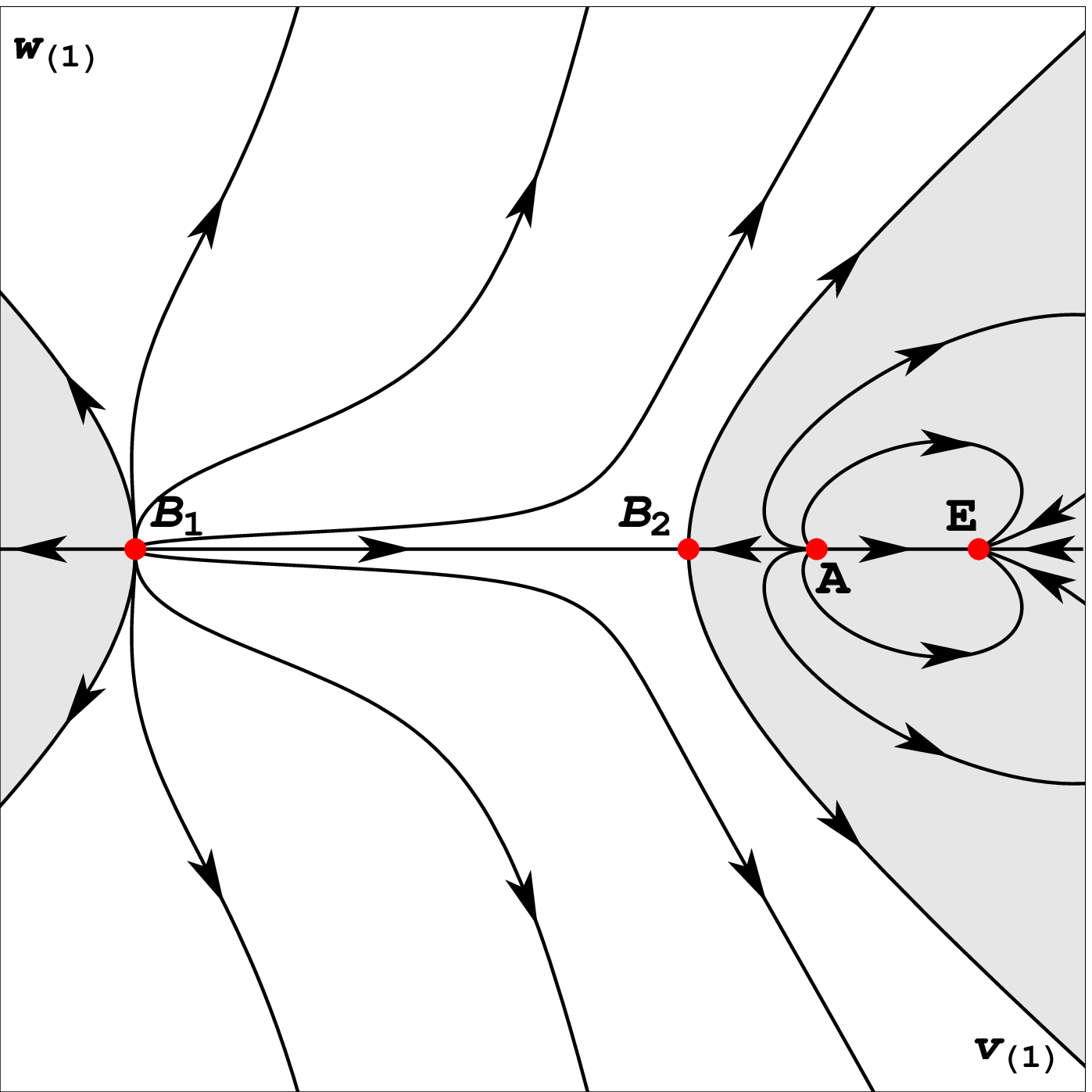}
		\caption{The phase space portraits for the dynamical system \eqref{eq:sys_inf_1} at the 
invariant manifold at infinity defined by $u_{(1)}\equiv0$. The following values of the model 
parameters were assumed: $\ve=-1$, $\frac{1}{6}<\xi<\frac{3}{16}$ (top left),
			             $\ve=+1$, $\frac{1}{6}<\xi<\frac{3}{16}$ (top right),
		                 $\ve=-1$, $\frac{3}{16}<\xi<\frac{1}{4}$ (bottom left), 
		                 $\ve=+1$, $\frac{3}{16}<\xi<\frac{1}{4}$ (bottom right).}
\label{fig:2}
\end{figure}

\begin{figure}
\centering
		\includegraphics[scale=0.5]{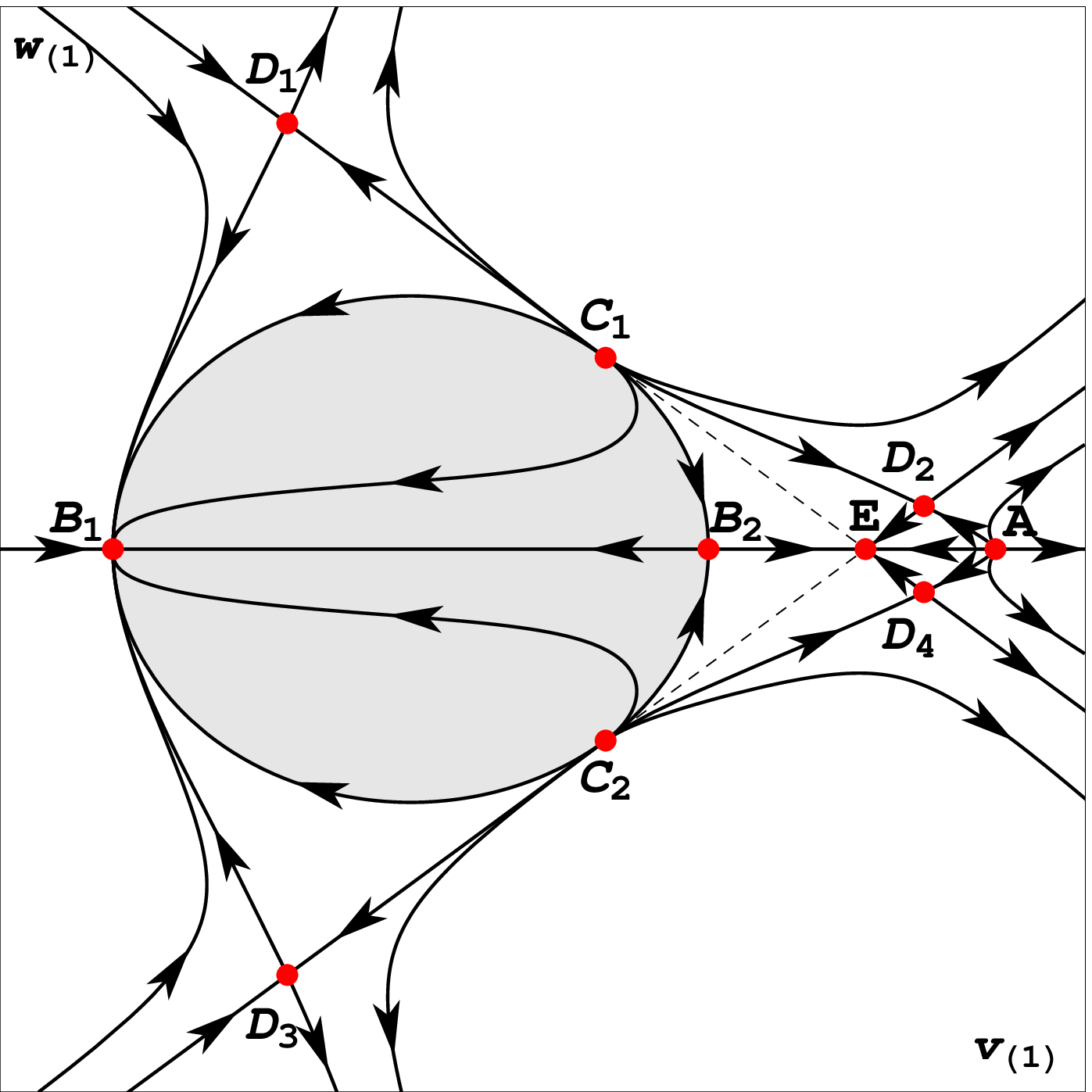}
		\includegraphics[scale=0.5]{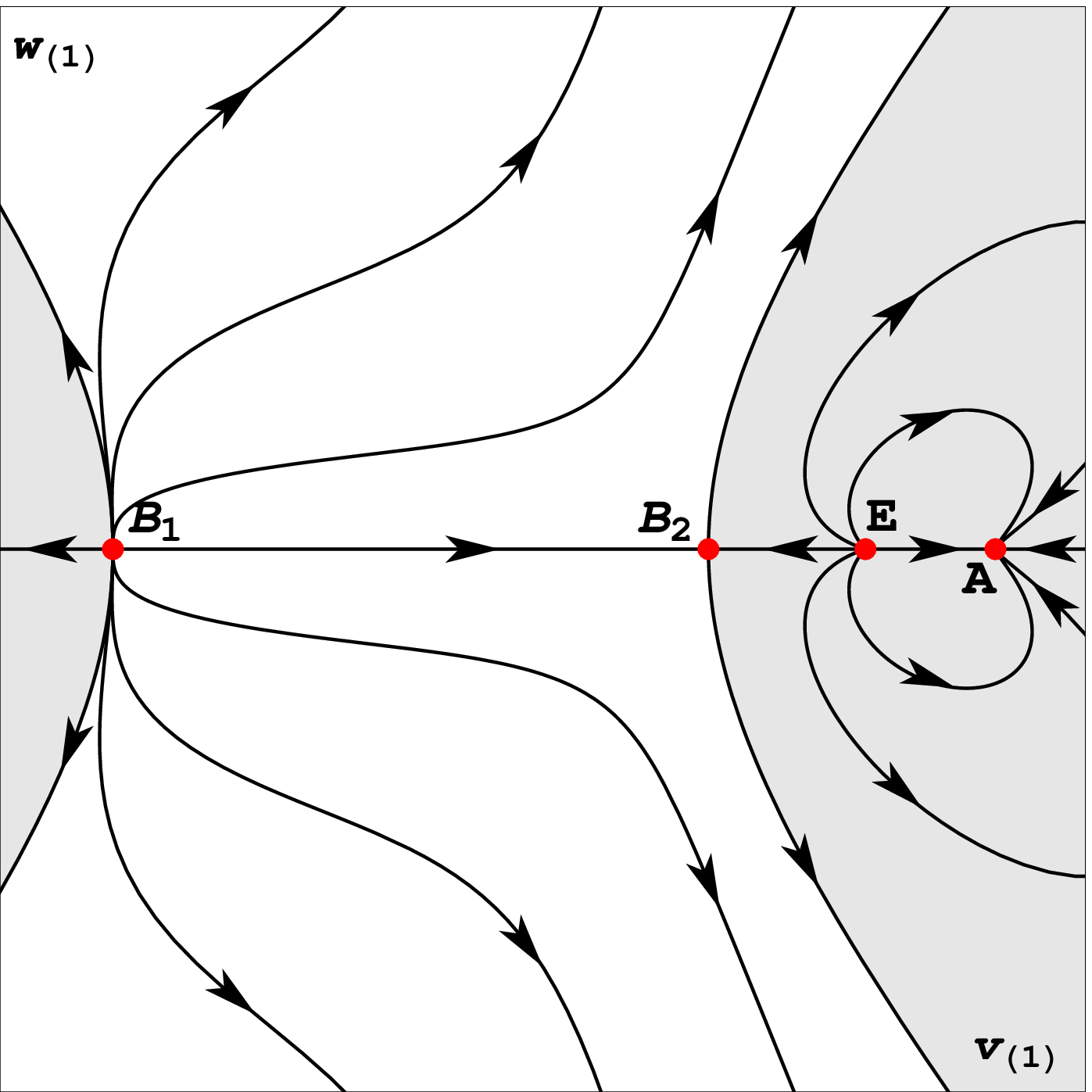}\\
		\includegraphics[scale=0.5]{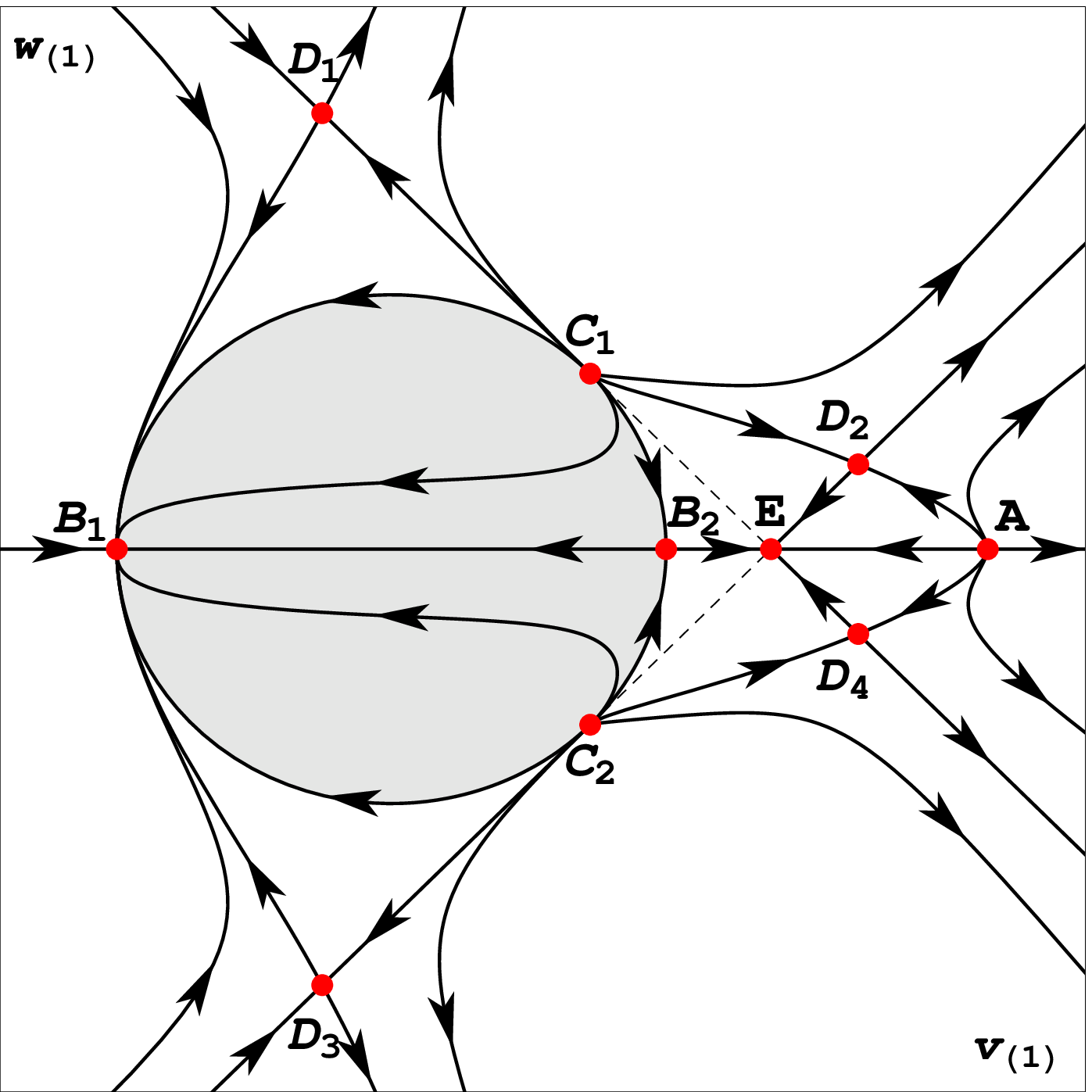}
		\includegraphics[scale=0.5]{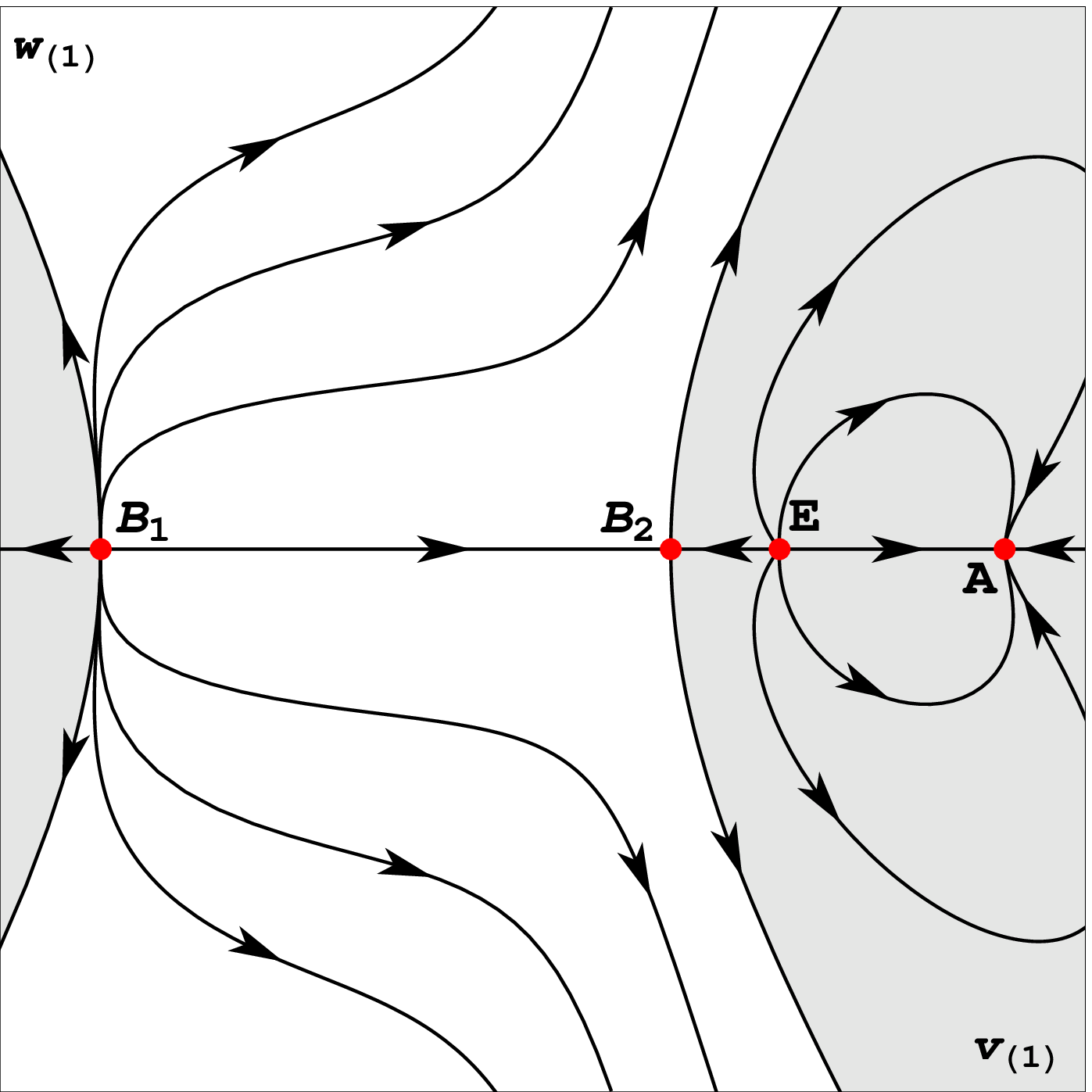}
		\caption{The phase space portraits for the dynamical system \eqref{eq:sys_inf_1} at the 
invariant manifold at infinity defined by $u_{(1)}\equiv0$. The following values of the model 
parameters were assumed: $\ve=-1$, $\frac{1}{4}<\xi<\frac{3}{10}$ (top left),
			 			 $\ve=+1$, $\frac{1}{4}<\xi<\frac{3}{10}$ (top right)
		                 $\ve=-1$, $\xi>\frac{3}{10}$ (bottom left), 
		                 $\ve=+1$, $\xi>\frac{3}{10}$ (bottom right). 
Note that the presented phase space diagrams for the given parameters are topologically 
equivalent both for canonical and phantom scalar fields. The difference comes from dynamical 
behaviour in the 3-dimensional phase space.}
\label{fig:3}
\end{figure}

In figures \ref{fig:1}, \ref{fig:2}, \ref{fig:3} we presented the phase space diagrams of the system \eqref{eq:sys_inf_1} at the invariant manifold at infinity. The diagrams represent all representative cases for all possible values of the non-minimal coupling constant $\xi$. The shaded regions denote nonphysical regions of the phase space. The dashed lines denote the singularity of the time transformation \eqref{eq:time_inf_1} where $w_{(1)}^{2}=\ve6\xi(1-6\xi)v_{(1)}^{2}$. On figures \ref{fig:2} and \ref{fig:3} these lines are  separatrices of the saddle type critical points $D_{1}$ , $D_{2}$ , $D_{3}$ , $D_{4}$ and they intersect at the critical point $E$.

In table \ref{tab:4} we gathered the eigenvalues of the linearisation matrix for the critical points which establish stability conditions for a given asymptotic state. They can be used to find the linearised solution in the vicinity of the critical points. The effective equation of state parameters give the physical interpretation of the solutions.

The second chart of the projective variables $U_{(2)}$ leads to the following dynamical system
\begin{equation}
\label{eq:sys_inf_2}
\begin{split}
\frac{\ud u_{(2)}}{\ud\eta_{(2)}} & = 
	-u_{(2)}(1+u_{(2)})\big(w_{(2)}^{2}-\ve6\xi(1-6\xi)\big)- \\ & \quad
	-\frac{3}{2}(u_{(2)}+6\xi)\Big( \Omega_{\Lambda,0}v_{(2)}^{2} +
	\frac{1}{3}w_{(2)}^{2} - \ve(1-6\xi)u_{(2)}^{2} - \ve2\xi(1+u_{(2)})^{2}\Big)\,,\\
\frac{\ud v_{(2)}}{\ud\eta_{(2)}} & =  - v_{(2)}\Big(
	(u_{(2)}-2)\big(w_{(2)}^{2}-\ve6\xi(1-6\xi)\big) + \\ &
	\hspace{1.55cm}+ \frac{3}{2}\Big( \Omega_{\Lambda,0}v_{(2)}^{2} +
        \frac{1}{3}w_{(2)}^{2} - \ve(1-6\xi)u_{(2)}^{2} -
	\ve2\xi(1+u_{(2)})^{2}\Big)\Big)\,,\\
\frac{\ud w_{(2)}}{\ud\eta_{(2)}} & = -u_{(2)}w_{(2)}\big(w_{(2)}^{2}-\ve6\xi(1-6\xi)\big)\,,
\end{split}
\end{equation}
where the time function along phase space trajectories is 
\begin{equation}
\label{eq:time_inf_2}
\frac{\ud}{\ud\eta_{(2)}} = v_{(2)}^{2}\frac{\ud}{\ud\tau} =
\big(w_{(2)}^{2}-\ve6\xi(1-6\xi)\big)\frac{\ud}{\ud\ln{a}}\,.
\end{equation}
The critical points of the system at the invariant manifold at infinity given by the condition $v_{(2)}\equiv0$ are gathered in table \ref{tab:2}.

\begin{table}
	\centering
	\begin{tabular}{|>{\footnotesize}c|>{\footnotesize}l|>{\footnotesize}l|>{\footnotesize}c|}
		\hline
		 & $u_{(2)}^{*}$ & $w_{(2)}^{*}$ & existence \\
		\hline
		A & $-\frac{2\xi}{1-4\xi}$ & $0$ & $\forall
		\xi\in\mathbf{R}\setminus\{\frac{1}{4}\}$\\
		B & $-6\xi\pm\sqrt{-6\xi(1-6\xi)}$ & $0$ & $\xi<0$ or
		$\xi>\frac{1}{6}$\\		
		C & $-6\xi$ & $\pm\sqrt{\ve6\xi(1-6\xi)}$ & $\ve=-1$:
		$\xi<0\lor\xi>\frac{1}{6}$ or $\ve=+1$: $0<\xi<\frac{1}{6}$\\
		D & $-\frac{2\xi}{1-4\xi}\Big(1\pm\sqrt{-2\xi(1-6\xi)}\Big)$ &
		$\pm\sqrt{\ve6\xi(1-6\xi)}$ & $\ve=-1$: $\xi>\frac{1}{6}$\\
		R & $0$ & $\pm\sqrt{\ve6\xi}$ & $\ve=-1$: $\xi<0$ or $\ve=+1$:
		$\xi>0$\\
		\hline
	\end{tabular}
	\caption{Critical points of the system \eqref{eq:sys_inf_2} in the map $U_{(2)}$ at infinity 
of the phase space defined by the condition $v_{(2)}\equiv0$.}
\label{tab:2}
\end{table}

In the projective coordinates of the chart $U_{(2)}$ the energy conservation condition \eqref{eq:constr} is now
\begin{equation}
\left(\Omega_{\Lambda,0}+\Omega_{m,0}\left(\frac{a}{a_{0}}\right)^{-3}\right) v_{(2)}^{2} = w_{(2)}^{2}-\ve(1-6\xi)u_{(2)}^{2}-\ve6\xi(1+u_{(2)})^{2}\,,
\end{equation} 
and at infinity of the phase space $v_{(2)}\equiv0$ it reduces to
\begin{equation}
\label{eq:cc_2}
w_{(2)}^{2} \ge \ve(1-6\xi)u_{(2)}^{2}+\ve6\xi(1+u_{(2)})^{2}\,.
\end{equation}

\begin{figure}
\centering
		\includegraphics[scale=0.5]{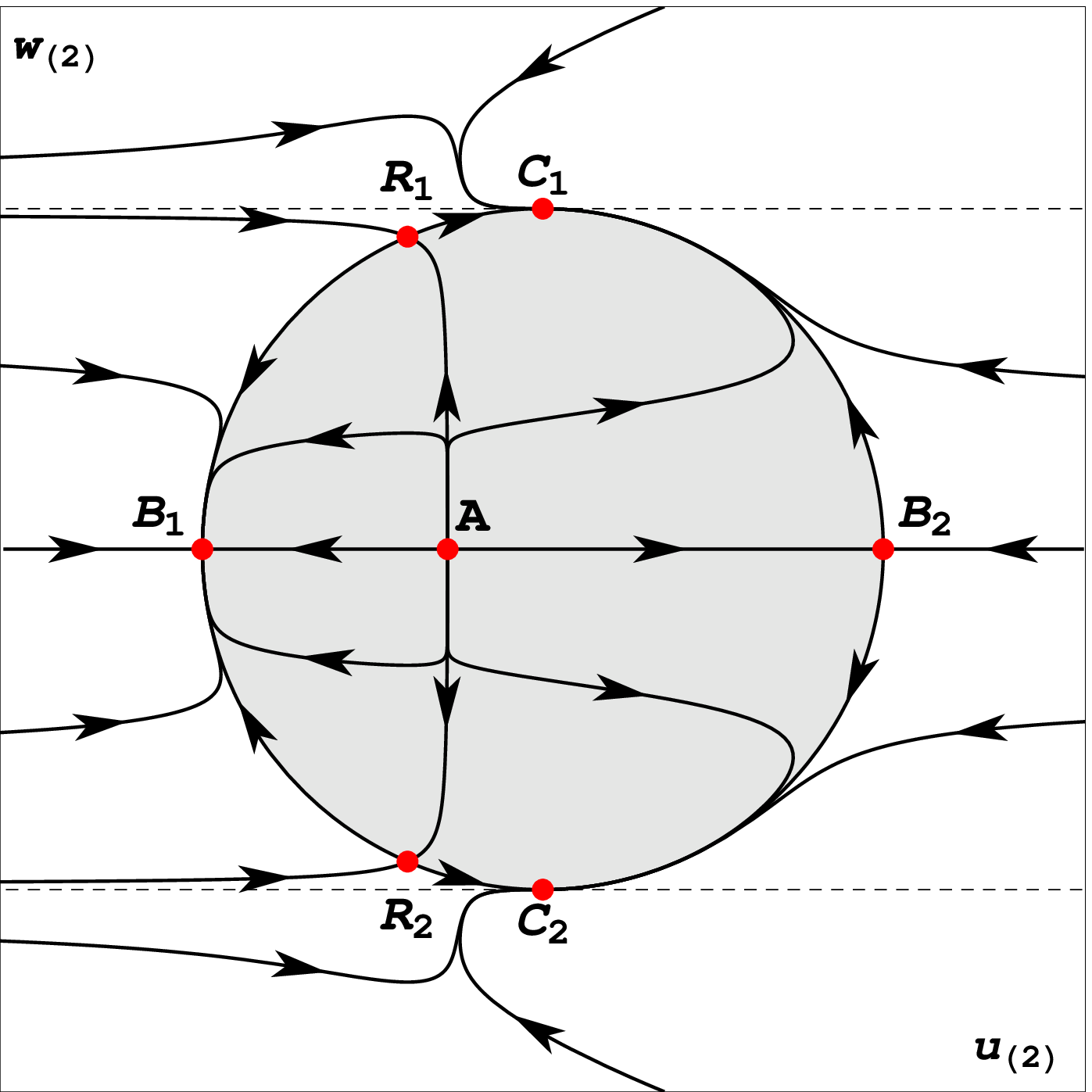}
		\includegraphics[scale=0.5]{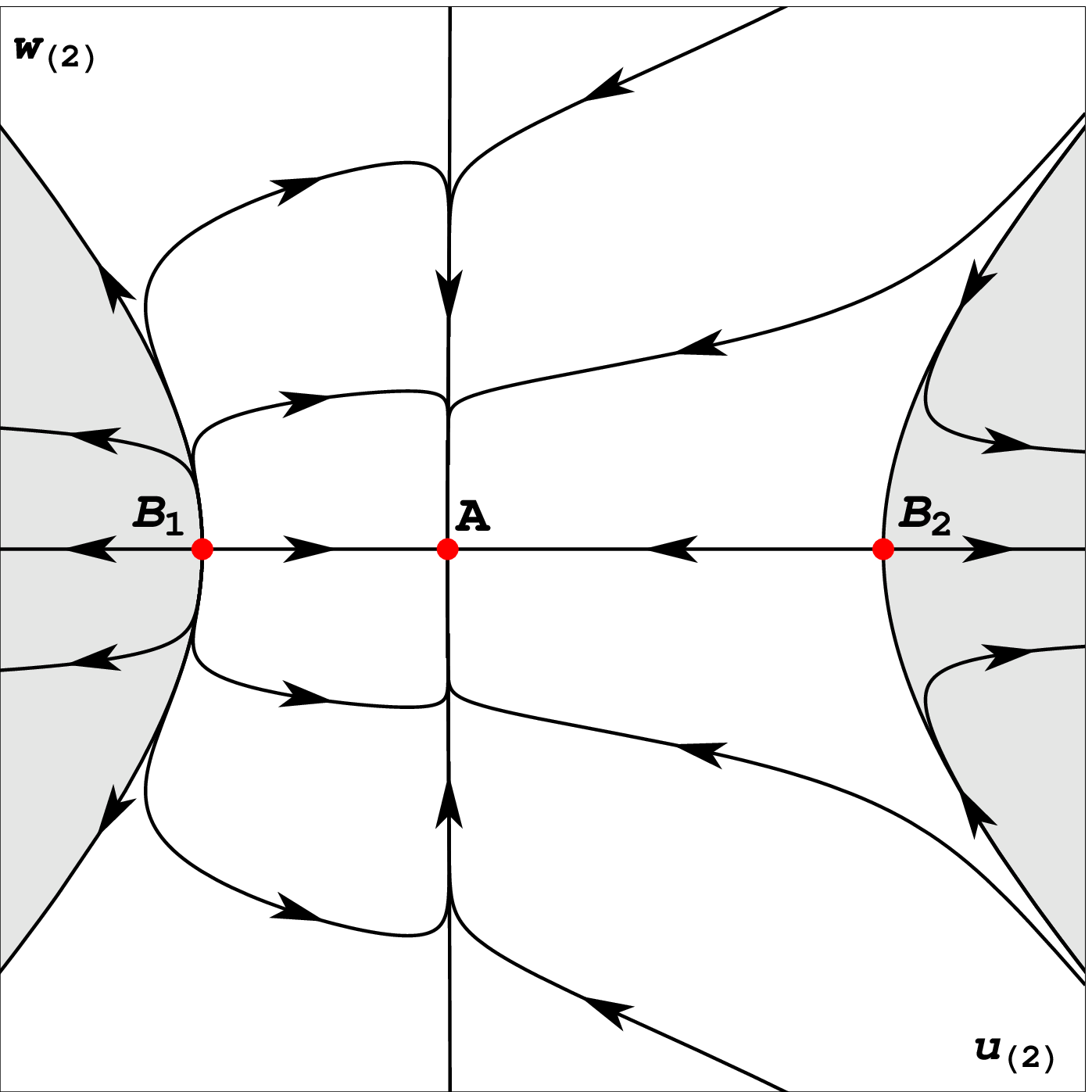}\\
		\includegraphics[scale=0.5]{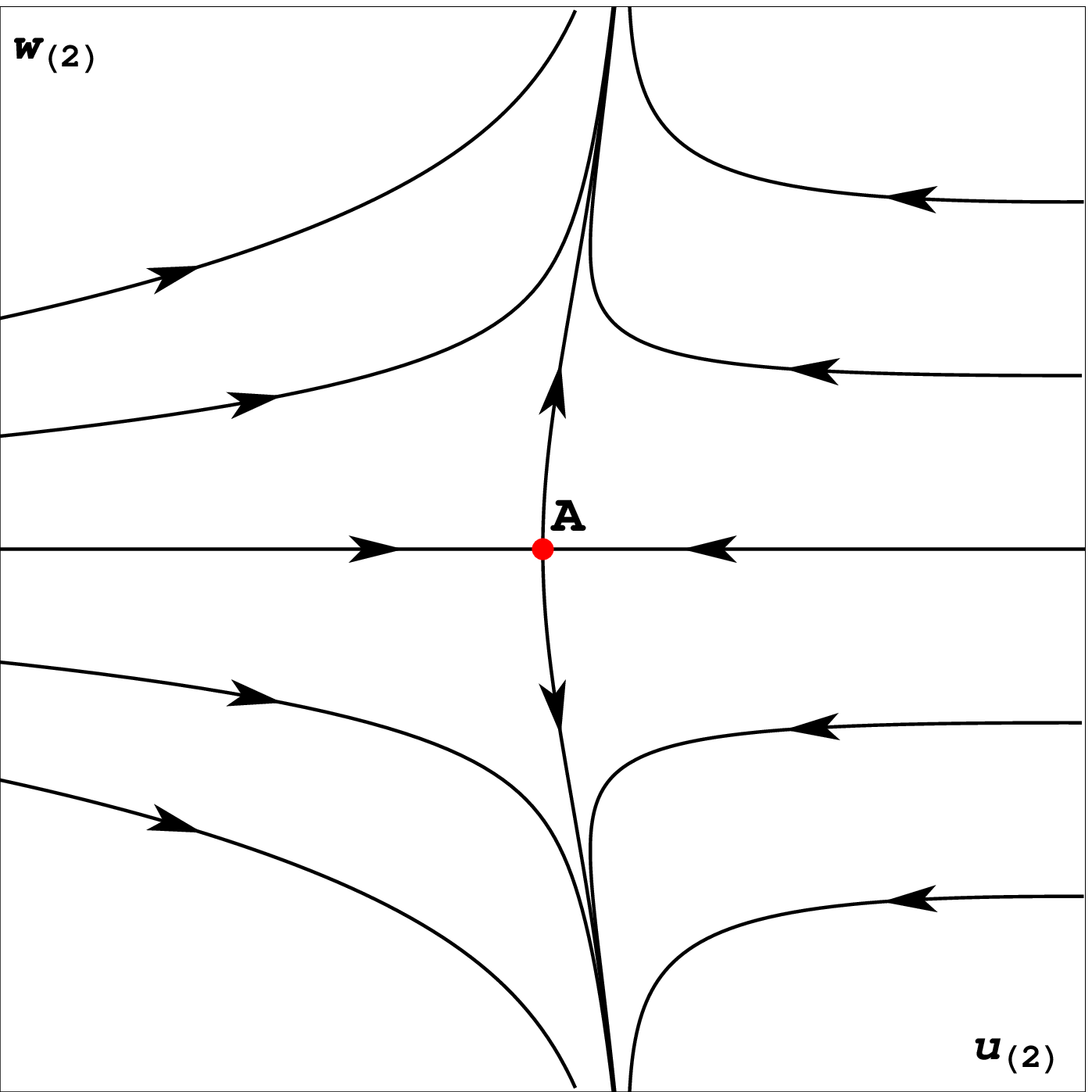}
		\includegraphics[scale=0.5]{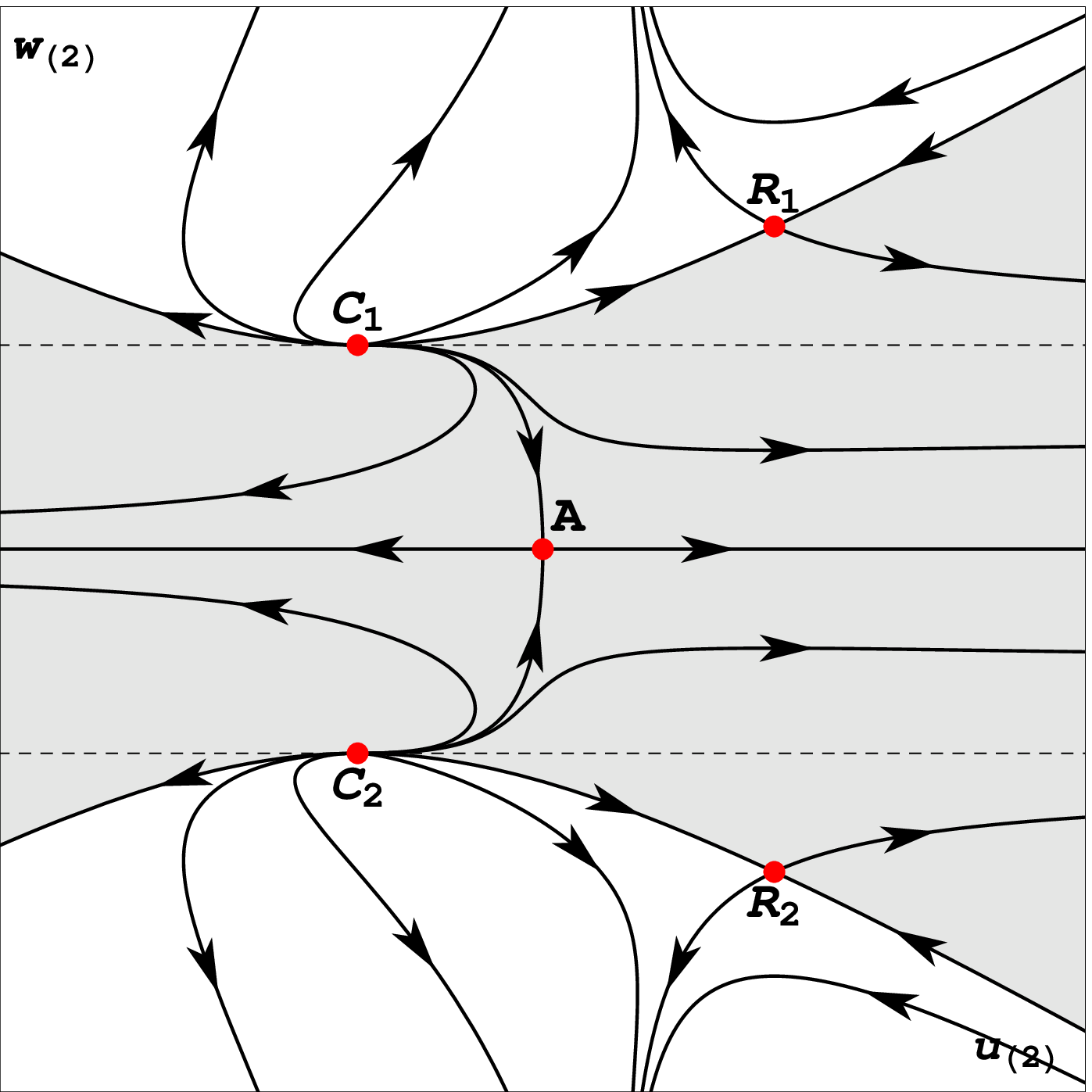}
		\caption{The phase space portraits for the dynamical system \eqref{eq:sys_inf_2} at the 
invariant manifold at infinity defined by $v_{(2)}\equiv0$. The following values of the model 
parameters were assumed: $\ve=-1$, $\xi<0$ (top left), $\ve=+1$, $\xi<0$
		(top right), $\ve=-1$, $0<\xi<\frac{1}{6}$ (bottom left), $\ve=+1$,
		$0<\xi<\frac{1}{6}$ (bottom right).}
\label{fig:4}
\end{figure}

\begin{figure}[h!]
\centering
		\includegraphics[scale=0.5]{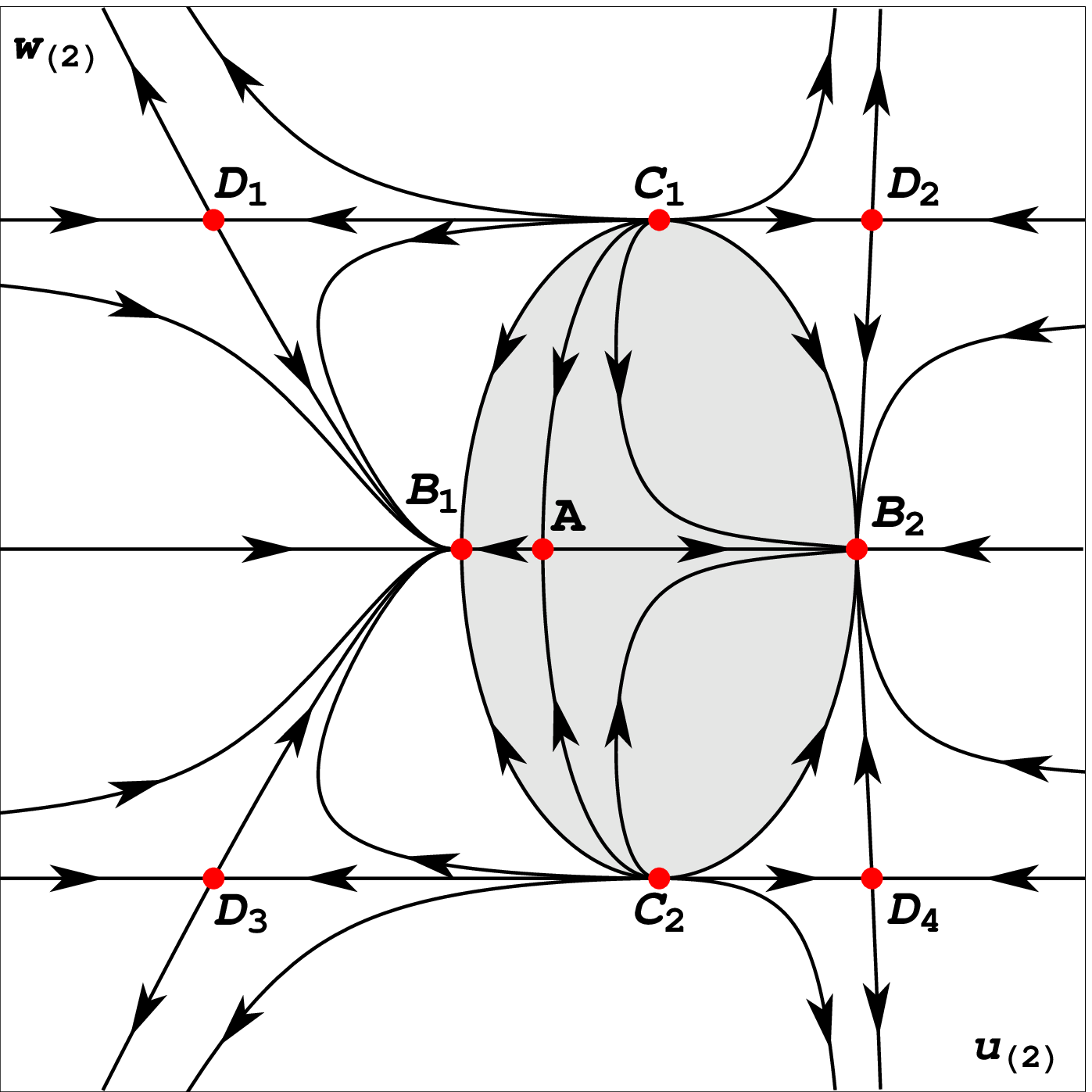}
		\includegraphics[scale=0.5]{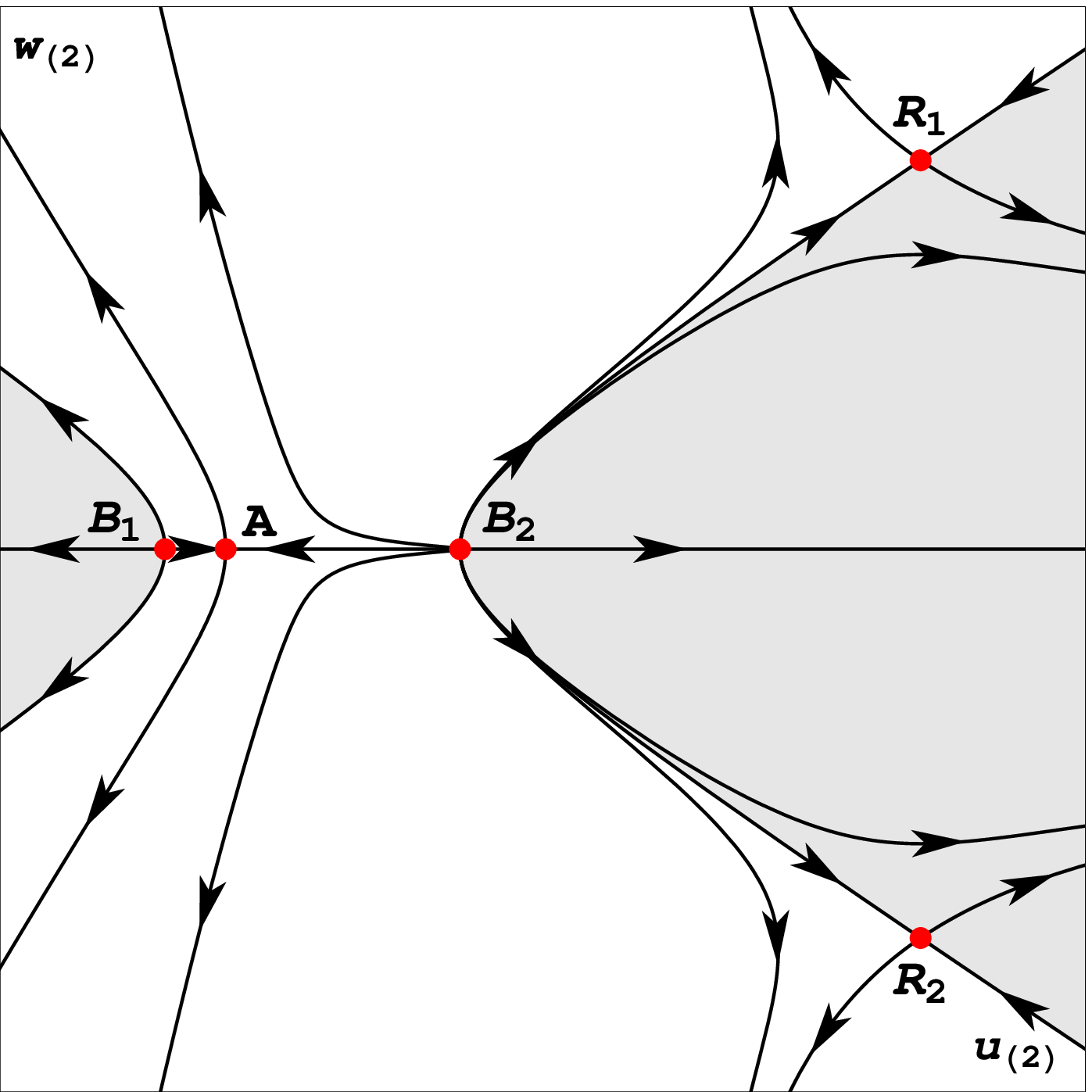}\\
		\includegraphics[scale=0.5]{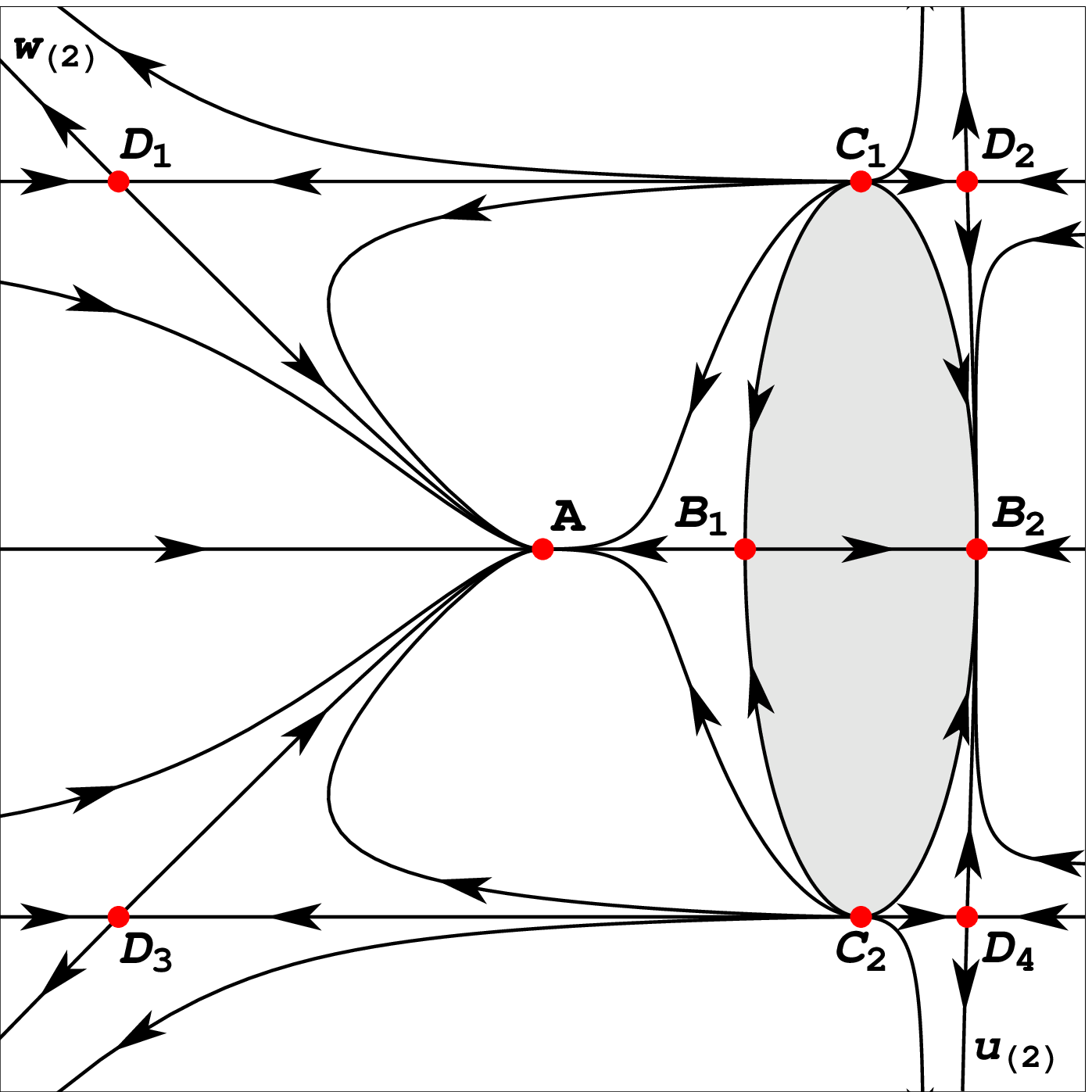}
		\includegraphics[scale=0.5]{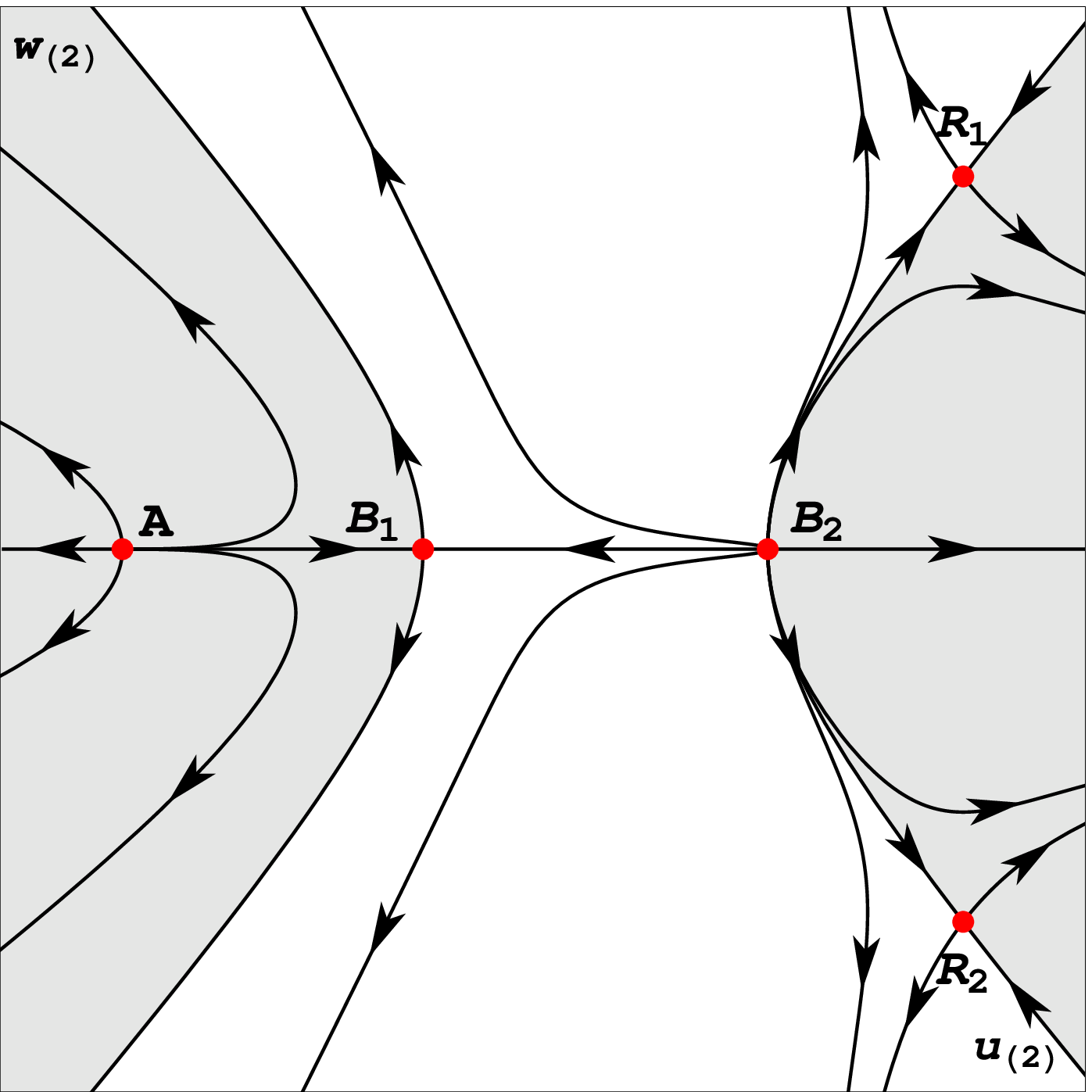}\\
		\includegraphics[scale=0.5]{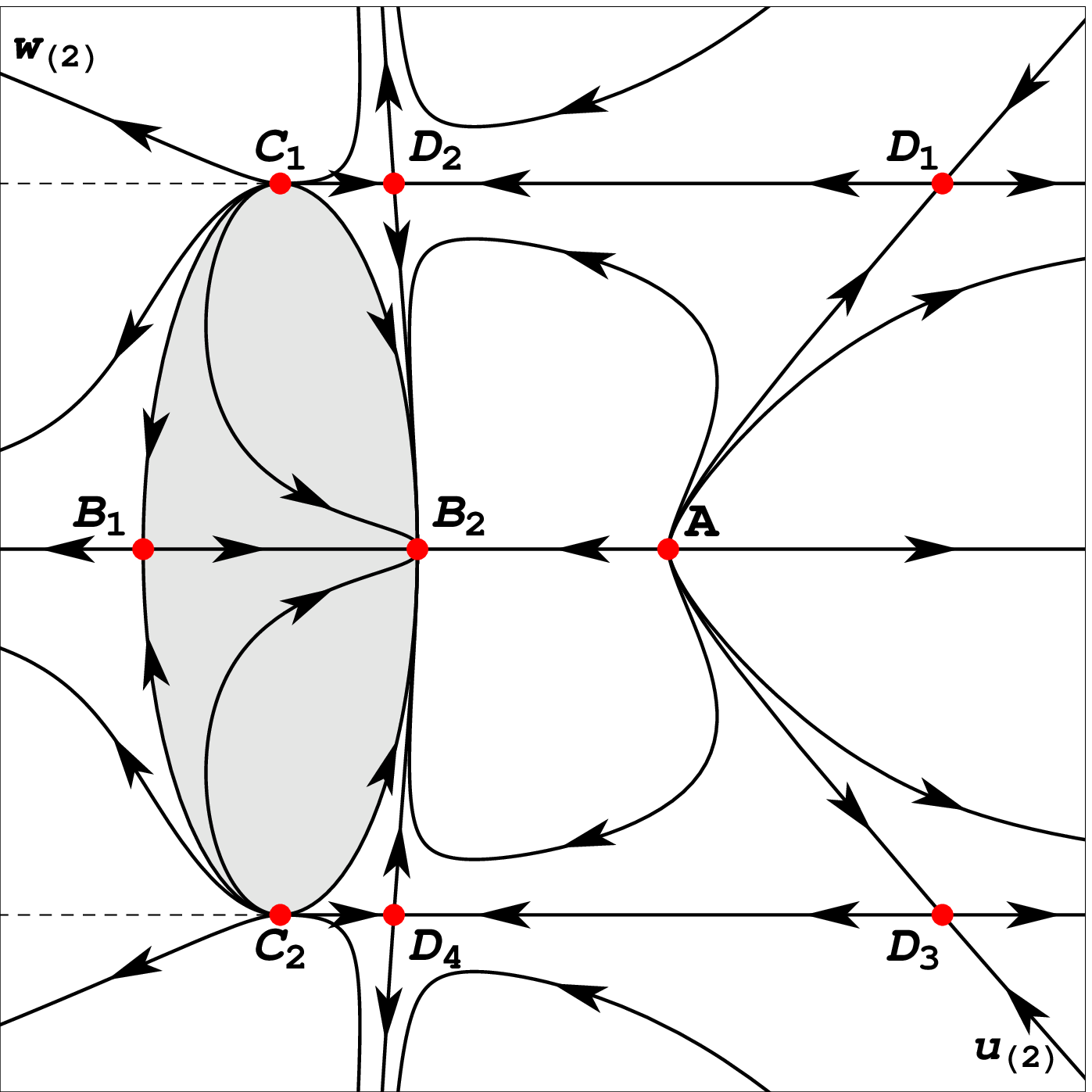}
		\includegraphics[scale=0.5]{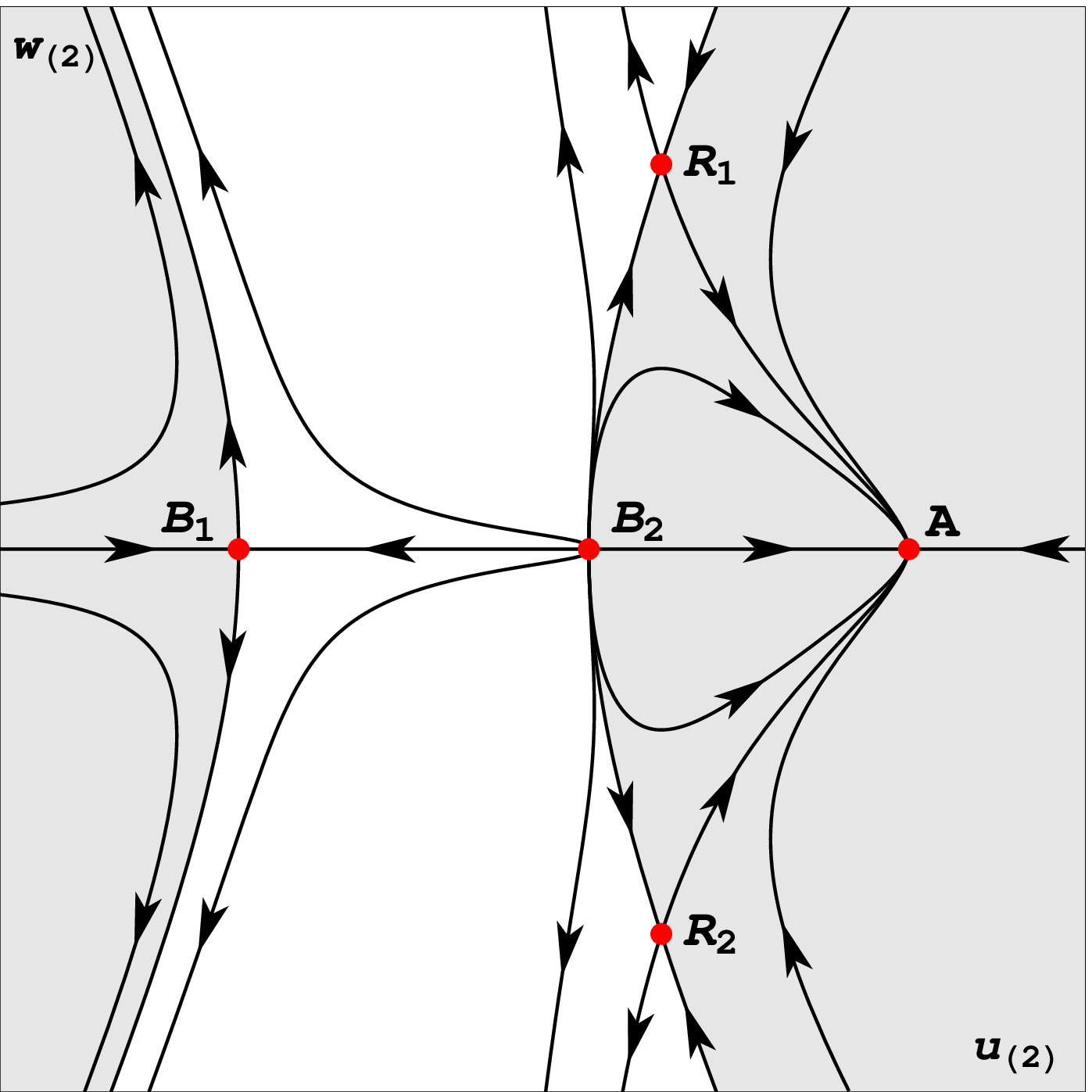}
		\caption{The phase space portraits for the dynamical system \eqref{eq:sys_inf_2} at the 
invariant manifold at infinity defined by $v_{(2)}\equiv0$. The following values of the model 
parameters were assumed: $\ve=-1$, $\frac{1}{6}<\xi<\frac{3}{16}$ (top left),
		$\ve=+1$, $\frac{1}{6}<\xi<\frac{3}{16}$ (top right),
		$\ve=-1$, $\frac{3}{16}<\xi<\frac{1}{4}$ (middle left),
			$\ve=+1$, $\frac{3}{16}<\xi<\frac{1}{4}$ (middle right), 
			$\ve=-1$, $\xi>\frac{1}{4}$ (bottom left), 
			$\ve=+1$, $\xi>\frac{1}{4}$ (bottom right).}
\label{fig:5}
	\vspace{-80pt}
\end{figure}

In figures \ref{fig:4} and \ref{fig:5} we presented the phase space diagrams of the system \eqref{eq:sys_inf_2} at the invariant manifold at infinity. The diagrams represent all representative cases for all possible values of the non-minimal coupling constant $\xi$. The shaded regions denote nonphysical regions of the phase space where the energy conservation condition \eqref{eq:cc_2} is not fulfilled. The dashed horizontal lines denote the singularity of the time transformation \eqref{eq:time_inf_2} where $w_{(2)}^{2}=\ve6\xi(1-6\xi)$.

In table \ref{tab:4} we gathered the eigenvalues of the linearisation matrix of the system \eqref{eq:sys_inf_2} calculated at the critical points. The stability conditions can be directly established for different values of the non-minimal coupling constant $\xi$. The values of the effective equation of state parameter give the physical interpretation of the given asymptotic state.

The information gathered in table \ref{tab:4} can be used in order to obtain linearised solutions in the vicinity of the critical points. In the next section we will concentrate our attention on the physical properties of the states. Here we give the linearised solutions in the vicinity of the critical points $B$, which as it turns out, represent very interesting asymptotic states from the physical point of view. 

The critical points $B$ in the chart $U_{(2)}$ are located at
$$
\left( u_{(2)}^{*}=-6\xi\pm\sqrt{-6\xi(1-6\xi)}\,,\quad v_{(2)}^{*} = 0\,,\quad w_{(2)}^{*}=0\,\right)\,
$$
and they exist only for $\xi<0$ or $\xi>\frac{1}{6}$. The eigenvalues of the linearisation matrix are
\begin{equation}
\begin{split}
\lambda_{1} & = -\ve6\xi(1-6\xi)(3+2u_{(2)}^{*})\,,\\
\lambda_{2} & = -\ve6\xi(1-6\xi)(3+u_{(2)}^{*})\,,\\
\lambda_{3} & = \ve6\xi(1-6\xi)u_{(2)}^{*}\,.
\end{split}
\end{equation}
Matching the existence conditions we obtain direct conditions for the stability of the given states. For the critical point denoted as $B_{1}$ on the phase space diagrams \ref{fig:4} and \ref{fig:5} with $u_{(2)}^{*}=-6\xi-\sqrt{-6\xi(1-6\xi)}$ we have the following conditions for a stable node with $\lambda_{i}<0$ 
$$
\ve=-1\, :\, \xi<0\,\vee \,\frac{1}{6}<\xi<\frac{3}{16}\,,
$$
and for an unstable node with $\lambda_{i}>0$
$$
\ve=+1\, :\, \xi<0\,\vee \,\frac{1}{6}<\xi<\frac{3}{16}\,.
$$
The state represented by the critical point $B_{1}$ is a transient one in the form of a saddle otherwise. The critical point denoted as $B_{2}$ on the phase space diagrams \ref{fig:4} and \ref{fig:5} with $u_{(2)}^{*}=-6\xi+\sqrt{-6\xi(1-6\xi)}$ is a stable node with $\lambda_{i}<0$ for
$$
\ve=-1\, :\, \xi>\frac{1}{6}\,,
$$
and an unstable node with $\lambda_{i}>0$ for
$$
\ve=+1\, :\, \xi>\frac{1}{6}\,,
$$
otherwise it represents a transient state.

Finally, the solutions in the vicinity of those points are in the plain form
\begin{equation}
\label{eq:lin_B}
\begin{split}
u_{(2)}(\eta_{(2)}) & = u_{(2)}^{*} +\Delta u_{(2)} e^{\lambda_{1}\eta_{(2)}}\,,\\
v_{(2)}(\eta_{(2)}) & = \Delta v_{(2)} e^{\lambda_{2}\eta_{(2)}}\,,\\
w_{(2)}(\eta_{(2)}) & = \Delta w_{(2)} e^{\lambda_{3}\eta_{(2)}}\,.
\end{split}
\end{equation}
The stability conditions are given for the time $\eta_{(2)}$ which is directly connected with the scale factor via \eqref{eq:time_inf_2}. In the next section when investigating the physical interpretation of the given states we will establish stability condition according to the expansion of the universe. 

In order to complete dynamical analysis at infinity of the phase space we need to investigate the last char of the projective coordinates. The dynamical variables of the chart $U_{(3)}$ lead to the following dynamical system
\begin{equation}
\label{eq:sys_inf_3}
\begin{split}
\frac{\ud u_{(3)}}{\ud\eta_{(3)}} & = 
	-u_{(3)}\big(1-\ve6\xi(1-6\xi)v_{(3)}^{2}\big) - \\ & \quad
	-\frac{3}{2}(u_{(3)}+6\xi v_{(3)}) 
	\Big(\Omega_{\Lambda,0}w_{(3)}^{2} +\frac{1}{3}-\ve(1-6\xi)u_{(3)}^{2} -
	\ve2\xi(u_{(3)}+v_{(3)})^{2}\Big)\,,\\
\frac{\ud v_{(3)}}{\ud\eta_{(3)}} & = 
	u_{(3)}\big(1-\ve6\xi(1-6\xi)v_{(3)}^{2}\big)\,,\\
\frac{\ud w_{(3)}}{\ud\eta_{(3)}} & = 
	-\frac{3}{2}w_{(3)}\Big( \Omega_{\Lambda,0}w_{(3)}^{2}
	-1-\ve(1-6\xi)u_{(3)}^{2} - \ve2\xi(u_{(3)}+v_{(3)})^{2} + \\ & \hspace{2cm}
	+\ve8\xi(1-6\xi)v_{(3)}^{2}\Big)\,,
\end{split}
\end{equation}
where the time function along the phase curves is now given by
\begin{equation}
\label{eq:time_inf_3}
\frac{\ud}{\ud\eta_{(3)}} = w_{(3)}^{2}\frac{\ud}{\ud\tau} =
\big(1-\ve6\xi(1-6\xi)v_{(3)}^{2}\big)\frac{\ud}{\ud\ln{a}}\,.
\end{equation}

In table \ref{tab:3} we gathered the critical points of the system \eqref{eq:sys_inf_3} located at the invariant manifold at infinity, together with conditions for their existence.

\begin{table}
	\centering
	\begin{tabular}{|>{\footnotesize}c|>{\footnotesize}l|>{\footnotesize}l|>{\footnotesize}c|}
		\hline
		& $u_{(3)}^{*}$ & $v_{(3)}^{*}$ & existence \\
		\hline
		M & $0$ & $0$ & $\forall\xi\in\mathbf{R}$ \\
		R & $0$ & $\pm\frac{1}{\sqrt{\ve6\xi}}$ & $\ve=-1$: $\xi<0$ or
		$\ve=+1$: $\xi>0$\\
		C & $-6\xi v_{(3)}^{*}$ & $\pm\frac{1}{\sqrt{\ve6\xi(1-6\xi)}}$
		  & $\ve=-1$: $\xi<0\lor\xi>\frac{1}{6}$ or $\ve=+1$:
		$0<\xi<\frac{1}{6}$\\
		D & $-\frac{\ve2\xi v_{(3)}^{*} \pm\sqrt{-\ve\frac{4}{3}\xi}}{\ve(1-4\xi)}$ &
		$\pm\frac{1}{\sqrt{\ve6\xi(1-6\xi)}}$ & $\ve=-1$:
		$\xi>\frac{1}{6}$\\ 
		\hline
	\end{tabular}
	\caption{Critical points of the system \eqref{eq:sys_inf_3} in the map $U_{(3)}$ at infinity 
of the phase space defined by the condition $w_{(3)}\equiv0$.}
\label{tab:3}
\end{table}

The physical regions of the phase space are determined from the energy conservation condition \eqref{eq:constr}. In the projective coordinates of the chart $U_{(3)}$ we have the following
\begin{equation}
\left(\Omega_{\Lambda,0}+\Omega_{m,0}\left(\frac{a}{a_{0}}\right)^{-3}\right) w_{(3)}^{2} = 1 -\ve(1-6\xi)u_{(3)}^{2}-\ve6\xi(u_{(3)}+v_{(3)})^{2}
\end{equation}
and at infinity of the phase space defined by the condition $w_{(3)}\equiv0$ we have
\begin{equation}
\ve(1-6\xi)u_{(3)}^{2}+\ve6\xi(u_{(3)}+v_{(3)})^{2}\le1\,.
\end{equation}

\begin{figure}
\centering
		\includegraphics[scale=0.5]{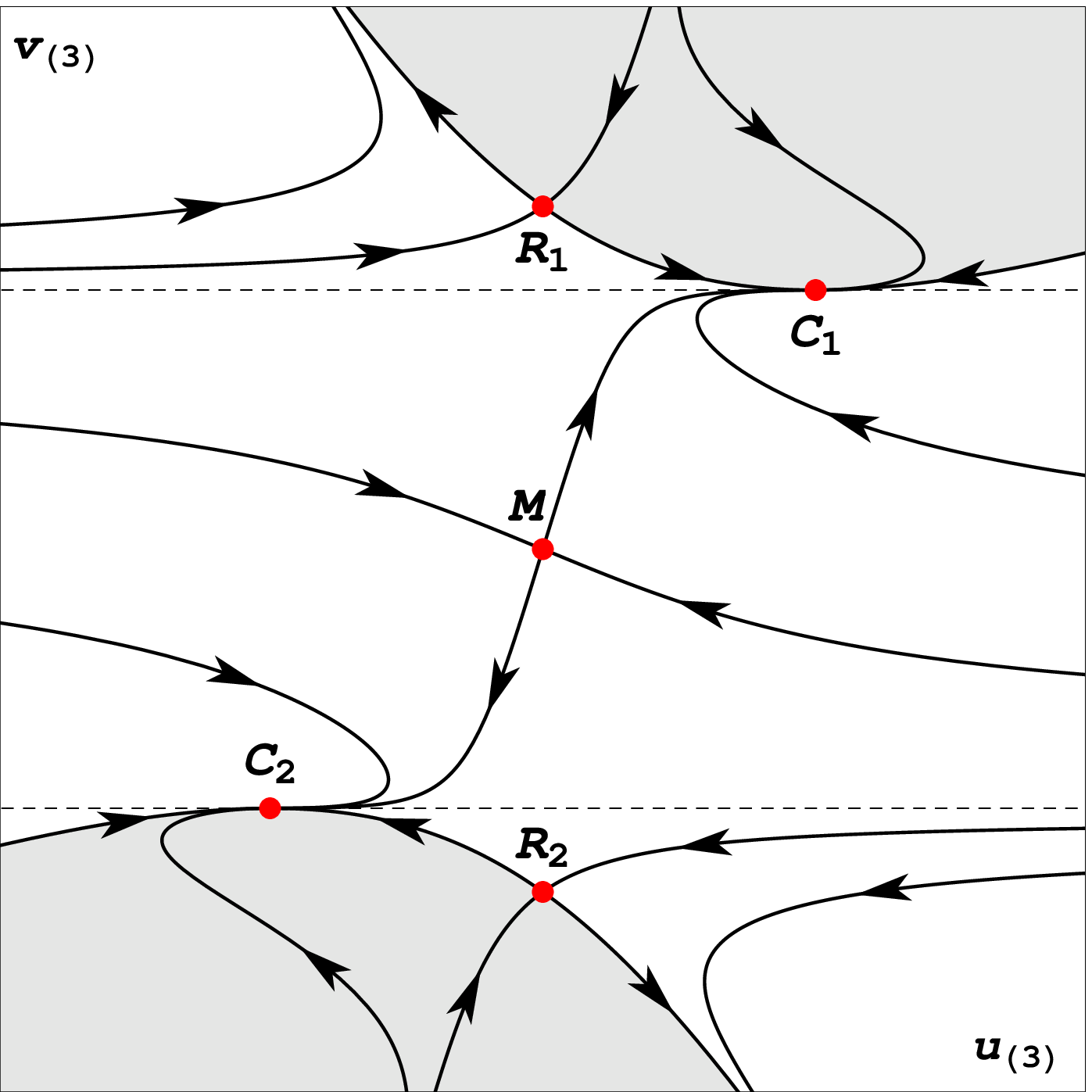}
		\includegraphics[scale=0.5]{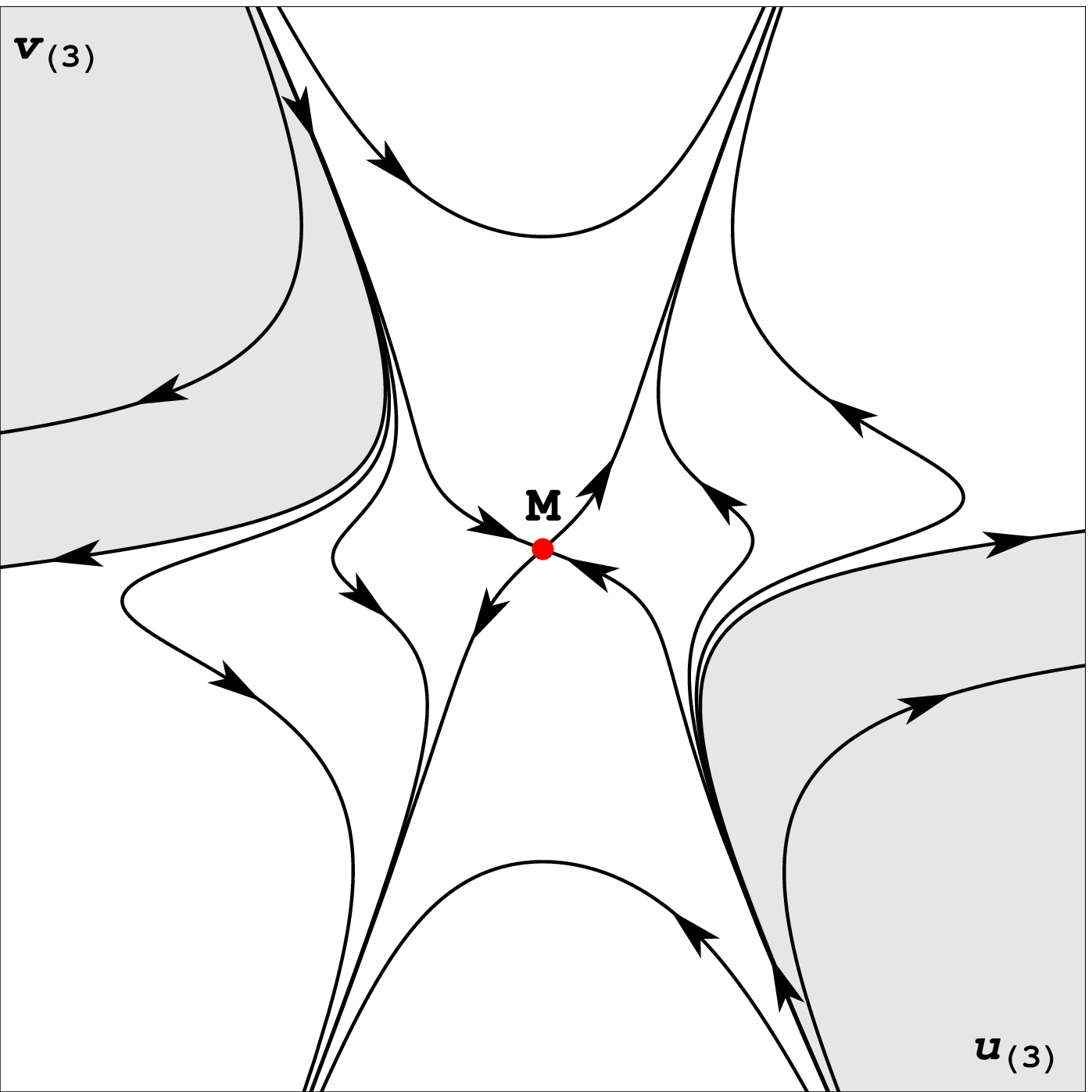}\\
		\includegraphics[scale=0.5]{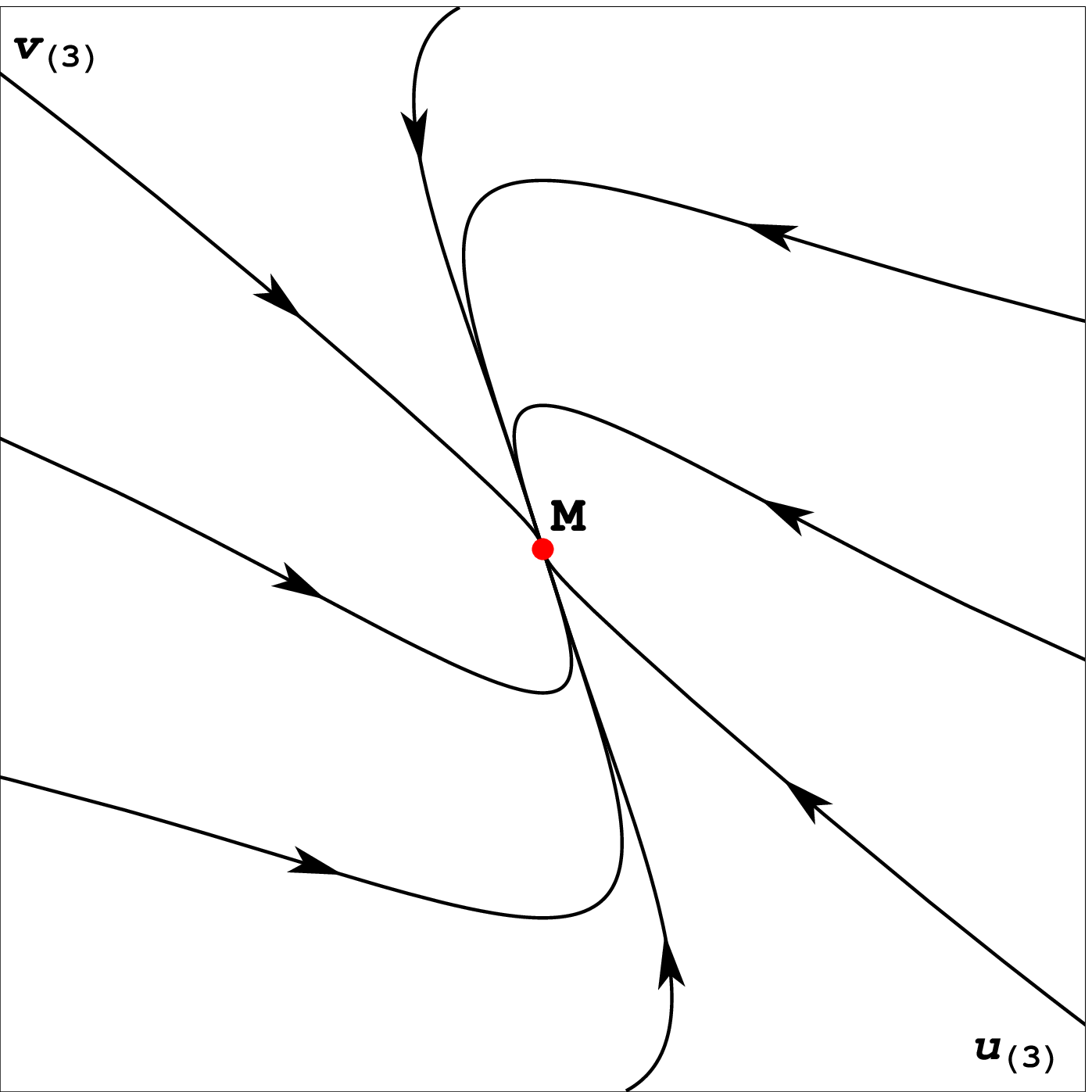}
		\includegraphics[scale=0.5]{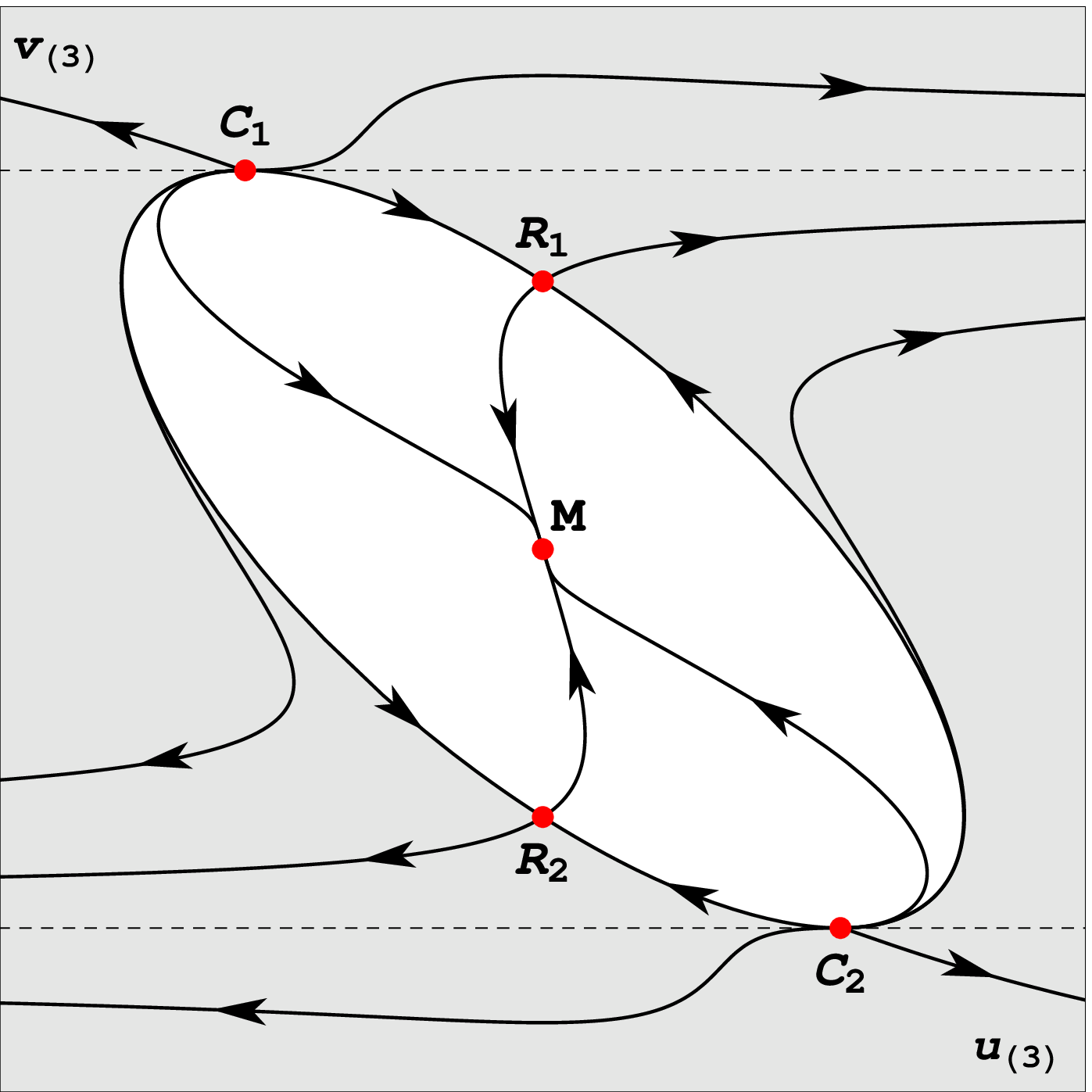}
		\caption{The phase space portraits for the dynamical system \eqref{eq:sys_inf_3} at the 
invariant manifold at infinity defined by $w_{(3)}\equiv0$. The following values of the model 
parameters were assumed: $\ve=-1$, $\xi<0$ (top left) and $\ve=+1$, $\xi<0$ (top right), $
\ve=-1$, $0<\xi<\frac{1}{6}$ (bottom left) and $\ve=+1$, $0<\xi<\frac{1}{6}$ (bottom right).}
\label{fig:6}
\end{figure}

\begin{figure}
\centering
		\includegraphics[scale=0.5]{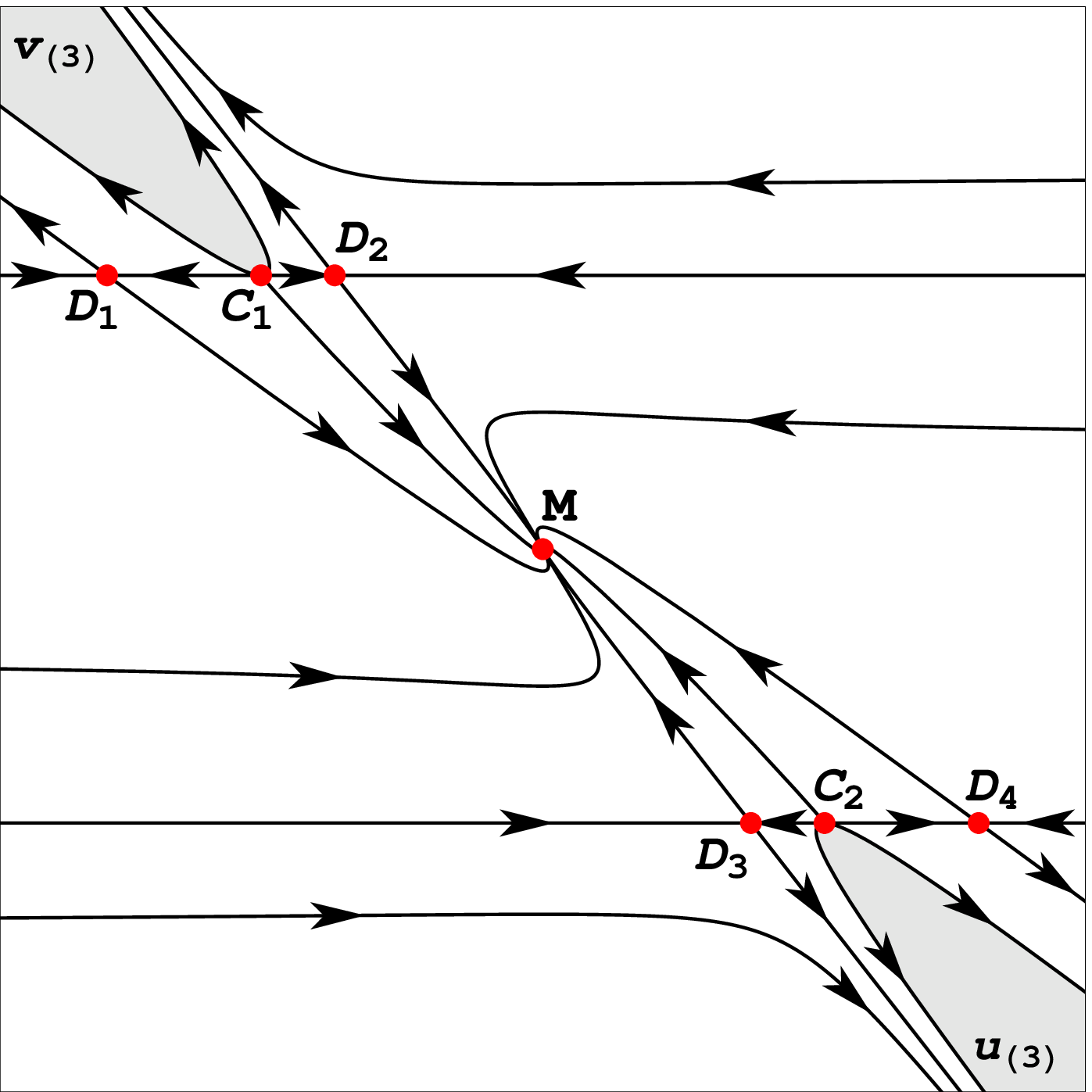}
		\includegraphics[scale=0.5]{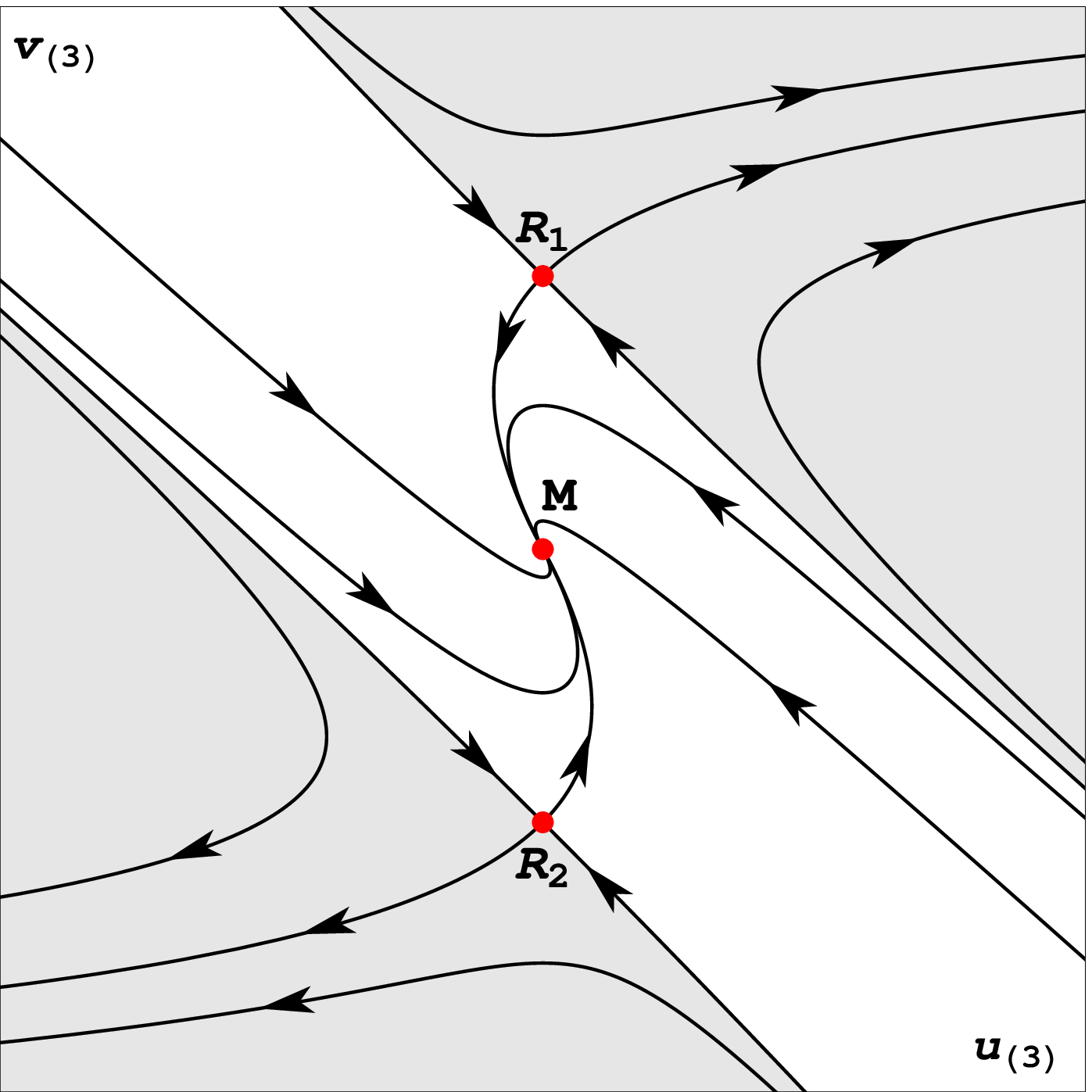}\\
		\includegraphics[scale=0.5]{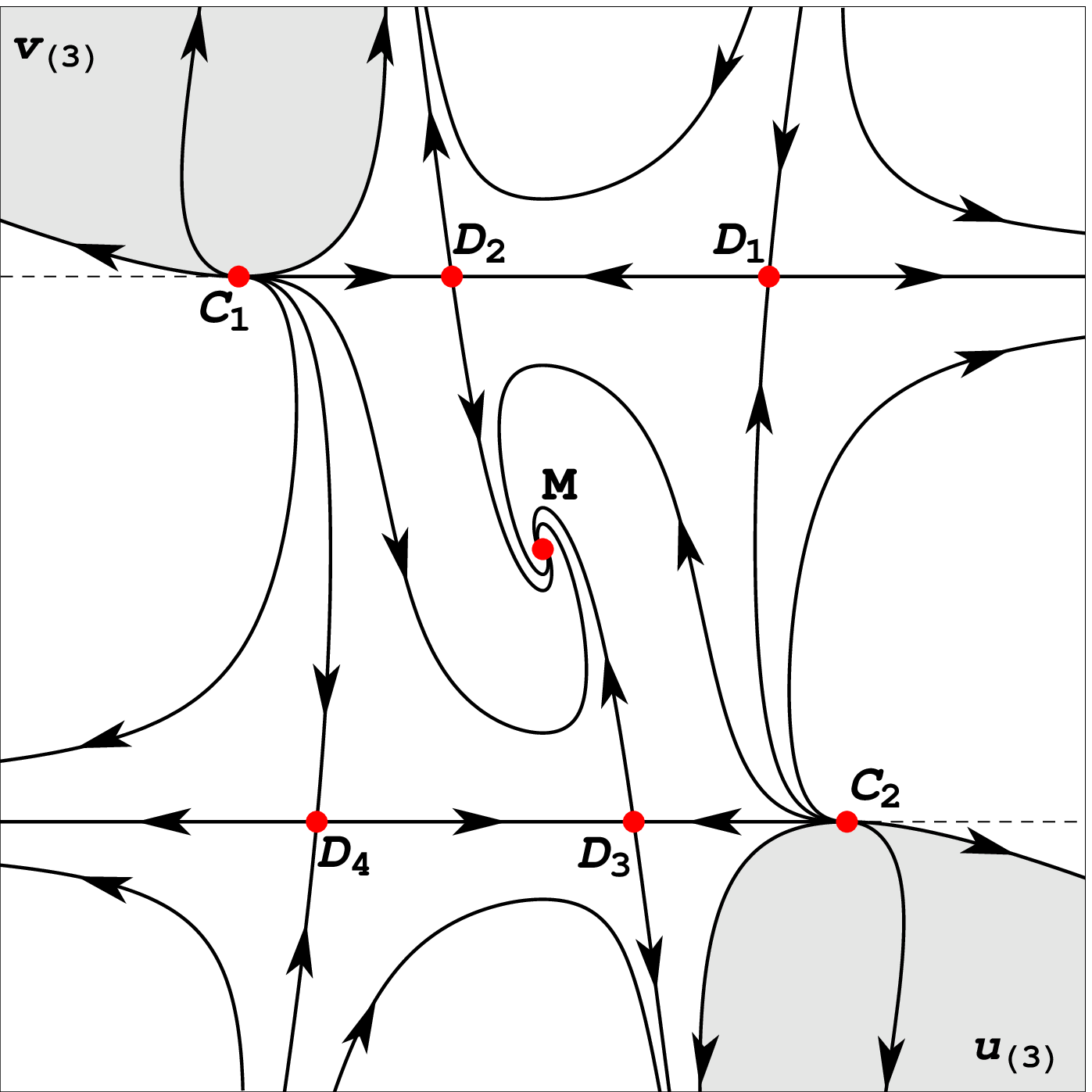}
		\includegraphics[scale=0.5]{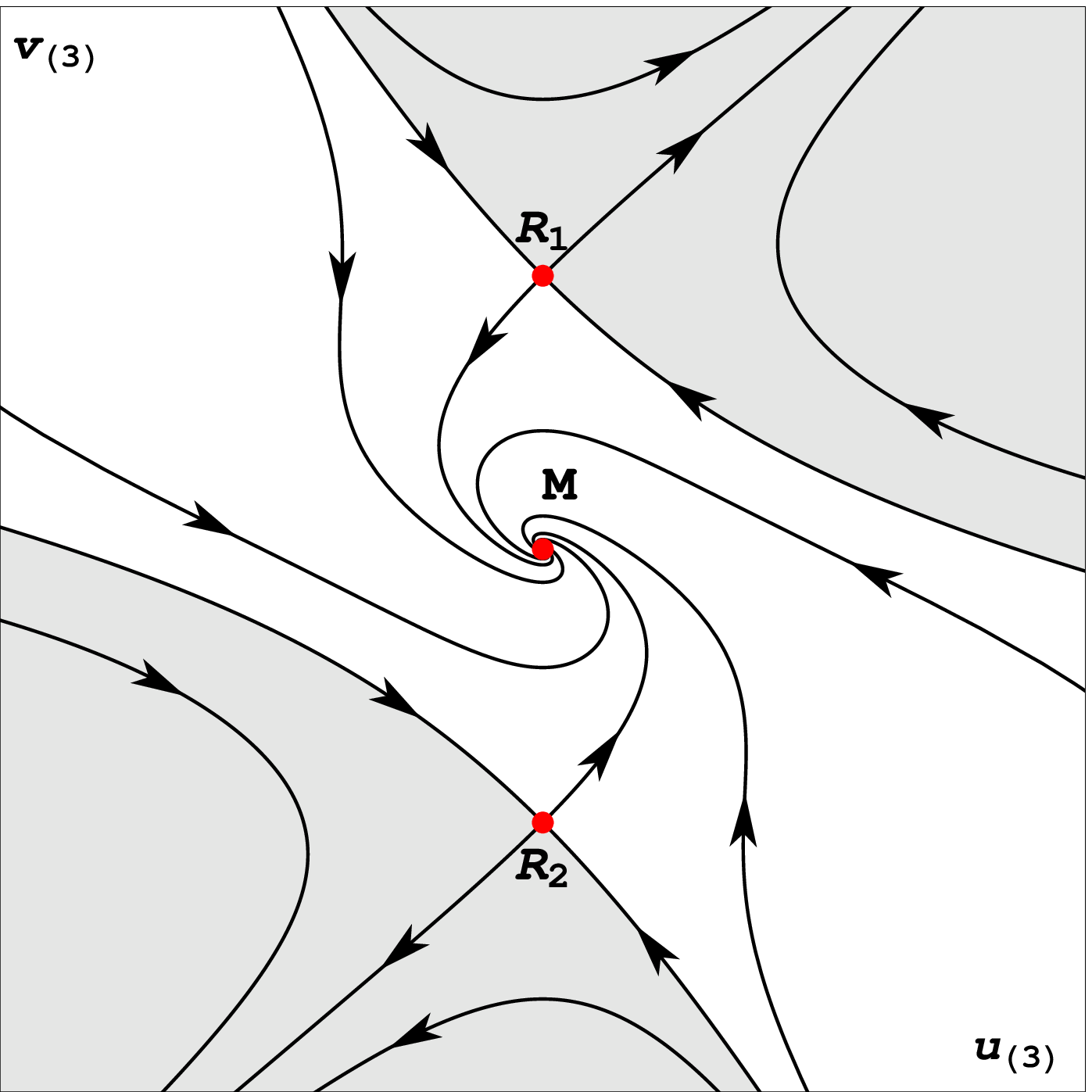}
		\caption{The phase space portraits for the dynamical system \eqref{eq:sys_inf_3} at the 
invariant manifold at infinity defined by $w_{(3)}\equiv0$. The following values of the model 
parameters were assumed:
		$\ve=-1$, $\frac{1}{6}<\xi<\frac{3}{16}$ (top left),
		$\ve=+1$, $\frac{1}{6}<\xi<\frac{3}{16}$ (top right),
		$\ve=-1$, $\xi>\frac{3}{16}$ (bottom right), 
		$\ve=+1$, $\xi>\frac{3}{16}$ (bottom right).}
\label{fig:7}
\end{figure}

In figures \ref{fig:6} and \ref{fig:7} we presented the phase space diagrams of the system \eqref{eq:sys_inf_3} at the invariant manifold at infinity. The shaded fields correspond to nonphysical regions of the phase space. The diagrams represent all the representative cases for all possible values of the non-minimal coupling constant $\xi$. The dashed lines again represent singularities of the time transformation \eqref{eq:time_inf_3} where $\ve6\xi(1-6\xi)v_{(3)}^{2}=1$. On figure \ref{fig:7} for the phantom scalar field with $\ve=-1$ singular lines form horizontal separatrices of the saddle type critical points $D_{i}$.

In table \ref{tab:6} we give eigenvalues of the linearisation matrix calculated at the given critical point together with the effective equation of state parameters. The eigenvalues give the stability conditions while the latter informs about physical behaviour at the asymptotic state. It what follows we present the linearised solutions in the vicinity of the critical points which represent the most interesting states from the physical point of view.

The critical point $M$ with the effective equation of state $w_{\text{eff}}=0$ located at
$$
\left(u_{(3)}^{*}=0\,,\quad v_{(3)}^{*}=0\,,\quad w_{(3)}^{*}=0\,\right)\,,
$$
exists for any value of the non-minimal coupling constant $\xi$ and has the following eigenvalues 
$$
\lambda_{1,2}=-\frac{1}{4}(3\pm\sqrt{3(3-16\xi)})\,,\qquad \lambda_{3}=\frac{3}{2}\,.
$$
In any possible case it represents a saddle type critical point. The linearised solutions are
\begin{equation}
\label{eq:lin_M}
\begin{split}
u_{(3)}(\eta_{(3)}) & = \frac{2}{\sqrt{3(3-16\xi)}}\Big(\big(3\xi\Delta v_{(3)}-\lambda_{1}\Delta u_{(3)}\big) e^{\lambda_{1}\eta_{(3)}}-\big(3\xi\Delta v_{(3)}-\lambda_{2}\Delta u_{(3)}\big) e^{\lambda_{2}\eta_{(3)}}\Big)\,,\\
v_{(3)}(\eta_{(3)}) & = \frac{2}{\sqrt{3(3-16\xi)}}\Big(\big(\lambda_{2}\Delta v_{(3)}-\Delta u_{(3)}\big) e^{\lambda_{1}\eta_{(3)}}-\big(\lambda_{1}\Delta v_{(3)}-\Delta u_{(3)}\big) e^{\lambda_{2}\eta_{(3)}}\Big)\,,\\
w_{(3)}(\eta_{(3)}) & = \Delta w_{(3)} e^{\frac{3}{2}\eta_{(3)}}\,,
\end{split}
\end{equation}
where $\Delta u_{(3)} = u_{(3)}^{(i)}$, $\Delta v_{(3)} = v_{(3)}^{(i)}$ and $\Delta w_{(3)} = w_{(3)}^{(i)}$ are the initial conditions.

The critical points $R$ with the effective equation of state parameter $w_{\text{eff}}=\frac{1}{3}$ are located at
$$
\left(u_{(3)}^{*}=0\,,\quad v_{(3)}^{*}=\pm\sqrt{\frac{1}{\ve6\xi}}\,,\quad w_{(3)}^{*}=0\,\right)\,.
$$
They exist only for $\ve \xi>0$, i.e., for the canonical scalar field with $\ve=+1$ only for $\xi>0$ while for the phantom scalar field with $\ve=-1$ only for $\xi<0$. The eigenvalues are 
$$
\lambda_{1}=-6\xi\,,\quad \lambda_{2}=6\xi\,,\quad \lambda_{3}=12\xi\,,
$$
and for any possible values of the non-minimal coupling constant $\xi$ it represents a saddle type critical point. The linearised solutions are
\begin{equation}
\label{eq:lin_R}
\begin{split}
u_{(3)}(\eta_{(3)}) & = 
\frac{1}{2}\left(\Delta u_{(3)}-\Delta v_{(3)}\right)\,e^{-6\xi\eta_{(3)}} +
\frac{1}{2}\left(\Delta u_{(3)}+\Delta v_{(3)}\right)\,e^{6\xi\eta_{(3)}}\,,\\
v_{(3)}(\eta_{(3)}) & = v^{*}_{(3)}
-\frac{1}{2}\left(\Delta u_{(3)}-\Delta v_{(3)}\right)\,e^{-6\xi\eta_{(3)}} +
\frac{1}{2}\left(\Delta u_{(3)}+\Delta v_{(3)}\right)\,e^{6\xi\eta_{(3)}}\,,\\
w_{(3)}(\eta_{(3)}) & = \Delta w_{(3)}\,e^{12\xi\eta_{(3)}}\,,
\end{split}
\end{equation}
where $\Delta u_{(3)} = u_{(3)}^{(i)}$, $\Delta v_{(3)} = v_{(3)}^{(i)}-v_{(3)}^{*}$ and $\Delta w_{(3)} = w_{(3)}^{(i)}$ are the initial conditions.

Note that this critical point denoted on the all relevant phase space diagrams as $R$ in the original variables corresponds to the condition $\ve\xi\kappa^{2}\phi^{2}=1$, which in the Einstein frame formulation of the theory leads to vanishing of the conformal factor and the conformal metric degenerates. For an isotropic evolution this is a regular point while a strong anisotropic curvature singularity forms there as it was show in \cite{Starobinsky:1981} (see also \cite{Figueiro:2009mm}).  In the Jordan frame, followed in this paper, this condition is represented by the critical point of a saddle type where two of the separatrices form the boundary between physical and nonphysical regions of the phase space.

The last two critical points which are interesting form the physical point of view are denoted on the phase space diagrams as $C_{1}$ and $C_{2}$. The effective equation of state parameter calculated at these points give an infinite value pointing to the presence of singularity. The coordinates of the asymptotic states under considerations are
$$
\left(u_{(3)}^{*}=-6\xi v_{(3)}^{*}\,,\quad v_{(3)}^{*}=\pm\frac{1}{\sqrt{\ve6\xi(1-6\xi)}}\,,\quad w_{(3)}^{*}=0\,\right)\,,
$$
with the eigenvalues
$$
\lambda_{1}=\lambda_{3}=6\xi\,,\quad \lambda_{2}=12\xi\,.
$$
Together with the existence condition from table \ref{tab:3} we have that the critical point is in the form of a stable node only for the phantom scalar field with $\ve=-1$ and for $\xi<0$ and is an unstable node for the phantom scalar field for $\xi>\frac{1}{6}$ while for the canonical scalar field $\ve=+1$ it exists and is in the form of an unstable node for $0<\xi<\frac{1}{6}$.
The linearised solutions are
\begin{equation}
\label{eq:lin_C}
\begin{split}
u_{(3)}(\eta_{(3)}) & = u^{*}_{(3)} +\big(\Delta u_{(3)}+2(1-3\xi)\Delta v_{(3)}\big)\,e^{6\xi\eta_{(3)}}
-2(1-3\xi)\Delta v_{(3)}\, e^{12\xi\eta_{(3)}}\,,\\
v_{(3)}(\eta_{(3)}) & = v^{*}_{(3)}+\Delta v_{(3)}\, e^{12\xi\eta_{(3)}}\,,\\
w_{(3)}(\eta_{(3)}) & = \Delta w_{(3)}\, e^{6\xi\eta_{(3)}}\,,
\end{split}
\end{equation}
where $\Delta u_{(3)} = u_{(3)}^{(i)}-u_{(3)}^{*}$, $\Delta v_{(3)} = v_{(3)}^{(i)}-v_{(3)}^{*}$ and $\Delta w_{(3)} = w_{(3)}^{(i)}$ are the initial conditions.

What remains is a discussion the other non-hyperbolic critical points with one eigenvalue vanishing. In order to obtain a qualitative behaviour in the vicinity of such critical points one is forced to use the centre manifold theorem \cite{Perko:book,Wiggins:book,Hrycyna:2010yv}. This is beyond the scope of the current study and will be presented elsewhere.  

\section{Physics from dynamics}

In the previous section we obtained the linearised solutions of the dynamics in the vicinity of the stationary states, i.e. critical points. Now we want to extract a physical meaning from these solutions in order to find a physical interpretation of specific states of the universe represented by the critical points. 

In tables \ref{tab:4}, \ref{tab:5} and \ref{tab:6} we have gathered the effective equation of state parameters and the eigenvalues of the linearisation matrix for all the critical points in all three charts covering the phase space at infinity. The eigenvalues represent stability conditions for a given state in the generic case, i.e. for a generic value of the non-minimal coupling constant $\xi$. 

\begin{table}
\centering
{\renewcommand{\arraystretch}{1.25}
	\begin{tabular}{|>{\small}c|>{\scriptsize}l|>{\scriptsize}c|}
		\hline
		& {\footnotesize eigenvalues } & {\footnotesize $w_{\textrm{eff}}$} \\
		\hline
		A & $\Big( -\ve\frac{9}{4\xi}(1-4\xi)^{2}(1-6\xi),\, \ve\frac{3}{4\xi}(1-4\xi)(1-6\xi)(3-16\xi),\,-\ve3(1-4\xi)(1-6\xi) \Big)$ & $-\frac{4\xi}{3(1-4\xi)}$\\
		B & $\Big(\ve3(1-6\xi)(1+10\xi v^{*}_{(1)}),\, \ve3(1-6\xi)(1+8\xi v^{*}_{(1)}),\, \ve6\xi(1-6\xi)v^{*}_{(1)})\Big)$ & $\frac{1+4\xi v^{*}_{(1)}}{1+12\xi v^{*}_{(1)}}$\\
		C & $\Big(\ve(1-6\xi),\, \ve(1-6\xi),\, \ve2(1-6\xi)\Big)$ & $\pm\infty$\\
		D & $\Big(0,\, -2(1-6\xi)\sqrt{-3(1-4\xi(1-v^{*}_{1}))},\, 2(1-6\xi)\sqrt{-3(1-4\xi(1-v^{*}_{1}))}\Big)$ & $-$\\
		E & $\Big(0,\,-\ve\frac{3}{2}(1-4\xi),\, -\ve\frac{3}{2}(1-4\xi)\Big)$ & $\pm\infty$\\
		\hline
	\end{tabular}
}
	\caption{The eigenvalues and the effective equation of state parameter for the critical points at infinity of the system \eqref{eq:sys_inf_1} in the map $U_{(1)}$ gathered in table \ref{tab:1}. }
\label{tab:4}
\end{table}

\begin{table}
\centering
{\renewcommand{\arraystretch}{1.25}
	\begin{tabular}{|>{\small}c|>{\scriptsize}l|>{\scriptsize}c|}
		\hline
		& {\footnotesize eigenvalues} & {\footnotesize $w_{\textrm{eff}}$} \\
		\hline
		A & $\Big(\ve3\xi\frac{(1-6\xi)(3-16\xi)}{1-4\xi},\,-\ve9\xi(1-6\xi),\,-\ve12\xi^{2}\frac{1-6\xi}{1-4\xi}\Big)$ & $-\frac{4\xi}{3(1-4\xi)}$\\
		B & $\Big(-\ve6\xi(1-6\xi)(3+2u^{*}_{(2)}),\,-\ve6\xi(1-6\xi)(3+u^{*}_{(2)}),\,\ve6\xi(1-6\xi)u^{*}_{(2)})\Big)$& $1+\frac{4}{3}u^{*}_{(2)}$\\
		C & $\Big(\ve36\xi^{2}(1-6\xi),\,\ve36\xi^{2}(1-6\xi),\,\ve72\xi^{2}(1-6\xi)\Big)$ & $\pm\infty$\\
		D & $\Big(\ve12\xi(1-6\xi)u^{*}_{(2)},\, 0,\,-\ve12\xi(1-6\xi)u^{*}_{(2)}\Big)$ & $-$\\
		R & $\Big(-6\xi\sqrt{\ve6\xi}w^{*}_{(2)},\,\ve72\xi^{2},\,6\xi\sqrt{\ve6\xi}w^{*}_{(2)}\Big)$ & $\frac{1}{3}$\\
		\hline
	\end{tabular}
}
	\caption{The eigenvalues and the effective equation of state parameter for the critical points at infinity of the system \eqref{eq:sys_inf_2} in the map $U_{(2)}$ gathered in table \ref{tab:2}. }
\label{tab:5}
\end{table}


We start our discussion from the two critical points in the chart $U_{(3)}$ which represent a barotropic matter domination and a radiation dominated epoch in evolution of the universe.  

The critical point with the effective equation of state parameter $w_{\text{eff}}=0$ denoted on the phase space diagrams as $M$ has at least one unstable direction. From the coordinates transformation in the chart $U_{(3)}$ \eqref{eq:charts} we have that
$$
\frac{H(a)}{H(a_{0})} = \frac{1}{w_{(3)}(\eta_{(3)})}\,,
$$ 
and using the last linearised solution form \eqref{eq:lin_M} we have that the Hubble function in the vicinity of this critical point behaves as
$$
\frac{H(a)}{H(a_{0})} = \frac{1}{\Delta w_{(3)}} e^{-\frac{3}{2}\eta_{(3)}}\,.
$$
The time transformation \eqref{eq:time_inf_3} up to linear terms in initial conditions gives
$$
\eta_{(3)} \approx \ln{\left(\frac{a}{a^{(i)}}\right)}\,,
$$
where $a$ is the scale factor and $a^{(i)}$ is the initial value of the scale factor in the vicinity of the critical point. Finally we obtain that the Hubble function as the function of the scale factor is
\begin{equation}
\label{eq:Hub_M}
\frac{H(a)}{H(a_{0})} \approx \frac{H(a^{(i)})}{H(a_{0})}\left(\frac{a}{a^{(i)}}\right)^{-\frac{3}{2}}\,,
\end{equation}
which is exactly the Hubble function in the barotropic dust matter domination phase of expansion of the universe.

The asymptotic states denoted on the phase space diagrams as $R$ have the effective equation of state parameter $w_{\text{eff}}=\frac{1}{3}$. For arbitrary values of the non-minimal coupling constant $\xi$ they are in the form of a saddle type critical points. Again from the coordinates of the chart $U_{(3)}$ \eqref{eq:charts} we have that
$$
\frac{H(a)}{H(a_{0})} = \frac{1}{w_{(3)}(\eta_{(3)})}\,.
$$
From the linearised solutions in the vicinity of this state \eqref{eq:lin_R} we have the Hubble function 
$$
\frac{H(a)}{H(a_{0})} = \frac{1}{\Delta w_{(3)}} e^{-12\xi\eta_{(3)}}\,.
$$
The time transformation \eqref{eq:time_inf_3} up to linear terms in initial conditions leads to
$$
\ud\ln{a} \approx \Big(6\xi+\ve6\xi(1-6\xi)v^{*}_{(3)}\big((\Delta u_{(3)}-\Delta v_{(3)})e^{-6\xi\eta_{(3)}}-(\Delta u_{(3)} -\Delta v_{(3)})e^{6\xi\eta_{(3)}}\big)\Big)\ud\eta_{(3)}\,,
$$
integrating this expression and expanding for small values of the time $\eta_{(3)}$ we obtain 
$$
\ln{\left(\frac{a}{a^{(i)}}\right)} \approx 6\xi\eta_{(3)}\big(1-\ve2(1-6\xi)v^{*}_{(3)}\Delta v_{(3)}\big)\,,
$$
where, again, $a$ is the scale factor and $a^{(i)}$ is the initial value of the scale factor in the vicinity of the critical point. The resulting Hubble function is
\begin{equation}
\label{eq:Hub_R}
\frac{H(a)}{H(a_{0})} \approx \frac{H^{(i)}}{H(a_{0})} \left(\frac{a}{a^{(i)}}\right)^{-2}\left(\frac{a}{a^{(i)}}\right)^{-\ve4(1-6\xi)v^{*}_{(3)}\Delta v_{(3)}}\,,
\end{equation}
which is the Hubble function during the radiation domination epoch in the history of the universe with the small correction proportional to initial conditions and the value of the non-minimal coupling constant.

\begin{table}
\centering
{\renewcommand{\arraystretch}{1.25}
	\begin{tabular}{|>{\small}c|>{\scriptsize}l|>{\scriptsize}c|}
		\hline
		& {\footnotesize eigenvalues} & {\footnotesize $w_{\textrm{eff}}$}\\
		\hline
		M & $\Big(-\frac{1}{4}\left(3+\sqrt{3(3-16\xi)}\right),\,
		-\frac{1}{4}\left(3-\sqrt{3(3-16\xi)}\right),\,\frac{3}{2}\Big)$ & $0$\\
		R & $\Big(-6\xi,\,6\xi,\,12\xi\Big)$& $\frac{1}{3}$\\
		C & $\Big(6\xi,\,12\xi,\,6\xi\Big)$ & $\pm\infty$\\
		D & $\Big(\ve12\xi(1-6\xi)u^{*}_{(3)}v^{*}_{(3)},\,
		-\ve12\xi(1-6\xi)u^{*}_{(3)}v^{*}_{(3)},\,0\Big)$ & $-$\\
		\hline
	\end{tabular}
}
	\caption{The eigenvalues and the effective equation of state parameter for the critical points at infinity of the system \eqref{eq:sys_inf_3} in the map $U_{(3)}$ gathered in table \ref{tab:3}. }
\label{tab:6}
\end{table}

Investigating dynamics of cosmological models within various theories of gravity one usually encounters some sort of singularities. They can be in the form of an initial finite scale factor singularity
\cite{Barrow:1990td, Cannata:2008xc}, future finite scale factor singularities:
the sudden future singularities \cite{Barrow:2004xh}, the Big Brake singularity
\cite{Gorini:2003wa} or the Big Boost singularity \cite{Barvinsky:2008rd}. The next two asymptotic state represent precisely a finite scale factor singularity.

The critical points denoted on the phase space diagrams as $C_{1}$ and $C_{2}$ in the chart $U_{(3)}$ have the following coordinates at the invariant manifold at infinity $w_{(3)}\equiv0$ 
$$
\left(u_{(3)}^{*}=-6\xi v_{(3)}^{*}\,,\quad v_{(3)}^{*}=\pm\frac{1}{\sqrt{\ve6\xi(1-6\xi)}}\right)
$$
are in the form of a stable or a unstable node depending on the sign of the non-minimal coupling constant $\xi$. Using the linearised solution \eqref{eq:lin_C} we obtain the following form of the Hubble function 
$$
\frac{H(a)}{H(a_{0})}=\frac{1}{w_{(3)}(\eta_{(3)})}=\frac{1}{\Delta w_{(3)}}e^{-6\xi\eta_{(3)}}\,.
$$
Clearly for $\xi>0$, i.e. when the asymptotic state is in the form of an unstable node, when approaching this state $\eta_{(3)}\to-\infty$ the Hubble function goes to infinity. The same is true for $\xi<0$, i.e. when the state is in the form of a stable node and at the limit $\eta_{(3)}\to+\infty$. Now we want to show that the scale factor at this state has a finite value. From the definition of time $\eta_{(3)}$ \eqref{eq:time_inf_3} and up to linear terms in initial conditions we have the following
$$
\ud\ln{a} \approx -\ve12\xi(1-6\xi)v^{*}_{(3)} \Delta v_{(3)} e^{12\xi\eta_{(3)}}\ud\eta_{(3)}\,,
$$
and after integration we obtain
$$
\ln{\left(\frac{a^{(f)}}{a^{(i)}}\right)} \approx \ve(1-6\xi)v^{*}_{(3)}\Delta v_{(3)}\,,
$$
where $a^{(i)}$ and $a^{(f)}$ are the initial and the final values of the scale factor.
Interestingly, the finite scale factor singularity under considerations can appear in two types. Namely in one case when $a^{(f)}>a^{(i)}$ and in opposite case when $a^{(f)}<a^{(i)}$, it means that it can be either in the future as well as in the past according to the expansion of the universe. Which scenario is realised crucially depends on the form of the scalar field as well as the value of the non-minimal coupling constant $\xi$. For example in the bottom right phase space diagram on figure \ref{fig:6} asymptotic state represented by the point $C_{1}$ for the canonical scalar field $\ve=+1$, $0<\xi<\frac{1}{6}$ and initial conditions taken in the physical domain (white region) we have $\ve(1-6\xi)v^{*}_{(3)}\Delta v_{(3)}<0$ and hence $a^{(f)}< a^{(i)}$, the finite scale factor singularity lies in the past according to the expansion of the universe. The same reasoning can be applied to the phantom scalar field $\ve=-1$ in the top left phase space diagram on figure \ref{fig:6} and the phase space diagrams in the left panel of figure \ref{fig:7}. 

The asymptotic states in the form of critical points discussed so far were previously found for a wide class of the scalar field potential functions \cite{Hrycyna:2010yv}. Now we can move on to investigations of the physical properties of the critical points previously missed due to lack of the dynamical analysis at infinity of the phase space.

Two critical points in the charts $U_{(1)}$ and $U_{(2)}$ denoted on the phase space diagrams as $B_{1}$ and $B_{2}$ possess very interesting and special properties. Namely for the specific value of the non-minimal coupling constant $\xi=\frac{3}{16}$ the effective equation of state parameter calculated at those points are $w_{\text{eff}}\big|_{B_{1}}=-1$ and $w_{\text{eff}}\big|_{B_{2}}=0$. We start from the latter one. The linearised solutions \eqref{eq:lin_B} specified for the critical point $B_{2}$ and for $\xi=\frac{3}{16}$ are
\begin{equation}
\begin{split}
u_{(2)}(\eta_{(2)}) & = -\frac{3}{4} +\Delta u_{(2)} e^{\ve\frac{27}{128}\eta_{(2)}}\,,\\
v_{(2)}(\eta_{(2)}) & = \Delta v_{(2)} e^{\ve\frac{81}{256}\eta_{(2)}}\,,\\
w_{(2)}(\eta_{(2)}) & = \Delta w_{(2)} e^{\ve\frac{27}{256}\eta_{(2)}}\,,
\end{split}
\end{equation}
where we have a stable node for the phantom scalar field $\ve=-1$ and an unstable node for the canonical scalar field $\ve=+1$.

From the coordinates of the chart $U_{(2)}$ \eqref{eq:charts} and using those linearised solutions we have
$$
\frac{H(a)}{H(a_{0})} = \frac{w_{(2)}(\eta_{(2)})}{v_{(2)}(\eta_{(2)})} = \frac{\Delta w_{(2)}}{\Delta v_{(2)}} e^{-\ve\frac{27}{128}\eta_{(2)}}\,,
$$
and from \eqref{eq:time_inf_2} up to linear terms in initial conditions
$$
\ln{\left(\frac{a}{a^{(i)}}\right)} \approx \ve\frac{9}{64}\eta_{(2)}\,.
$$
Finally the Hubble function in the vicinity of the asymptotic state is given by
\begin{equation}
\frac{H(a)}{H(a_{0})}\approx\frac{H(a^{(i)})}{H(a_{0})}\left(\frac{a}{a^{(i)}}\right)^{-\frac{3}{2}}\,.
\end{equation}
It represents unstable during expansion the Einstein-de Sitter solution for both types of the scalar field.

The critical point $B_{1}$ with the effective equation of state parameter $w_{\text{eff}}\big|_{B_{1}}=-1$ for $\xi=\frac{3}{16}$ has the vanishing eigenvalue in the direction of the variable $u_{(2)}$ and is of a non-hyperbolic type. Its detailed discussion deserves a special treatment and will be presented in the next section. Nevertheless, based on the linearised solutions in the remaining directions we can find the behaviour of the Hubble function in the vicinity of this point. From the last two linearised solutions of \eqref{eq:lin_B} and for $\xi=\frac{3}{16}$ we have the following 
\begin{equation}
\begin{split}
v_{(2)}(\eta_{(2)}) & = \Delta v_{(2)} e^{\ve\frac{27}{127}\eta_{(2)}}\,,\\
w_{(2)}(\eta_{(2)}) & = \Delta w_{(2)} e^{\ve\frac{27}{128}\eta_{(2)}}\,,
\end{split}
\end{equation}
and from the coordinates of the chart $U_{(2)}$ \eqref{eq:charts} we directly find that the Hubble function
$$
\frac{H(a)}{H(a_{0})} = \frac{w_{(2)}(\eta_{(2)})}{v_{(2)}(\eta_{(2)})} = \frac{\Delta v_{(2)}}{\Delta w_{(2)}} = \text{const.}
$$
is constant during the evolution in the vicinity of this state and corresponds to the de Sitter type of evolution. Because of the non-hyperbolicity of this state we are unable to determine stability conditions on the basis of the linearisation procedure and we are forced to take a different path in order to find specific terms leading to the unstable de Sitter state during expansion of the universe.

\section{Unstable de Sitter state at the bifurcation value $\xi=\frac{3}{16}$}

The asymptotic states of dynamical systems of cosmological origin correspond to different phases of evolution of universe. In analysis presented so far we found that there is a special bifurcation value of the parameter $\xi$ for which the de Sitter state exists at infinity of the phase space. The critical point corresponding to this state is of a non-hyperbolic type with one vanishing eigenvalue. In order to find solutions in the vicinity of critical points of such type one can use the standard approach, the centre manifold theorem \cite{Perko:book,Wiggins:book,Hrycyna:2010yv}.  We will take a different approach to emphasise and extract the physical significance of resulting dynamical behaviour in the vicinity of this critical point. 

We begin with the system \eqref{eq:sys_inf_2} together with the following transformation of the phase space variables  
\begin{equation}
\bar{u}=u_{(2)}+\frac{3}{2}\,,\,\, \bar{v}=v_{(2)}^{2} \,,\,\, \bar{w}=w_{(2)}^{2}\,,
\end{equation}
and the dynamical system is
\begin{equation}
\begin{split}
\frac{\ud \bar{u}}{\ud\eta_{(2)}} & = 
	-\bigg(\bar{u}-\frac{3}{2}\bigg)\bigg(\bar{u}-\frac{1}{2}\bigg)\bigg(\bar{w}+\ve\frac{9}
{64}\bigg)
	-\frac{3}{2}\bigg(\bar{u}-\frac{3}{8}\bigg)\bigg( \Omega_{\Lambda,0}\bar{v} +
	\frac{1}{3}\bar{w} - \ve\frac{1}{4}\bar{u}^{2} + \ve\frac{3}{16}\bigg)\,,\\
\frac{\ud \bar{v}}{\ud\eta_{(2)}} & =  -2 \bar{v}\Bigg(
	\bigg(\bar{u}-\frac{7}{2}\bigg)\bigg(\bar{w}+\ve\frac{9}{64}\bigg)
	+ \frac{3}{2}\bigg( \Omega_{\Lambda,0}\bar{v} +
	\frac{1}{3}\bar{w} - \ve\frac{1}{4}\bar{u}^{2} + \ve\frac{3}{16}\bigg)\Bigg)\,,\\
\frac{\ud \bar{w}}{\ud\eta_{(2)}} & = -2\bar{w}\bigg(\bar{u}-\frac{3}{2}\bigg)\bigg(\bar{w}+\ve
\frac{9}{64}\bigg)\,,
\end{split}
\end{equation}
where
\begin{equation}
\frac{\ud}{\ud\eta_{(2)}} = \bar{v}\frac{\ud}{\ud\tau} =
\bigg(\bar{w}+\ve\frac{9}{64}\bigg)\frac{\ud}{\ud\ln{a}}\,.
\end{equation}
Now the critical point under investigations is located at the centre of the coordinate system and we were able to lower the degree of dynamical equation in variables $\bar{v}$ and $\bar{w}$.

\begin{figure}[ht!]
\centering
		\includegraphics[scale=0.75]{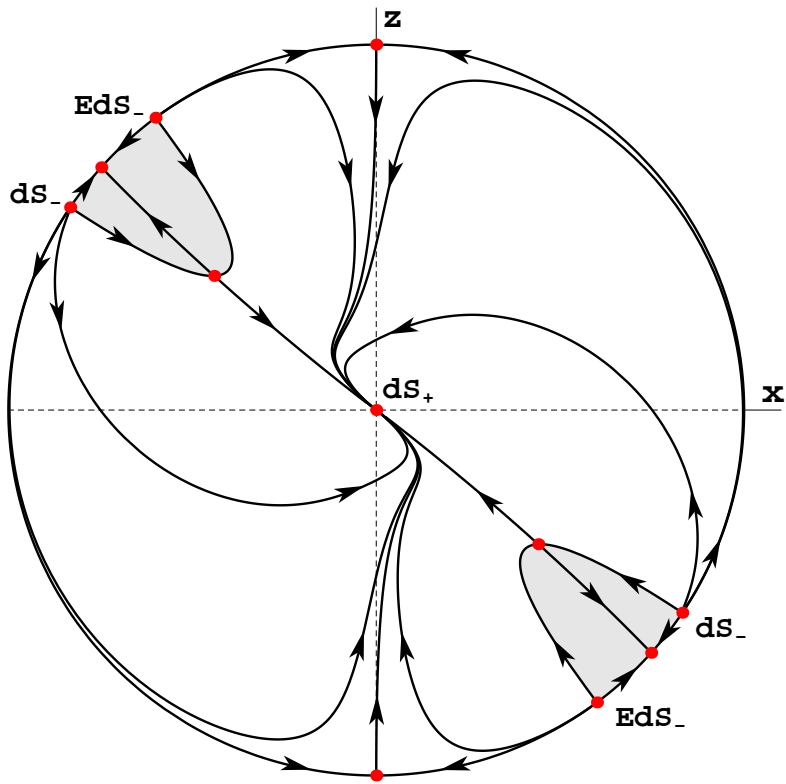}
		\caption{The phase space portrait compactified with a circle at infinity of the model without the barotropic matter and the non-minimal coupling constant $\xi=\frac{3}{16}$. The de Sitter state at the centre dS$_{+}$ is a global attractor for the dynamics. Four critical points at infinity describe unstable de Sitter states dS$_{-}$ and Einstein-de Sitter states EdS$_{-}$. All the trajectories connecting two de Sitter states correspond to possible singularity free evolutional paths.}
\label{fig:8}
\end{figure}

Up to linear terms in the variables we obtain the following dynamical system
\begin{equation}
\begin{split}
\bar{u}' & \approx \frac{9}{16}\Omega_{\Lambda,0}\,\bar{v} -\frac{9}{16}\,\bar{w}\,,\\
\bar{v}' & \approx \ve\frac{27}{64}\,\bar{v}\,,\\
\bar{w}' & \approx \ve\frac{27}{64}\,\bar{w}\,,
\end{split}
\end{equation}
where the prime denotes differentiation with respect to the $\eta_{(2)}$ time and the last two 
equations are now decoupled. This linearised system can be easily solved, resulting in
\begin{equation}
\begin{split}
\bar{u}(\eta_{(2)}) & = \bar{u}^{(i)}-\ve\frac{4}{3}\big(\Omega_{\Lambda,0}\bar{v}^{(i)}-\bar{w}
^{(i)}\big)\left(1-\exp{\left(\ve\frac{27}{64}\eta_{(2)}\right)}\right)\,,\\
\bar{v}(\eta_{(2)}) & = \bar{v}^{(i)}\exp{\left(\ve\frac{27}{64}\eta_{(2)}\right)}\,,\\
\bar{w}(\eta_{(2)}) & = \bar{w}^{(i)}\exp{\left(\ve\frac{27}{64}\eta_{(2)}\right)}\,.
\end{split}
\end{equation}
We need to note that the critical point corresponding to the de Sitter state is located at the centre of the coordinate system. Above solutions for a generic set of initial conditions do not lead to this point. We need to take some open and dense subset of possible initial conditions satisfying  
\begin{equation}
\label{eq:inicond}
\bar{u}^{(i)}=\ve\frac{4}{3}\big(\Omega_{\Lambda,0}\bar{v}^{(i)}-\bar{w}^{(i)}\big)\,.
\end{equation}
The effective equation of state parameter up to linear terms in the phase space variables is
\begin{equation}
w_{\textrm{eff}} \approx \frac{1}{3}-\frac{\Omega_{\Lambda,0}\bar{v}+\frac{1}{3}\bar{w}+\ve
\frac{3}{16}}{\bar{w}+\ve\frac{9}{64}} = -1 - \frac{\Omega_{\Lambda,0}\bar{v}-\bar{w}}{\bar{w}+
\ve\frac{9}{64}}\,,
\end{equation}
and inserting above solutions and taking the limit $\ve\eta_{(2)}\to-\infty$ we obtain pure de Sitter state. From the other hand the energy conservation condition
\begin{equation}
\Omega_{m,0}\left(\frac{a}{a_{0}}\right)^{-3} = h^{2} - \Omega_{\Lambda,0}+\ve\frac{1}{8}x^{2}-
\ve\frac{9}{8}(x+z)^{2} \ge 0\,
\end{equation}
calculated for the specific initial conditions \eqref{eq:inicond}
indicate that the effects of the barotropic dust matter must be negligible in order to obtain the de Sitter state in form of an unstable node. In the upcoming paper \cite{Hrycyna:2015b} we investigate in details stability conditions and asymptotic forms of scalar field potential functions giving rise to a singular free type of cosmological evolution. On figure \ref{fig:8} we presented the phase space portrait of the dynamics of the model for the canonical scalar field $\ve=+1$ without the barotropic matter and $\xi=\frac{3}{16}$. There is open and dense subspace of initial conditions leading to a non-singular evolutional path from the unstable de Sitter state to the stable one.

The obtained possible evolutional path can indicate for some generalisations of the seminal Starobinsky type of evolution \cite{Starobinsky:1980te}, as well as obtained evolution was not expected nor designated from the very beginning \cite{Mukhanov:1991zn,Brandenberger:1993ef} and may suggest some additional symmetry in the matter sector of the theory.
 
\section{Beyond linear approximation}

Dynamical systems theory gives us great opportunity to investigate qualitative behaviour of 
dynamical systems in vicinity of stationary states. The linearisation procedure applied to 
hyperbolic critical points enables us to obtain description of dynamics up to some 
characteristic time interval assuming some initial conditions. We were using this 
approach extensively in the previous sections. Unfortunately sometimes it is not enough to 
recover physical description of the processes under investigations. We need to go beyond the 
linear approximation. 

Within local dynamical systems theory in addition to the linearisation procedure we are equipped with the notion of invariant manifolds \cite{Perko:book,Wiggins:book}. 

In section \ref{sec:2} we stated that the system \eqref{eq:dynsys1} is equipped with two invariant manifolds which reduce dimensionality of the system. One invariant manifold corresponds to the static Einstein universe with the vanishing Hubble function and the second one corresponds to the de Sitter universe with the constant value of the Hubble function. In what follows we use the latter in order to obtain the exact solutions of reduced dynamics and then we obtain the Hubble function beyond the linear approximation.

Let us begin with the dynamical system \eqref{eq:dynsys1} in full form
\begin{equation}
\label{eq:dynsysnew}
\begin{split}
	\frac{\ud x}{\ud \ln{a}} &= -3x-6\xi
	z\left(\frac{\dot{H}}{H^{2}}+2\right)\,,\\
	\frac{\ud z}{\ud\ln{a}} &= x+z\frac{\dot{H}}{H^{2}}\,, \\
     \frac{\ud (h^{2})}{\ud\ln{a}} &= 2 (h^{2}) \frac{\dot{H}}{H^{2}}\,,
\end{split}
\end{equation}
where now we treat $h^{2}$ as the phase space variable and the acceleration equation is
\begin{equation}
        \frac{\dot{H}}{H^{2}} = -2
	+\frac{3}{2}\frac{\Omega_{\Lambda,0}+\frac{1}{3}h^{2}-
	\ve(1-6\xi)x^{2}-\ve2\xi(x+z)^{2}}{h^{2}-\ve6\xi(1-6\xi)z^{2}}\,.
\end{equation}

On the invariant manifold $\frac{\dot{H}}{H^{2}}=0$ corresponding to the de Sitter state the 
dynamical system \eqref{eq:dynsysnew} reduces to a simple linear system
\begin{equation}
\label{eq:dyn_lin}
\begin{split}
	\frac{\ud x}{\ud \ln{a}} & = -3x -12\xi z\,, \\
	\frac{\ud z}{\ud \ln{a}} & =  x\,.\\
\end{split}
\end{equation}

This linear system of equations can be directly integrated. We make the following change of 
variables
\begin{equation}
\begin{split}
	x & = \lambda_{1}p + \lambda_{2}q\,,\\
	z & = p + q\,,
\end{split}
\end{equation}
where $\lambda_{1}=-\frac{3}{2}-\frac{3}{2}\sqrt{1-\frac{16}{3}\xi}$,
$\lambda_{2}=-\frac{3}{2}+\frac{3}{2}\sqrt{1-\frac{16}{3}\xi}$ are the eigenvalues of the constant 
matrix of the linear system \eqref{eq:dyn_lin}. Then we obtain
\begin{equation}
	\begin{split}
		\frac{\ud p}{\ud \ln a} & = \lambda_{1} p\,,\\
		\frac{\ud q}{\ud \ln a} & = \lambda_{2} q\,.
	\end{split}
\end{equation}
and the solutions are
\begin{equation}
\label{eq:sol_inv}
	\begin{split}
		p(a)& = p(a_{0})\left(\frac{a}{a_{0}}\right)^{\lambda_{1}}\,,\\
		q(a)& = q(a_{0})\left(\frac{a}{a_{0}}\right)^{\lambda_{2}}\,.
	\end{split}
\end{equation}
These solutions are the exact solutions of the reduced system, thus the initial conditions can 
be arbitrary chosen and we choose them at the present epoch. The quantities $p(a_{0})$ and 
$q(a_{0})$ denote values of the phase space variables at $a_{0}=1$ value of the scale factor.  

The third equation in the system \eqref{eq:dynsysnew} expanded in to the Taylor series up to 
linear terms in $z$ variable is
\begin{equation}
\label{eq:accel_fin}
\frac{\ud h^{2}}{\ud N} \approx 3\,\Omega_{\Lambda,0} - 3 h^{2} - \ve3\Big((1-4\xi)x^{2}+4\xi x 
z\Big)\,,
\end{equation}
where $\ud N = \ud\ln{a}$.

The condition $z^{2}\approx0$ expressed in the variables $p$ and $q$ leads to
$$
z^{2}=(p+q)^{2}=p^{2}+2 pq +q^{2}\approx0\,.
$$
For the remaining variables appearing in equation \eqref{eq:accel_fin} we obtain
\begin{equation}
\begin{split}
x^{2}&=\frac{9}{4}\left((p+q)+\sqrt{1-\frac{16}{3}\xi}(p-q)\right)^{2} \\
&\approx \frac{9}{2}\sqrt{1-\frac{16}{3}\xi}\left(p^{2}-q^{2}\right)+
\frac{9}{2}\left(1-\frac{16}{3}\xi\right)\left(p^{2}+q^{2}\right)\,,
\end{split}
\end{equation}
and
\begin{equation}
xz = -\frac{3}{2}\left((p+q)+\sqrt{1-\frac{16}{3}\xi}(p-q)\right)\left(p+q\right) 
\approx -\frac{3}{2}\sqrt{1-\frac{16}{3}\xi}\left(p^{2}-q^{2}\right)\,.
\end{equation}
Finally, inserting those approximation in to the equation \eqref{eq:accel_fin}, the dynamical 
equation for the Hubble function is given by
\begin{equation}
\label{eq:dyn_hub}
\frac{\ud h^{2}}{\ud N} \approx -3h^{2}+3\Omega_{\Lambda,0}+
\ve\,9\left(1-\frac{16}{3}\xi\right) 
\Big((\lambda_{1}+6\xi)\,p(N)^{2}+(\lambda_{2}+6\xi)\,q(N)^{2}\Big)\,,
\end{equation}
where the functions $p(N)$ and $q(N)$ are the exact solutions of the system \eqref{eq:dyn_lin} 
reduced to the invariant manifold of the de Sitter state and are given by \eqref{eq:sol_inv}.

Now we are ready to integrate the dynamical equation \eqref{eq:dyn_hub} and the result is
\begin{equation}
\begin{split}
\left(\frac{H(a)}{H(a_{0})}\right)^{2} 
\approx & \,\,\Omega_{\Lambda,0}+\Omega_{m,0}\left(\frac{a}{a_{0}}\right)^{-3}-\\
&-\ve\,3\sqrt{1-\frac{16}{3}\xi}\left(\frac{a}{a_{0}}\right)^{-3}
\Bigg((\lambda_{1}+6\xi)\,p_{0}^{2}\left(\frac{a}{a_{0}}\right)^{-3\sqrt{1-\frac{16}{3}\xi}}+\\
&\hspace{4.3cm}+(\lambda_{2}+6\xi)\,q_{0}^{2}\left(\frac{a}{a_{0}}\right)^{+3\sqrt{1-\frac{16}{3}\xi}}\Bigg)\,.
\end{split}
\end{equation}
The first two terms in this solution correspond to the standard $\Lambda$CDM model and remaining 
term can be interpreted as an extension depending on the non-minimal coupling $\xi$ and the current values of the phase space variables $p(a_{0})$ and $q(a_{0})$. For $\xi>\frac{3}{16}$ this additional term reduces to an oscillatory behaviour with dumping.
We need to note that in the first approximation beyond linear investigated in this section, the 
obtained Hubble function reduces to the $\Lambda$CDM model with the assumption 
$\xi=\frac{3}{16}$.

\section{Conclusions}

In this paper the global dynamics of the simplest case of the non-minimally coupled scalar field cosmology was investigated in the 
details in the framework of dynamical system methods. The dynamics is simple in this sense that in the flat Friedmann-Robertson-Walker model with the dust barotropic matter can be reduced to the three-dimensional dynamical system. The complexity of the dynamics is revealed while 
considering different values of the non-minimal coupling constant $\xi$. The bifurcation 
values of this parameter were identified and the global phase analysis for representative cases was performed.

It is interesting that dynamics with the minimal $\xi=0$ and the conformal $\xi=\frac{1}{6}$ coupling constants are equivalent to the $\Lambda$CDM model with additional terms which behave like a stiff matter and a radiation matter, respectively. In the case of conformal coupling for the canonical scalar field $\ve=+1$ this additional term behaves like an ordinary radiation while for the phantom scalar field $\ve=-1$ like a dark radiation with a negative energy density. The corresponding Hubble functions were obtained in the exact form are the models are therefore integrable by elliptic functions.

We demonstrated that for the case of the vanishing potential function of the scalar field the right hand sides of the dynamical system are 
homogeneous functions of the third degree which gives the possibility of lowering the dimension of the phase space by one using the projective coordinates.

We have shown that the constant $\xi$ describing the non-minimal coupling between the curvature and the scalar field is the crucial ingredient not only during the inflationary epoch \cite{Faraoni:2000gx} but also is describing the current accelerated expansion of the universe. 
While the special bifurcation value of the non-minimal coupling $\xi=3/16$ is distinguished by the dynamics of the model and can lead to a singularity free cosmological evolution may suggest some additional symmetry in the matter sector of the theory.

\acknowledgments{O.H. is grateful to organisers of the XIIth School of Cosmology, September 15--20, 
2014 held at Institut d'{\'E}tudes Scientifiques de Carg{\`e}se, for invitation and opportunity 
to present a part of this work and to Kei-ichi Maeda for comments. 

The research of O.H. was funded by the National Science Centre
through the postdoctoral internship award (Decision No.~DEC-2012/04/S/ST9/00020), M.S. 
was supported by the National Science Centre through the OPUS 5 funding scheme 
(Decision No.~DEC-2013/09/B/ST2/03455).}

\bibliographystyle{JHEP}
\bibliography{../moje,../darkenergy,../quintessence,../quartessence,../astro,../dynamics,../standard,../inflation,../sm_nmc,../singularities,../math}

\end{document}